\RequirePackage{fix-cm}
\documentclass{svjour3}                     \smartqed  \usepackage{graphicx}
\usepackage[pdftex,svgnames,dvipsnames]{xcolor}
\usepackage{enumitem}
\usepackage{tikz}
\usepackage{todonotes}
\usepackage{pgfplots}
\usepackage{sparklines}
\pgfplotsset{compat=newest}
\usepgfplotslibrary{groupplots}
\usepackage{tikzscale}
\pgfplotsset{
    every tick label/.append style={scale=0.75},
    every axis/.append style={
}
}
\usetikzlibrary{positioning,fit,shapes,decorations.shapes,decorations.markings}
\usepackage{xstring}
\usepackage{xfp}
\usepackage{expl3}
\usepackage{xparse}
\usepackage[normalem]{ulem}
\usepackage{wasysym}
\usepackage{amssymb}
\usepackage{adjustbox}
\usepackage{natbib}
\def\:{\hskip0pt}
\usepackage{booktabs,tabularx}

\newcommand*{\priority}[1]{{\small{
\begin{tikzpicture}[scale=0.10] \draw (0,0) circle (1); \fill[fill opacity=1,fill=black] (0,0) -- (90:1) arc (90:90-#1*3.6:1) -- cycle;
\end{tikzpicture}}}}

\newcommand*{\priorityss}[1]{{\scriptsize{
\begin{tikzpicture}[scale=0.10]\draw (0,0) circle (1);
    \fill[fill opacity=1,fill=black] (0,0) -- (90:1) arc (90:90-#1*3.6:1) -- cycle;
\end{tikzpicture}}}}

\usetikzlibrary{positioning,fit,shapes,decorations.shapes,decorations.markings}

\usepackage[colorlinks=true, allcolors=blue]{hyperref}
\usepackage{multirow}
\usepackage[dvipsnames]{xcolor}

\newcounter{cccnt}
\newcounter{bdscnt}
\newcounter{tzcnt}
\newcounter{bccnt}
\newcounter{wmcnt}

\newcommand{\TITLE}{Empirical Studies on Adversarial Reverse Engineering with Students}
    
\begin{document}

\title{\TITLE }

\author{Tab (Tianyi) Zhang \and Bjorn De Sutter \and Christian Collberg \and Bart Coppens \and Waleed Mebane}

\institute{Tab Zhang \at
              Technologiepark-Zwijnaarde 126, 9052 Gent, Belgium\\
              Tel.: +32 498018989\
              \email{tab.zhang@ugent.be}           \and
           Bjorn De Sutter \at
            \email{Bjorn.DeSutter@UGent.be}  
}

\date{Article submitted to Empirical Software Engineering}

\maketitle

\begin{abstract}
Empirical research in reverse engineering and software protection is crucial for evaluating the efficacy of methods designed to protect software against unauthorized access and tampering. However, conducting such studies with professional reverse engineers presents significant challenges, including access to professionals and affordability. This paper explores the use of students as participants in empirical reverse engineering experiments, examining their suitability and the necessary training; the design of appropriate challenges; strategies for ensuring the rigor and validity of the research and its results; ways to maintain students’ privacy, motivation, and voluntary participation; and data collection methods. We present a systematic literature review of existing reverse engineering experiments and user studies, a discussion of related work from the broader domain of software engineering that applies to reverse engineering experiments, an extensive discussion of our own experience running experiments ourselves in the context of a master-level software hacking and protection course, and recommendations based on this experience. Our findings aim to guide future empirical studies in RE, balancing practical constraints with the need for meaningful, reproducible results.

\end{abstract}

\section{Introduction}
\label{sec:intro}

Empirical research into reverse engineering and software protection is essential for understanding and evaluating the effectiveness and robustness of methods used to protect software from unauthorized access, tampering, and piracy.  The significance of this research area grows as software continues to play a critical role in both industry applications and areas requiring stringent security measures. Although there is extensive theoretical research on software protection and reverse engineering, empirical studies frequently encounter methodological difficulties, and studies with human subjects are relatively limited: Among the 571 papers covered in their survey on evaluation methodologies in software protection research, \cite{desutter2024evaluation} identified only 21 papers reporting on human subject studies, and they discussed numerous recurring methodological challenges with the used sample sets, sample treatments, and measurements.

In this paper, we focus on empirical reverse engineering experiments using students as participants, aiming to provide insights into the strengths, limitations, and challenges of using such participants to approximate the performance of professional reverse engineers. To contextualize our work, we define the following key terms that will be used throughout this paper:

\begin{itemize}

\item \textbf{Machine At The End (MATE)} will refer to the attack model where an adversary has complete control over a software implementation in the form it was distributed to its users or targets. For example, the adversary could have gained access to a binary executable that a vendor ships to its customers, or to a piece of malware with which victims are targeted. The adversary is then able to examine and/or tamper with that piece of software at will on their own machine, over which they have full control.

\item \textbf{Software Protection} (SP) will refer to the transformation of software to enhance confidentiality
and integrity of its components under the MATE attack model.\footnote{For the purposes of this study, we also consider standard compilation as a protection technique. While compilation is primarily used
to convert source code into an executable binary, many software providers
distribute compiled binaries rather than source code explicitly to maintain
confidentiality. Therefore, compiled binary code, even if not intentionally obfuscated,
is treated as protected.} SPs can be deployed defensively, e.g., by a vendor wanting to protect a software license management component in a benign commercial software application (a.k.a.\ goodware), as well as offensively, e.g., by a malware author trying to defeat detection and analysis of their malicious software.

\item Unless explicitly mentioned otherwise, \textbf{Reverse Engineering} (RE) will refer to adversarial examination of such a software implementation to identify its components and their interrelationships and to create representations at a higher level of abstraction. In this adversarial context, the goal of the RE activities is to obtain unauthorized access to, and use of, components (a.k.a.\ assets) of the software implementation.  In other words, they apply RE techniques to compromise the confidentiality and/or integrity of assets embedded in software. Just like in the case of SP, these RE activities can be defensive, e.g., in case of a security analyst RE a piece of malware~\citep{Wong2021inside,Wong2024}, or offensive, e.g., in case of a criminal wanting to circumvent a license enforcement mechanism in a piece of goodware~\citep{Ceccato2019}. As observed by \cite{schrittwieser2016protecting}, the concrete goals of such RE activities are (i) finding the location of data, (ii) finding the location of program functionality, (iii) extracting code fragments, or (iv) understanding the program. The examination techniques used by such adversarial reverse engineers include targeted attacks explicitly aimed at defeating SPs, as well as many code analysis techniques they share with friendly RE practices in support of, e.g., software certification or security auditing starting from binaries.

In this paper, we will hence use the term RE for only a subset of the activities considered by \cite{chikofsky2002reverse}. They focus almost exclusively on benign RE objectives such as aiding maintenance and new development starting from existing software. Their scope hence includes examination starting from more abstract software representations, including source code. By contrast, we focus exclusively on adversarial RE in the MATE scenario.

\item \textbf{Software Engineering} (SE) will be used as a shorthand for \textit{forward engineering} activities such as requirements specification, design, programming, testing, etc., as well as for RE activities that start from source code. This includes, e.g., source code comprehension, source code restructuring, design recovery, etc. In other words, when we will use the term SE, we will use it to denote all aspects of SE except what falls under the above definition of RE.
\end{itemize}

SE and RE are obviously related. For example, reverse engineers can use decompilers to transform binary executables into source code representations and then use code comprehension skills and methods that regular software developers also use to understand them. The required skills and experience and the cognitive processes are certainly not identical~\citep{Votipka2020}, in particular in the case when the executables have been obfuscated~\citep{hansch2018programming}, yet they overlap sufficiently to make insights from empirical SE research valuable for empirical RE research as well. This intersection is especially important given the limited body of empirical work dedicated specifically to RE.

Previous studies, referenced in Table~\ref{tab:literature_review_part1} and discussed in detail in Section~\ref{sec:literature}, indicate that there is a substantial interest in empirical RE experimentation, and that these experiments are driven by a diverse set of motives. However, there are multiple issues in how to conduct such studies, in particular with respect to the selection of student participants rather than professionals. Student participants offer significant advantages such as accessibility, affordability, and ease of management~\citep{falessi2018empirical,feitelson2022considerations}, yet introduce concerns about external validity due to their differing experience levels and competencies compared to professional reverse engineers~\citep{hansch2018programming,falessi2018empirical}.

With this paper, we offer four contributions beyond existing work:
\begin{enumerate}
\item We present a systematic literature review on empirical RE experiments and RE user studies, focusing on different motivations for such research; how students were relied upon as participants; how the students were motivated and trained; how privacy and ethics were handled; which data collection methods were used; which human competencies appeared as relevant for software RE; and how the experiments were designed. 
\item We discuss related work, in particular how existing experience and insights from researchers conducting empirical SE experiments with and without students, also applies to RE experiments. 
\item We report our own experience in designing and running RE experiments with student participants, which we did rather differently than how it has been done in the research reported in the reviewed literature. We discuss extensively how our experiments were aligned with the activities in a master level course on software RE; how we tried to motivate students to participate in our experiments; how we designed our experiments to manage the heterogeneity of participating students; how we designed our experiments to minimize the potential threats to external validity due to potential gaps between student competencies and those of experts; how we trained and prepared the students for our experiments; how we collected extensive and detailed data during our experiments; and how to handle privacy and ethics requirements properly to ensure that students provide truly voluntary (and informed) consent. 
We also show what outcomes we obtained, in particular that running such experiments successfully remains difficult. 
\item Finally, we present a number of recommendations that, together with the reporting on our own experiments, we hope will enhance the rigor, reproducibility, and external validity of empirical research in software RE.
\end{enumerate}

The remainder of this paper is structured as follows. Section~\ref{sec:literature} presents a systematic literature review on human experiments and user studies in RE. Next, Section~\ref{sec:related_work} discusses considerations expressed by other authors regarding empirical SE and RE experiments, focusing mostly but not exclusively on the use of students. This is followed by a presentation of our own experience and opinions on designing and running RE experiments with students in Section~\ref{sec:our-experience}. Section~\ref{sec:recommend} provides recommendations for future RE experiments involving students, after which Section~\ref{sec:conclusions} draws conclusions.

 \section{Reverse Engineering Experiments and Studies - Literature Review}
\label{sec:literature}

In this section we report the results of a literature study on RE user studies and experiments using human subjects. In Section~\ref{sec:literature:methodology} we discuss how we chose which papers to include in the survey, how data was extracted from these papers, and how this data is presented. In subsequent sections we discuss various aspects of those experiments and studies as reported by their authors.

Relevant related work, such as a more general discussion on empirical research with students conducted outside the field of RE will be discussed separately in Section~\ref{sec:related_work}. This separation aims to make a clear distinction between research conducted in RE on the one hand, and claims made outside the field on the other.

\subsection{Methodology}
\label{sec:literature:methodology}
We first discuss how we selected publications, how we extracted data, and how we will present the results.

\subsubsection{Literature Selection}

The goal of our systematic literature review is to capture all relevant publications that report how researchers have been trying to answer research questions relevant to adversarial RE with empirical experiments and user studies, and then to learn from those publications.

As a first information source for this review, we started with \textbf{21} protected software experiments that \cite{desutter2024evaluation} listed in their 2025 571-paper survey on evaluation methodologies of SP research. We refer to that survey for a detailed description of their 571 paper retrieval and selection process~\citep{desutter2024evaluation}. Here we summarize it as consisting of three subsets: (i) papers on SPs included in three pre-existing surveys~\citep{schrittwieser2016protecting,2018_diversification_and_obfuscation,ebad2021measuring}; (ii) papers in the ACM Digital Library, IEEE Xplore, and SpringerLink in the period 2016-2022 including the term ``obfuscation'' and variants thereof in the title; (iii) papers by 28 authors known for their contributions in the field of SP. 

Only 19 of those 21 papers reported on RE activities. The other two involve humans for other purposes~\citep{2019resilient,obf_googleplay}, so we excluded them entirely from our survey. 

Because adversarial reverse engineers rely to a large degree on techniques also used for other binary RE goals, we expanded our literature review beyond those 19 papers using a protocol aligned with Kitchenham et al.’s three stages (planning, conducting, reporting) for systematic literature reviews~\citep{kitchenham2009systematic}.

As \emph{information sources} for this extension, we thus combined the 19 papers from \cite{desutter2024evaluation} with an indexed database search on Web of Science (WoS), and with forward citation chasing on Google Scholar.

For our \emph{search strategy} we queried WoS with the following topic string to capture user studies and human experiments in RE and malware analysis:

{
\small
\begin{verbatim}
   ((
       "reverse engineer"        OR "reverse engineers" OR "reverse engineering"  
    OR "malware analysis"     OR "malware analyst"      OR "malware analysts"
   ) AND (
       "controlled experiment".  OR "controlled experiments" 
    OR "user study"              OR "user studies"            
    OR "user assessment"         OR "user assessments" 
    OR "user feedback"           OR "user research"
    OR "user observation".       OR "interview"
    OR "interviews"              OR "questionnaire"    OR "questionnaires"
   ))
\end{verbatim}
}

The query returned \textbf{81} records. After deduplication against the 19‑paper seed set and screening (criteria below), we retained \textbf{9} papers.

Using the combined list of \textbf{26} papers (19 seed + 9 WoS additions), we performed forward citation chasing (a.k.a. snowballing) in Google Scholar. We screened all citing items; this resulted in \textbf{8} additional papers not in the prior sets.

Screening for \emph{study selection and eligibility} followed Kitchenham’s guidance in two passes~\citep{kitchenham2009systematic}:
\begin{enumerate}
  \item \textit{Title/abstract screen:} exclude (a) non‑empirical papers (algorithms/tools without human participants; position/vision papers), (b) papers not about human-\:in-the‑loop RE or malware analysis, or (c) non–peer‑reviewed items (theses, blog posts, preprints without peer review, extended abstracts without a peer-\:reviewed companion).
  \item \textit{Full‑text screen:} include studies that (a) involve human participants (students and/or experts) and (b) report a controlled experiment or user/field study—or a practitioner survey—focused on RE tasks, RE tools, SP strength, or analyst workflows. Where duplicate versions existed (e.g., tech report + published paper), we kept the publication that appeared in the peer‑reviewed venue.
\end{enumerate}
We imposed no publication‑year limit and accepted peer‑reviewed journals, conferences, and workshops. Malware analysis studies were included if they involved RE activities (e.g., code comprehension, decompilation, code or trace inspection).

In summary, starting from the 19 seed papers our WoS search added \textbf{9} and forward snowballing \textbf{8} papers. The resulting \textbf{36} papers were kept for data extraction and analysis.

\subsubsection{Data Extraction}

Following the \emph{data extraction} strategy of \cite{kitchenham2009systematic}, we designed a structured extraction sheet and a multi‑person coding workflow. For each paper, at least three authors individually extracted data into an extraction sheet and an agreed upon coding workflow. A senior author compared and aggregated the individually extracted data. Where divergences in the extracted data were identified, they were resolved by discussion between all authors to consensus.

\subsubsection{Result Presentation}

Five of the 36 analyzed papers~\citep{2008twoardsexperimental,2009assessment,obf_optvialangmods,Ceccato2017,exp_eval_obf_against_reveng} reported on experiments also reported in other papers~\citep{2014afamily,liu2017stochastic,Ceccato2019,2019impact}. Since we are interested in surveying conducted empirical RE experiments and how papers report on them rather than how many papers report on them, we will omit those five papers when we list papers in tables in which we count the occurrence of different aspects of experiments. In other words, we will de-duplicate these papers. If the information retrieved from a de-duplicated paper complements the information retrieved from a still included paper on the same experiment, we obviously still extracted that information, and report it in the tables. 

The \textbf{31} papers remaining after de-duplication and the extracted information are listed in Tables~\ref{tab:literature_review_part1},~\ref{tab:literature_review_part2}, and~\ref{tab:literature_review_part3}, to which tables~\ref{tab:literature_review_part1_legend},~\ref{tab:literature_review_part2_legend}, and~\ref{tab:literature_review_part3_legend} are the respective legends. Table~\ref{tab:literature_review_part1} lists the types of studies, the researchers' motivations, the types and numbers of participants, the file format considered in the studies, and, for student participants, the training they got. Table~\ref{tab:literature_review_part2} lists more detailed features of the targeted programs and the tasks of the experiments. Table~\ref{tab:literature_review_part3} lists the data collection methods used in each study. The next sections discuss various aspects of the surveyed experiments and studies.

\begin{table}
\scriptsize
\centering
\setlength{\tabcolsep}{1.5pt} \begin{tabular}{l|c|c|c|c|p{4.7cm}}
\textbf{Publication} & \rotatebox{90}
{\textbf{Type}} & \rotatebox{90}
{\textbf{Motivation}} & \rotatebox{90}
{\textbf{Participants}} & \rotatebox{90}
{\textbf{Format}} & \textbf{For students: received training} \\ 
\toprule
\multicolumn{6}{c}{}\\[-0.5em]
\multicolumn{6}{l}{\emph{(A) Experiments to which bachelors/masters students participated}}\\[-1.7em]
\multicolumn{6}{c}{}\\
\multicolumn{6}{c}{}\\

\cite{sutherland2006reverse} & E & M4,5 & 9\!\priority{0} 1\!\priority{25} & N & $*$ School of computing program + unspecified reading list\\ 
\cite{2014afamily} & $\textrm{E}_p$ & M3,4 & 22\!\priority{25} 52\!\priority{0} & $\textrm{B}_d$ & $*$ SE course + obfuscation lectures\\ 
\cite{2014another} & $\textrm{E}_p$ & M3 & 12\!\priority{0} & S & $*$ had completed a RE course \\ 
\cite{viticchie2016assessment} & $\textrm{E}_p$ & M1,3 & 1\!\priority{25} 14\!\priority{0} & S & $*$ C programming and SE, no tampering or RE training\\ 
\cite{yakdan2016} & $\textrm{E}_m$ & M2,4 & 21\!\priority{25} 9 \!\priority{100} & M & $*$  completed malware boot camp\\ 
\cite{liu2017stochastic} & $\textrm{E}_p$ & M3 & 10\!\priority{25} 10\!\priority{0} & S & $>2$ years JS experience \\ 
\cite{hansch2018programming} & $\textrm{E}_p$ & M1,3,4 & 2\!\priority{25} 64\!\priority{0} & $\textrm{B}_d$ & $*$ basic knowledge of Java and Eclipse, 24\% had obfuscation course \\ 
\cite{kuang2018enhance} & $\textrm{E}_p$ & M3 & 2\!\priority{25} 13\!\priority{0} & N & $*$ cybersec. program + hands-on RE experience \\ 
\cite{2020creative} & $\textrm{E}_p$ & M3,4 & 22\!\priority{50}\!\priority{100}\!\priority{0} & S & IT/CS with Java knowledge \\
\cite{2020splitting} & $\textrm{E}_p$ & M1,3,4 & 87\!\priority{0} & S & $*$ taking security course with C  \\ 
\cite{henry2020sensorre} & S,E & M2 & 5\!\priority{100} 11\!\priority{50} & N & $*$ passed graduate course on binary RE with IDA Pro, 10 minute guided training on Binary Ninja\\
\cite{2021inputoutput} & $\textrm{E}_p$ & M2,3 & 5\!\priority{00} & N,$\textrm{N}_d$ & $*$ hands-on experience in software reverse analysis\\ 
\cite{Mantovani2022} & E & M1 & 72\!\priority{0}\!\priority{50}\!\priority{75}\!\priority{100} & N & had taken binary analysis or RE course \\
\midrule

\multicolumn{6}{c}{}\\[-0.5em]
\multicolumn{6}{l}{\emph{(B) Experiments without students}}\\[-1.7em]
\multicolumn{6}{c}{}\\
\multicolumn{6}{c}{}\\
\cite{dongleprot_secmeasure} & $\textrm{E}_p$ & M3 & 3\!\priority{100} & N \\ 
\cite{bryant2012eliciting} & E & M1 & 4\!\priority{100} & N \\ 
\cite{2016comparing} & $\textrm{E}_p$ & M3 & 1\!\priority{75} & N \\ 
\cite{wagner2017knowledge} & $\textrm{E}_m$ & M2 & 6 \!\priority{100} & M\\
\cite{Ceccato2019} & $\textrm{E}_p$ & M1,3 & 6\!\priority{100} 1\!\priority{50} & N \\ 
\cite{2019impact} & $\textrm{E}_p$ & M3 & 14\!\priority{75} & B \\
\cite{matzen2021effects} & E & M1,2 & 13\!\priority{100} 7\!\priority{75} & $\textrm{N}_d$, S \\
\cite{nyre2022task} & E & M1 & 12\!\priority{100} & N \\
\cite{Yamagishi2025} & $\textrm{E}_m$ & M1 & 6\!\priority{50} 5\!\priority{100} & M\\ 
\midrule

\multicolumn{6}{c}{}\\[-0.5em]
\multicolumn{6}{l}{\emph{(C) Surveys/interviews with professionals}}\\[-1.7em]
\multicolumn{6}{c}{}\\
\multicolumn{6}{c}{}\\
\cite{bryant2013top} & S & M1 & 5\!\priority{100} & N \\ 
\cite{baldwin2016} & S & M2,6 & 21\!\priority{100} & N+M\\
\cite{Votipka2020} & S & M1 & 16\!\priority{50}  & N \\ 
\cite{henry2020exploring}  & S & M1,2 & 5\!\priority{100} & N+M \\
\cite{Wong2021inside} & S & M1 & 21\!\priority{100} & M \\ 
\cite{Wong2024} & S & M1 & 24\!\priority{100} & M\\ 
\midrule

\multicolumn{6}{c}{}\\[-0.5em]
\multicolumn{6}{l}{\emph{(D) Description of participants is vague}}\\[-1.7em]
\multicolumn{6}{c}{}\\
\multicolumn{6}{c}{}\\
\cite{malware_visualcomp} & $\textrm{E}_m$ & M2 & 6 & M \\ 
\cite{burk2022} & E & M1,2,5 & 425 & N \\ 
\cite{bergmann2025potential} & $\textrm{E}_p$ & M3 & 10 & N \\

\bottomrule

\end{tabular}

\caption{Results of the literature review, part 1. The legend is given in Table~\ref{tab:literature_review_part1_legend}.}
\label{tab:literature_review_part1}
\end{table}

\begin{table}
\begin{tabular}{lp{9cm}}
\textbf{Column} & \textbf{Legend} \\\toprule
\textbf{Type} & 
   The paper reports results from a study where participants
   
   \begin{tabular}{lp{8cm}}
      $E_p$ & performed an assignment on protected goodware samples. \\
      $E_m$ & performed an assignment on malware samples. \\
      $S_p$ & answered questions on their practices, competences, etc. on protected goodware samples;\\
      $S_m$ & answered questions on their practices, competences, etc. on protected goodware samples;
   \end{tabular}
   
   When both $E$ and $S$ are indicated, this means that an experiment was done and a separate survey or interview (i.e., not about the experiment).
   \\
    \textbf{Motiva\-tion} & 
       Motivations M1--4 as defined in Section~\ref{sec:motivation}, i.e., classes of research questions the paper aimed to respond.
    \\
     \textbf{Partici\-pants} & 
     How many people participated at various levels of expertise to the reported experiment or survey: 
     
     \begin{tabular}{lp{8cm}}
         \protect\priorityss{0} & 
            Bachelors and masters students (appearing at top) \\ 
         \protect\priorityss{25} &
            PhD and postgraduate students not expert in SP or RE \\
         \protect\priorityss{50} &
            Experienced amateurs/students in SP or RE, or professionals with less than one year of experience \\
         \protect\priorityss{75} &
            Professional programmers \\
        \protect\priorityss{100} &
           Professional security experts and pen testers with more than one year of experience.
    \end{tabular}
    \\
     \textbf{Format} & 
     Program format on which the experiments were conducted or that was the subject of the practice studied in the user study:
     
     \begin{tabular}{lp{8cm}}
        N & Native binary \\
        S & Source code \\
        B & (Java, CIL, ...) Bytecode \\
        M & (native) Malware.
    \end{tabular}
    An index $d$ indicates that the code given to the participants had already been decompiled by the researchers. 
    \\
     \textbf{Training} & 
        For experiments involving students, the last column describes what training bachelors/masters student participants in experiments received, as described in the papers. A $*$ indicates that the students where found, and hence trained, at the authors' host institutions. No $*$ means that the paper is unclear about whether the students come from those host institutions.\\\bottomrule
\end{tabular}
\caption{Legend for Table~\ref{tab:literature_review_part1}.}
\label{tab:literature_review_part1_legend}
\end{table}

\subsection{Motivations of Empirical Experiments and User Studies}
\label{sec:motivation}
Many different types of research questions can be answered with empirical RE experiments or by surveying practitioners. From the research questions, motivations, and conclusions expressed in the surveyed papers, we distilled six major motivations, corresponding to six classes of research questions that researchers aimed to answer, or, in other words, six classes of knowledge that researchers aimed to obtain. The Motivation column in Table~\ref{tab:literature_review_part1} lists which motivations we attributed to each paper. The six motivations are:

\begin{itemize}
    \item \textbf{M1 - Discover Reverse Engineering Strategies:} to obtain knowledge on RE techniques and strategies: which ones are deployed (more often), which ones work best, how adaptive they are, etc. This was the focus of 15 studies. 
    \item \textbf{M2 - Assess Reverse Engineering Tools:} to assess the effectiveness and efficiency of RE tools and to improve them, e.g., by eliciting requirements. This was the focus of 9 studies. 
    \item \textbf{M3 - Assess Software Protection Strength:} to assess the strength of man-at-the-end SPs such as obfuscations~\citep{collbergbook}, i.e., to assess their impact on RE tasks. This was the focus of 14 studies. 
    \item \textbf{M4 - Discover Required Reverse Engineering Competencies:} to learn which human competencies, experience, expertise, and knowledge are useful or required for productive RE. Only 6 studies answer questions in this regard, of which only the 2 studies by~\cite{sutherland2006reverse} and by~\cite{hansch2018programming} introduce it as their main motivation. The other 4 only report results obtained via co-factor analysis of their participant's profiles and their performance.
    \item \textbf{M5 - Evaluate Complexity Metrics:} to learn how program complexity metrics (e.g., cyclomatic complexity~\citep{mccabe1976complexity} and various length measures~\citep{halstead1977elements}) of the challenge binaries affect RE difficulty. Only 2 papers studied this, both on unprotected software samples.
    \item \textbf{M6 - Acquire Meta-Knowledge:} to learn the best ways to conduct research motivated by any of the previous motivations. This was studied in only 1 paper, namely the work by~\cite{baldwin2016} on the best approach to elicit RE tool requirements during group sessions.
\end{itemize}

\subsection{Student Participation in Experiments}
\label{subsec:student_participation}
In the remainder of this section we will expound on the information extracted from the papers in our literature review, summarized in Tables~\ref{tab:literature_review_part1},~\ref{tab:literature_review_part2}, and~\ref{tab:literature_review_part3}. We will start by considering how RE researchers rely on students in their research as proxies for experts.

The Participants column in Table~\ref{tab:literature_review_part1} shows the level of expertise of the subjects that took part in each study, and the number of participants in each category. Part (A) of the table lists the 13 studies that involved bachelor and master students as subjects. None of those 13 papers motivate their use of student participants positively, i.e., by pointing out advantages of relying on students. Most of them do not discuss the disadvantages either.

Proximity and accessibility clearly appears to have mattered in the researchers' choice of subjects, as 10 of the 13 papers (indicated with a $*$ in the last column of Table~\ref{tab:literature_review_part1}) reveal that the authors found the student participants at their own host institutions. In fact, one paper explicitly mentioned their ``convenience sampling'' with volunteer students. The other 3 papers are unclear on this.

\cite{2014afamily} argue that students might not be representative of experts, but that ``hackers are not easily available and the only pragmatic possibility is resorting to students.'' They also claim to mitigate the threat to validity by relying on students by considering students with different level of experience, analyzing a worst-case scenario attack (i.e., the fastest observed attacks, probably by the highest skilled participants), and by performing a co-factor analysis by experience. They claim (without data) that ``many hackers are not that different from best students.'' They mention that their study needs confirmation or contradiction by more experienced participants. 

\cite{viticchie2016assessment} argue that their student participants ``are probably not the best choice to model attack exploitation'' and point to the difficulty of involving professional hackers. They extensively discuss this threat to external validity, and claim that it is mitigated by (i) the greater motivation of volunteer student participants, (ii) by the best students' comparable problem solving skills, (iii) by the selection of target applications that are attackable within a limited time frame. Furthermore, they argue that because they draw conclusions by comparing the performance of participants on clear and obfuscated versions of the application, ``students with homogeneous expertise [the participating students are all taking the same course in the same study program at the same university] are expected to lead to the same validity of comparative results as hackers with homogeneous expertise.'' 
\cite{2020splitting} provide the same arguments. 

\cite{hansch2018programming} mention the limited representativeness of their samples and students, and  ``professionals using their own analysis equipment and a more realistic scenario [...] would be desirable, but would be hard to pursue given the scarcity of obfuscation analysis resources [human experts] in the professional market.''

\cite{henry2020sensorre} hint at an issue of using students with insufficient RE skills, but provide no concrete description and point out that while their participant students were familiar with RE processes, they lacked practical experience outside of coursework. 

Few papers discuss how they managed varying skill levels across students. In fact, in most experiments with students, no active management of their skill levels took place, meaning that no balancing of skill levels across treatments was done. \cite{2014afamily} categorized skill levels based on academic performance (past exam grades) and balanced skill levels across treatments and used academic background (Bachelor, Master, PhD) but assigned each participant pool to separate experiments. By contrast, \cite{hansch2018programming} gathered pre-study data on skill levels in a pre‐study questionnaire but randomly assigned participants to one of four treatment groups, rather than stratifying or balancing them beforehand. This introduces a threat to internal validity, as observed performance differences could arise from skill disparities rather than from treatment effects~\citep{wohlin2012experimentation}. Other studies, such as those by \cite{Mantovani2022} and \cite{viticchie2016assessment}, gave all participants the same binaries which may avoid confounding skill levels across different treatments but still left the random heterogeneity of participants uncontrolled \citep{wohlin2012experimentation}.

Of the 13 studies with student participants, only 4 also feature more experienced reverse engineers among their participants. Of those 4, \cite{Mantovani2022} do not discuss the difference in performance between students and non-students. \cite{2020creative}, which present preliminary results of an experiment, only speculate about a potential difference. Few papers hence report on how measured student performance compares to professionals, or how the performance of different levels of students compares. \cite{yakdan2016} observed that overall, experts outperformed students. Similarly, \cite{2014afamily} observed that subjects' experience had a marginal positive effect on code comprehension.

Of the 4 experiments that include students and more experienced reverse engineers, only 1 targeted protected goodware (marked with $E_p$ in the Type column in Table~\ref{tab:literature_review_part1}). In addition, 5 experiments without (clearly identified) student participants targeted protected goodware 
(parts (B) and (D) of Table~\ref{tab:literature_review_part1}).

At first sight, the existence of these 6 (1+5) experiments might appear to contradict the opinion that some authors of experiments with students expressed regarding the difficulty of finding professional, experienced reverse engineers to participate in experiments on protected goodware. A deeper analysis of papers targeting protected goodware reveals that there is no such contradiction. 

In that regard, we first note that several authors expressing that opinion were motivated (at least in part) by learning about reverse engineers' strategies (M1) vis-\`a-vis protected goodware, for which they needed to observe the behavior of the participants in action. 

By contrast, 5 out of the 6 experiments on protected goodware with non-student participants do not have that motivation and did not need to observe the participants in action on binaries.
\cite{2020creative} are unclear about how many professionals participated, and their experiment was performed on Java source code, not a binary. \cite{dongleprot_secmeasure} performed the experiment assignments themselves. \cite{2016comparing} have only one participant, who they describe as a ``knowledgeable engineer'', the second lowest skill-level (novice, knowledgeable, skilled, ninja) in their considered class of adversaries, so definitely no expert. Furthermore, no details of that engineer's methods are described in the paper. \cite{2019impact} has a very short running experiment (as listed in the Time column in Table~\ref{tab:literature_review_part2}), in which only task execution times and correct answers to tasks were recorded, nothing else. \cite{bergmann2025potential} similarly ran a very short experiment. They report vaguely on the expertise of their participants, who were not asked to perform RE tasks but instead were asked to grade the perceived influence of several obfuscations on their understanding of the samples (M1). In short, none of these 5 experiments required observing expert reverse engineers showcasing their expertise and capabilities. 

What remains then, is the work by \cite{Ceccato2019}. This includes the only experiment with professional pen-testers targeting natively compiled, protected software that runs long enough to allow them to use their advanced capabilities, and from which their methods and strategies are documented (M1). That experiment was coordinated and executed in part by co-authors of this paper as part of the European research project ASPIRE (\url{https://www.aspire-fp7.eu}). The industrial partners only agreed to let their preferred professionals participate under the conditions that (i) the researchers' intrusion in their modus operandi was minimal, (ii) no video recording or screen capturing tools or other intrusive measures were used to collect information, (iii) they would not have to fill in structured questionnaires, instead falling back to free-form reporting, (iv) their reports would remain confidential, (v) the names and numbers of individual hackers per team would not be reported, (vi) they got a right to veto the publication of operational details they deemed sensitive, resulting in 11 ``omissis'' placeholders in the paper.

The underlying reason for these restrictions was those companies' managements fear that the leakage into the public domain of information about their experts' skills and experience, such as the analysis methods they are most familiar with, could become a weapon in the hands of their adversaries.\footnote{When we launched the Grand Reverse Engineering Challenge (\url{https://grand-re-challenge.org/}) where 10 000 USD could be won by participants that installed data collection software to allow us to observe their operation, we got the following reaction on Reddit: ``$<$10k to sell my techniques to denuvo [a protection tool vendor] by uploading a screen recording of my workflow? dont think so...'' (\url{https://www.reddit.com/r/ReverseEngineering/comments/nezxd1/grand_reverse_engineering_challenge_10000_prize/})}

\subsection{Options to Motivate Participants}
\label{subsec:literature_motivation}
Few papers discussed how they motivated students to participate. 
\cite{2020splitting} offered bonus points for the course, based on diligent participation, not on attack success. They considered this necessary to find students willing to participate outside of their official course hours. To reduce the risk of introducing noise into the collected data from participants that only participate for the bonus but are not motivated or capable to deliver, they included checks on their C knowledge and on the quality of their reports. The students were assured that the results of those checks would not be shared or used outside of the experiment. 

\cite{viticchie2016assessment} similarly considered it necessary to encourage students in order to find enough participants, and set up a lottery of a couple of gift cards among the participating students, regardless of the success of their execution, to encourage participation and a larger commitment. They also point to the possibility that this introduces noise, but they do not mention checks on the participants' knowledge nor on the quality of their reports.

\cite{2014another} mention gift vouchers as an option for future experiments. \cite{yakdan2016} offered the participating students 40 Euro compensation. \cite{hansch2018programming} only gave 10 Euro compensation. This is in line with the shorter duration of the experiment, and the use of less trained/experienced students, but the authors do not discuss the reasons for choosing particular amounts. None of these works discusses threats to validity because of the offered compensation. 

\cite{Mantovani2022} offered the participants, including the students, no compensation, but instead told them ``There is no ranking or prize, this is just an experiment: so please do not cheat.''

For professionals or experienced amateurs, other motivations were typically offered. In contrast to their student participants, professional participants did not get a monetary reward from \cite{yakdan2016}. They conjectured that professionals would be motivated intrinsically to help progress in their domain, and that a small amount of money would not motivate them, as they consider their time is more valuable than money. Their participants instead were given early access to the research results, including free access to the developed decompiler upon its release. Similarly, \cite{Yamagishi2025} shared early survey results in debriefing session with the participants. In contrast, \cite{Votipka2020} did award 40 Euro gift cards to their expert participants. 

In the study of \cite{Ceccato2019}, the participating professionals were hired one month to pen test the given applications, similarly to how they would be hired by their customers. In addition, the participants to their public bounty could have their self-chosen user names listed next to that challenge on the challenge
web site and the first successful attacker of each challenge was rewarded with a monetary prize of 200 Euro. Also \cite{matzen2021effects} compensated their participants (who worked for their own institute) for their time at their normal hourly rate. \cite{nyre2022task} compensated for 2 hours of time using an undisclosed, project-specific charge code.

\cite{henry2020exploring} note that their participants were under no internal or external pressure to participate. \cite{bryant2012eliciting} mention that their participants were not remunerated. Two groups participated in the study by \cite{baldwin2016}. For one of them, participation in the exercise was made mandatory by their manager, whereas in the other group, participation was voluntary. This explained a difference in response rate.

\newcommand{\Y}{\checkmark &}
\newcommand{\N}{&}
\newcommand{\Java}{\texttt{Java}}
\newcommand{\Clang}{\texttt{C}}
\newcommand{\CIL}{\texttt{CIL}}
\newcommand{\Javascript}{\texttt{JS}}
\newcommand{\XEightySix}{\texttt{X86}}
\newcommand{\ARM}{\texttt{ARMv7}}

\newcommand{\CNT}[1]{#1 &}
\newcommand{\NONE}{-&}
\newcommand{\ManyPrograms}{\checkmark}
\newcommand{\SameProgram}{}

\begin{table}[t]
\scriptsize
\centering
\setlength{\tabcolsep}{2pt}
\begin{tabular}{l|c|c@{}cc|c|c@{}c@{}c@{}c@{}c|c}
\textbf{Paper} & \textbf{Nr.\ of} &
   \multicolumn{3}{c}{\textbf{Format}} & 
   \textbf{Size} & 
   \multicolumn{5}{c}{\textbf{Task}} &
   \textbf{Time}\\
 & \textbf{samples} &                                                               \textbf{U} & \textbf{O} & \textbf{Lang} &
&                                 \textbf{L} & \textbf{M} & \textbf{C} & \textbf{U} & \textbf{S} \\\toprule
\multicolumn{11}{c}{}\\[-0.5em]
\multicolumn{11}{l}{\emph{(A) Experiments to which bachelors/masters students participated}}\\[-1.7em]
\multicolumn{11}{c}{}\\
\multicolumn{11}{c}{}\\

\cite{sutherland2006reverse}
& 6
& \Y\N\XEightySix 
& 6--665 SLoC
& \CNT{8}\CNT{8}\NONE\NONE\NONE 5h
\\
\cite{2014afamily} 
& 2
& \Y\Y\Java     & 1215, 1030 SLoC                                    & \CNT{4}\CNT{4}\NONE\NONE\NONE 4h
\\
\cite{2014another} 
& 1
& \Y\Y\Java     & 1215 SLoC                                           & \CNT{2}\CNT{2}\NONE\NONE\NONE 1h
\\
\cite{viticchie2016assessment} 
& 2
& \Y\Y\Clang      & 238, 62 SLoC                                          & \NONE\CNT{2}\CNT{2}\NONE\NONE 3.5h
\\
\cite{yakdan2016} 
& 6
& \N\Y\Clang
& 12--69 SLoC 
& \CNT{6}\NONE\NONE\NONE\NONE 2h, 3h
\\
\cite{liu2017stochastic} 
& 10
& \N\Y\Javascript      & 17--42 SLoC                                         & \NONE\NONE\NONE\CNT{10}\NONE ?
\\
\cite{hansch2018programming} 
& 2
& \Y\Y\Java      & 1215, 1030 SLoC                                      & \CNT{4}\NONE\NONE\NONE\NONE 1.5h
\\
\cite{kuang2018enhance} 
& 1
& \N\Y\XEightySix      & 1357 ins                                            & \CNT{3}\NONE\NONE\NONE\NONE 72h
\\
\cite{2020creative} 
& 2
& \Y\Y\Java      & 1215, 1030 SLoC                                    & \CNT{4}\NONE\NONE\NONE\NONE 1h
\\
\cite{2020splitting} 
& 1
& \Y\Y\Clang      & 1873 SLoC                                           & \NONE\CNT{1}\NONE\NONE\NONE 2h
\\
\cite{henry2020sensorre}
& 1
& \Y\N\XEightySix
& 30k ins
& \CNT{2}\NONE\NONE\NONE\NONE 2h/50m
\\
\cite{2021inputoutput} 
& 506
& \N\Y\XEightySix
& 100--300 ins, ?
& \NONE\CNT{6}\NONE\NONE\NONE 24h
\\
\cite{Mantovani2022} 
& 2
& \Y\N\XEightySix      & 146,207 SLoC                                          & \NONE\NONE\CNT{2}\NONE\NONE 941m \\

\midrule

\multicolumn{11}{c}{}\\[-0.5em]
\multicolumn{11}{l}{\emph{(B) Experiments without students}}\\[-1.7em]
\multicolumn{11}{c}{}\\
\multicolumn{11}{c}{}\\

\cite{dongleprot_secmeasure} 
& 8
& \N\Y ? 
& ? 
&\NONE\NONE\NONE\NONE\CNT{6}
 320m,24h \\
\cite{bryant2012eliciting} 
& 1
& \Y\N\XEightySix
& 27 BBLs
& \NONE\NONE\CNT{1}\NONE\NONE
1h
\\
\cite{2016comparing} 
& 4
& \N\Y\XEightySix
& 0.03MB--4.2MB files
&\CNT{6}\CNT{8}\NONE\NONE\NONE
?
\\
\cite{wagner2017knowledge} 
& 16
& \N\N\XEightySix
& ?
& \NONE\NONE\NONE\NONE\CNT{?}
30m (30m)
\\
\cite{Ceccato2019} 
& 11
& \N\Y\ARM
& 7k, 52k, 338k SLoC;
& \NONE\NONE\CNT{8}\NONE\CNT{3}
30d
\\
&
& & &
& 12kB--5MB code
& & & & 
\\
\cite{2019impact} 
& 2
& \Y\Y\CIL
& ?
& \NONE\CNT{22}\NONE\CNT{22}\NONE
36m
\\

\cite{matzen2021effects} 
& 24
& \Y\N\Clang
& ?  
& \NONE\NONE\NONE\CNT{24}\NONE
45m
\\
\cite{nyre2022task} 
& 1
& \Y\N ?  
& ?
& \NONE\NONE\CNT{1}\NONE\NONE
2h
\\

\cite{Yamagishi2025}  
& 1
&\N\Y\XEightySix
& ? 
& \NONE\NONE\NONE\NONE\CNT{1}
735m (90m)
\\

\midrule

\multicolumn{11}{c}{}\\[-0.5em]
\multicolumn{11}{l}{\emph{(C) Surveys/Interviews with professionals}}\\[-1.7em]
\multicolumn{11}{c}{}\\
\multicolumn{11}{c}{}\\

\cite{bryant2013top}
& --
& \multicolumn{3}{c|}{--}
& --
& \multicolumn{5}{c|}{--} &
2h
\\
\cite{baldwin2016} 
& --
& \multicolumn{3}{c|}{--}
& --
& \multicolumn{5}{c|}{--} &
3h (1h)
\\
\cite{Votipka2020}  
& --
& \multicolumn{3}{c|}{--}
& --
& \multicolumn{5}{c|}{--} &
75m
\\
\cite{henry2020exploring}  
& --
& \multicolumn{3}{c|}{--}
& --
& \multicolumn{5}{c|}{--} &
2h
\\
\cite{Wong2021inside}
& --
& \multicolumn{3}{c|}{--}
& --
& \multicolumn{5}{c|}{--} &
1h
\\
\cite{Wong2024}  
& --
& \multicolumn{3}{c|}{--}
& --
& \multicolumn{5}{c|}{--} &
70m
\\

\midrule

\multicolumn{11}{c}{}\\[-0.5em]
\multicolumn{11}{l}{\emph{(D) Description of participants is vague}}\\[-1.7em]
\multicolumn{11}{c}{}\\
\multicolumn{11}{c}{}\\
\cite{malware_visualcomp} 
& 2
& \N\Y\XEightySix
& ? 
&\CNT{6}\NONE\NONE\NONE\NONE
? \\
\cite{burk2022}
& 23
& \Y\N ?
& 1486 SLoC (total)
& \NONE\NONE\NONE\CNT{23}\NONE
24h
\\
\cite{bergmann2025potential} 
& 1
& \Y\Y ?  
& ?
& \NONE\NONE\NONE\CNT{1}\NONE
36m\\
\bottomrule

\end{tabular}
\caption{Results of the literature review, part 2. The legend is given in Table~\ref{tab:literature_review_part2_legend}.}
\label{tab:literature_review_part2}
\end{table}

\begin{table}
\setlength{\tabcolsep}{3pt} \begin{tabular}{lp{10.7cm}}
\textbf{Column} & \textbf{Legend} \\\toprule
\textbf{Format} & Checkmarks in the \textbf{U} and \textbf{O} columns indicate that participants were given Unobfuscated and/or Obfuscated code, respectively. The \textbf{Lang} column lists the programming language in which code was provided to the participants to reverse engineer in experiments. X86 implies that executable binaries were provided.\\
\textbf{Size} & This column lists the sizes of the samples provided to participants of experiments for RE, as reported in the papers.

\begin{tabular}{lp{9cm}}
    SLoC &  source lines of code\\
    Ins & number of instructions in the target binary\\
    BBLs & number of basic blocks in the samples\\
    B code & bytes of code in the executable\\
    \end{tabular}
    \\
\textbf{Task} & For experiments, this column lists which types of tasks, and how many such tasks,  participants were asked to perform:

\begin{tabular}{lp{10cm}}
        L & localization tasks \\
        M & code modification tasks \\
        C & crackme tasks, i.e., finding input that triggers some behavior \\
        U & code understanding tasks such as determining correct variable names or function summaries\\
        S & tasks specific for the target application and its assets: breaking a dongle protection, a digital rights management (DRM), a one-time password generator, or a software license manager; extracting malicious behaviour fingerprints from malware; countering malware dynamic analysis evasion tactics.
    \end{tabular}
The values in the table indicate the number of different tasks given to participants, counting identical assignments on different samples as different tasks.\\
\textbf{Time} & We list the duration of the experiment/interview sessions in the case of time-limited assignments/interviews, or the total time that participants could work on them at home at their own choice, or the expected duration of the assignment/interview as communicated to the participants that were expected to continue until task/interview completion, or the maximum time or average time that a participant required to complete the assignment/interview, whichever is applicable and reported in the paper. When multiple, comma-separated times are listed, this implies that different expected durations were communicated to different types of participants. If additional time was required from the participants to fill in pre or post questionnaires or give an interview and the duration thereof is reported, we add that between parenthesis.\\
\bottomrule

\end{tabular}
\caption{Legend for Table~\ref{tab:literature_review_part2}.}
\label{tab:literature_review_part2_legend}
\end{table}

\begin{table}
\scriptsize
\centering
\setlength{\tabcolsep}{4pt} \begin{tabular}{l|c|c|c|c|c|c|c|c|c|c|c|c|c|c|c|}
\textbf{Publication} 
& \rotatebox{90}{\textbf{Screen recording}} 
& \rotatebox{90}{\textbf{Voice recording}} 
& \rotatebox{90}{\textbf{Event logging}} 
& \rotatebox{90}{\textbf{Eye tracking}} 
& \rotatebox{90}{\textbf{Live observation}}
& \rotatebox{90}{\textbf{Video recording}}
& \rotatebox{90}{\textbf{Structured report}}
& \rotatebox{90}{\textbf{Free-form report}}
& \rotatebox{90}{\textbf{Questionnaire}}
& \rotatebox{90}{\textbf{Unstructured interview}}
& \rotatebox{90}{\textbf{(Semi-)structured interview}}
& \rotatebox{90}{\textbf{Group session}}
& \rotatebox{90}{\textbf{Test}}
& \rotatebox{90}{\textbf{Self-executed}}
& \rotatebox{90}{\textbf{Unclear}}
\\

\toprule

\multicolumn{10}{c}{}\\[-0.5em]
\multicolumn{10}{l}{\emph{(A) Experiments to which bachelors/masters students participated}}\\[-1.7em]
\multicolumn{10}{c}{}\\
\multicolumn{10}{c}{}\\
\cite{sutherland2006reverse}   & X &   & X &   &   &   & X &   & X &   &   &   &   &   &   \\ 
\cite{2014afamily}             &   &   &   &   &   &   & X &   & X &   &   &   &   &   &   \\ 
\cite{2014another}             &   &   &   &   &   &   & X &   & X &   &   &   &   &   &   \\ 
\cite{viticchie2016assessment} &   &   &   &   &   &   & X &   & X &   &   &   &   &   &   \\ 
\cite{yakdan2016}              &   &   &   &   &   &   & X &   & X &   &   &   &   &   &   \\ 
\cite{liu2017stochastic}       &   &   &   &   &   &   &   &   &   &   &   &   &   &   & X \\ 
\cite{hansch2018programming}   &   &   & X &   &   &   & X &   & X &   &   &   &   &   &   \\ 
\cite{kuang2018enhance}        &   &   &   &   &   &   &   &   &   &   &   &   &   &   & X \\ 
\cite{2020creative}            & X &   &   &   &   &   &   &   & X &   &   &   & X &   &   \\
\cite{2020splitting}           &   &   &   &   & X &   & X &   & X &   &   &   & X &   &   \\ 
\cite{henry2020sensorre}       & X & X & X &   & X &   &   &   & X &   & X &   &   &   &   \\ 
\cite{2021inputoutput}         &   &   &   &   &   &   &   &   &   &   &   &   &   &   & X \\ 
\cite{Mantovani2022}           &   &   & X &   &   &   & X &   & X &   &   &   &   &   &   \\
\midrule

\multicolumn{10}{c}{}\\[-0.5em]
\multicolumn{10}{l}{\emph{(B) Experiments without students}}\\[-1.7em]
\multicolumn{10}{c}{}\\
\multicolumn{10}{c}{}\\
\cite{dongleprot_secmeasure}   &   &   &   &   &   &   &   &   &   &   &   &   &   & X &   \\ 
\cite{bryant2012eliciting}     & X & X &   &   & X &   & ? &   &   &   & ? &   &   &   &   \\ 
\cite{2016comparing}           &   &   &   &   &   &   &   &   &   &   &   &   &   &   & X \\ 
\cite{wagner2017knowledge}     & X &   & X &   & X & X &   &   & X &   &   &   &   &   &   \\
\cite{Ceccato2019}             &   &   &   &   &   &   &   & X &   & X &   &   &   &   &   \\ 
\cite{2019impact}              &   &   &   &   &   &   & X &   &   &   &   &   &   &   &   \\
\cite{matzen2021effects}       &   &   & X & X &   &   & X &   & X &   &   &   &   &   &   \\
\cite{nyre2022task}            & X & X &   &   &   &   &   &   & X &   &   &   &   &   &   \\
\cite{Yamagishi2025}           & X &   &   &   &   &   & X &   & X &   & X &   &   &   &   \\ 
\midrule

\multicolumn{10}{c}{}\\[-0.5em]
\multicolumn{10}{l}{\emph{ (C) Surveys/interviews with professionals}}\\[-1.7em]
\multicolumn{10}{c}{}\\
\multicolumn{10}{c}{}\\
\cite{bryant2013top}           &   &   &   &   &   &   &   &   &   &   & X &   &   &   &   \\ 
\cite{baldwin2016}             & X & X &   &   & X &   &   &   & X &   & X & X &   &   &   \\
\cite{Votipka2020}             & X & X &   &   & X &   &   &   &   &   & X &   &   &   &   \\ 
\cite{henry2020exploring}      &   & X &   &   &   &   &   &   &   &   & X &   &   &   &   \\
\cite{Wong2021inside}          &   & X &   &   &   &   &   &   & X &   & X &   &   &   &   \\ 
\cite{Wong2024}                &   & X &   &   &   &   &   &   & X &   & X &   &   &   &   \\ 
\midrule

\multicolumn{10}{c}{}\\[-0.5em]
\multicolumn{10}{l}{\emph{(D) Description of participants is vague}}\\[-1.7em]
\multicolumn{10}{c}{}\\
\multicolumn{10}{c}{}\\
\cite{malware_visualcomp}     &   &   &   &   &   &   & ? &   &   &   & ? &   &   &   &   \\ 
\cite{burk2022}               &   &   & X &   &   &   & X &   & X &   &   &   &   &   &   \\ 
\cite{bergmann2025potential}  &   &   &   &   &   &   & X &   &   &   &   &   &   &   &   \\
\bottomrule
\end{tabular} 

\caption{Results of the literature review, part 3. The legend is given in Table~\ref{tab:literature_review_part3_legend}.}
\label{tab:literature_review_part3}
\end{table}

\begin{table}
\setlength{\tabcolsep}{3pt} \begin{tabular}{p{3.6cm}p{8.3cm}}
\textbf{Column} & \textbf{Legend} \\\toprule
\textbf{Screen recording}  & Video recording or frequent screenshots were made during experiment.\\
\textbf{Voice recording} & Thinking aloud experiment or other activity\\
\textbf{Event logging} & Events were logged during the experiment for later analysis: mouse clicks, shell commands, produced files, which functionality of tools is being used, etc. \\
\textbf{Eye tracking} & Eye tracking was monitored during the experiment.\\
\textbf{Live observation} & Experiment supervisors/facilitators observed and checked the performance and obtained results (for correctness), and/or took notes about observed participants, possibly asking them questions on the fly\\
\textbf{Video recording} & Participant is filmed, e.g., with webcam.\\
\textbf{Structured report} & Experiment participants reported in writing on specific aspects of their execution of the assignments, such as start and end times, outcomes, solutions, code fragment readability scores, etc. \\
\textbf{Free-form report} & Experiment participants reported on their activities in free-form reports.\\
\textbf{Questionnaire} & Participants answered questions about their experience and knowledge in writing\\ 
\textbf{(Semi-)structured interview} & Researchers interviewed participants following a predetermined format. In structured interviews, this is followed rigidly. In semi-structured interviews, some flexibility is allowed, e.g., to change the order of the questions or to insert follow-up questions. \\
\textbf{Unstructured interview} & Researchers interviewed participants informally and conversational (by email)\\
\textbf{Group session} & Managed, structured group discussion of activities and requirements\\
\textbf{Test} & Participants underwent test, for creativity, intelligence, C knowledge, ...\\
\textbf{Self-executed} & The authors themselves performed the experiments and logged their activities.\\
\textbf{Unclear} & The paper does not report how data was collected from participants.\\
\midrule
\multicolumn{2}{p{12cm}}{Note: question marks in multiple columns in Table~\ref{tab:literature_review_part3} indicate that the exact method was not clear, e.g., because the paper does not clarify whether participants were asked to respond in writing or orally to questions.}\\
\bottomrule
\end{tabular}
\caption{Legend for Table~\ref{tab:literature_review_part3}.}
\label{tab:literature_review_part3_legend}
\end{table}

\subsection{Training of Student Participants}
The last column of Table~\ref{tab:literature_review_part1} lists the training students received before participating in experiments. While the descriptions in the table are tersely rephrased versions of the descriptions in the papers, we made sure they convey almost all information available in the papers. For most papers, this is rather limited.

Because professional organizations such as ACM reflect on SE curricula and regularly publish visions on them, there is some level of standardization in such curricula, and descriptions such as ``final year master's students in Software Engineering'' do not require the reader to guess which basic SE skills those students are likely to have obtained.

With regards to RE and SP, however, there is much less standardization. Several curricula guidelines mention the need for educating students in RE, but only prescribe what needs to be learned at a coarse granularity. For example, ACM's Cybersecurity Curricula 2017 Curriculum Guidelines for Post-Secondary Degree
Programs in Cybersecurity~\citep{ACMguidelines} mention that the essentials and learning outcomes need to entail ``List reasons why someone would reverse engineer a component'' and ``Explain the difference between static and dynamic analysis in
reverse engineering software.'' Additionally, the report specifies the need to introduce students to anti-reverse engineering: ``[The anti-reverse engineering topic] covers techniques such as design obfuscation and camouflaging for making component designs and implementations difficult to reverse engineer.'' These excerpts are the most concrete descriptions in that 88-page document. Other security education guidelines are even less concrete~\citep{CybersecurityCurricularGuidelines}. So descriptions such as those in Table~\ref{tab:literature_review_part1} in our eyes do not provide enough information about the training received by the students to assess whether that training was adequate for the students to carry out the experiments. 

Consider, for example, the five studies that had the student participants reverse engineer binary goodware executables (as indicated with an N in the Format column 
in part (A) of Table~\ref{tab:literature_review_part1}).
In those studies, the participants are described as having some RE experience, which probably includes some level of tool usage. This helps to reduce the interaction between the participant selection and the treatment, but those studies do not discuss in sufficient detail to what extent the skills acquired by the students overlap with the skills that professionals would build on for the same assignments. The same holds for the obfuscation training that students had been given in most of the experiments with obfuscated code: the papers contain no more details on that training than what is listed in Table~\ref{tab:literature_review_part1}. For this reason alone, it is hence next to impossible to assess the external validity of the experiments' results accurately.

\subsection{Privacy and Ethics Considerations}
\label{sec:literature-privacy-ethics}

Few of the papers presenting results obtained with students discuss the ethical side or privacy considerations.

\cite{hansch2018programming} mention that their data protection office approved of the experiment, and that the participant students participated under ``anonymous'' IDs and had been informed about the used data collection methods. \cite{yakdan2016} also clarified that they only collected ``anonymous'', not personally identifying information. Neither of these works clarify whether the researchers knew the participating students' identities, however, i.e., how they knew who to give the financial compensation. Note that the European General Data Protection Regulation (GDPR) (enacted in 2018) only considers collected data \emph{anonymous} if the researchers cannot in any technical way identify the participants. Strictly speaking, as soon as someone (e.g., a financial department) collects the names of participants to pay them the compensation, even if they promise they will not share those names with the researchers, the participation is not to be considered anonymous~\citep{SRBvEDPS_T557_20}.  

\cite{Mantovani2022} also collected anonymous data only, stating that they ``ensured that all methods and experiments performed for this work are in line with our institutions’ research ethics guidelines and our country regulations on data collection and retention.'' They had no need to know identities, as they did not compensate their participants. 

\cite{2014afamily} only clarify that students were told that they would not be evaluated on their performance during the experiment. \cite{kuang2018enhance} clarify that their experiment was approved by their host institution's research ethics board, but provide no details. 

Those are the only discussion of ethics and privacy in the papers reporting experiments with students. Of those 13 papers, only one \citep{Mantovani2022} provides sufficient clarity on how they resolve the potential ethical problem of putting students under pressure to participate in scientific experiments that in many cases were conducted by those students' own professors. 

By contrast, in the experiments by \cite{2014afamily}, only the voluntarily participating students attended the experiments, but their course instructors could observe exactly who they were. How many of the other experiments with students were conducted similarly is unclear. What is clear, is that none of the studies that handed out gift cards to students address how they distributed those rewards pseudonymously, thus again not revealing how they safeguarded the privacy of participants and ensured that their consent was given freely.

In experiments without students, the participants can of course not be dependent on the researchers to obtain their academic grades. The authors of those experiments and studies do not discuss privacy and ethics in much detail either.

\cite{Yamagishi2025} declared that they designed their study in line with the Menlo report \citep{menlo-report}, and that they did not get formal approval because their institute did not have an Institutional Review Board (IRB) for Computer Science research. Participants gave their informed consent to the data collection, including of personally identifiable information. 

\cite{Wong2024}, \cite{nyre2022task}, and \cite{matzen2021effects} got approval from an IRB board (or similar) and obtained informed consent from participants. \cite{bryant2013top}, \cite{Wong2021inside}, \cite{Votipka2020}, and \cite{burk2022} also got approval from an IRB board (or similar organ). The latter declared to use no personally identifiable information. \cite{baldwin2016} clarifies that participant received and signed ethics forms. \cite{henry2020exploring} obtained consent for their data collection, and declare that their participants were under no internal or external pressure to participate. 

Other authors, such as \cite{Ceccato2019} do not report institutional approval or consent forms. For their public bounty experiment, they mention that participants could register and participate anonymously (i.e., via an anonymous email address), but in order to receive the monetary prizes, the winners had to reveal their true identity. For their experiment with hired professionals, several legally binding agreements between all involved parties were in place such as a Consortium Agreement and a Grant Agreement regarding, e.g., the rights and obligations of the parties and how confidentiality of information was to be handled. In other studies, participants were recruited from within the organizations conducting the research (employees or project staff rather than external participants) \citep{dongleprot_secmeasure,bryant2012eliciting,baldwin2016}.

\subsection{Data Collection Methods}
\label{subsec:survey_data_collection}

The data collection methods to which participants consented, varied significantly from one experiment to another. Table~\ref{tab:literature_review_part3} provides an overview of the used methods.

In research relying on students, structured reports and questionnaires are by far the most popular forms of data collection. Questionnaires were most often used to collect demographic information such as the participants prior experience, and sometimes sought to record the participants' perception of different aspects of the experiments. Structured reports were mostly used to record the relevant data for inferring the effects of treatments by, e.g., hypothesis testing. The recorded data then included start and end times of different tasks, and outcomes of the executed tasks that allowed the researchers to assess them as successful or not. Other than \cite{2020splitting}, who note that letting students report time invested in different tasks is potentially inaccurate, no other authors in our survey discuss potential threats to validity of their use of structured reports. 
Some structured reports collected additional data. For example, \cite{2020splitting} recorded which functions participants modified to complete certain assignments.

More intrusive data collection methods, such as screen and voice recording, eye tracking, event logging, and live observation were used much less frequently, namely in only 4 out of the 13 papers, 3 of which also involved expert participants in addition to the student participants. Experiments without students are more likely to rely on such intrusive data collection methods: 5/9 experiments listed in the part (B) of Table~\ref{tab:literature_review_part3} relied on one or more such intrusive methods.

From this, one might conclude that professional experts often allow intrusive data collection on how they put their competencies into practice. However, note that none of those 5 papers involved obfuscated or otherwise protected goodware. Also of the three papers with students and expert participants, two did not target protected goodware. Only~\cite{2020creative} involved experts in their experiments, and report screen captures from ``selected participants''. They do not clarify, however, whether those were taken from experts or students. Moreover, their experiments lasted for 1 hour only, and the participants targeted small academic software samples that had previously been used in three other experiments and that consisted of Java source code (as documented in Table~\ref{tab:literature_review_part2}). These experiments and screen captures hence were not of a nature that risked uncovering or leaking of advanced competencies of the participants. 

The used data collection methods hence do not contradict what we reported more anecdotally on professional obfuscated software pen testers' demands about data collection in the last two paragraphs of Section~\ref{subsec:student_participation}, i.e., on their unwillingness to see their expert skills become public.

From the 6 papers on surveys/interviews among professionals (part C of Table~\ref{tab:literature_review_part1}), 5 recorded the interviews, which is not surprising. In those interviews, the participating experts control what they choose to reveal about their competencies. In addition, two papers reported screen recording and live observation \citep{baldwin2016,Votipka2020}.\footnote{Even though the two papers recorded screens and observed RE activities live, we did not classify those papers as reporting experiments, because the recorded activities were demonstrations of participants of their activities on self-chosen samples. So they did not perform tasks designed by the researchers to measure effects of treatments.} Those two papers did not consider obfuscated or otherwise protected goodware either, and hence also do not contradict the experience reported in Section~\ref{subsec:student_participation}. 

As for event logging, most researchers logged only the interactions with specific tools that allow them to do so, e.g., via plugins. \cite{hansch2018programming} used Eclipse plugins to record Eclipse functionality usage, while \cite{Mantovani2022} created a customized browser-based disassembler that logged which function control-flow graphs (CFGs) and basic blocks participants viewed. \cite{henry2020sensorre} used plugins for the Binary Ninja interactive disassembler. \cite{matzen2021effects} used a tool to collect answers from participants, with which they could also collect the timing of the answers. The prototype visual malware analysis system evaluated by \cite{wagner2017knowledge} logged interactions, and so did the decompilation aid tool evaluated by \cite{burk2022}. Only \citep{sutherland2006reverse} logged events without relying on custom tool functionality, namely by collecting Bash histories as well as other temporary and history files from the participants' machines.

\subsection{Human Competencies}
\label{subsec:survey_human_competencies}
The Participants column in Table~\ref{tab:literature_review_part1} lists the experience and levels of expertise of the participants in the experiments and studies. For students, the Training column adds the specific training they got according to the authors of the papers. Beyond that demographic data, we extracted six categories of competencies and knowledge that expert reverse engineers have according to the papers in our study. Numbered C1--C6, they are the following: 

\paragraph{C1 - Code Reading} Professional reverse engineers can be expected to be highly familiar with executable file formats and low-level code representations such as assembly code, instruction set architectures, idioms and patterns that occur in such code (possibly after heavy compiler optimizations), and how those map onto corresponding idioms and patterns in higher-level languages such as C or C++. 
    
For example, in the taxonomy of constructs used by reverse engineers assembled by~\cite{Ceccato2019} (from here on named the ASPIRE taxonomy after the project in which it was created), the top-level category \emph{Software elements} encompasses 45 constructs (e.g., \emph{Disassembled code}, \emph{Decompiled code}, \emph{Trace}, \emph{Socket}, \emph{Basic Block}, \emph{Function Call}, \emph{Control Flow Graph}, ...) that correspond to artifacts in the program representations that reverse engineers ``read'' and reason about.

\paragraph{C2 - Software Protection} Professional pen testers of protected software and malware analyst can be expected to be highly familiar with SPs, including conceptual knowledge about their design and main features, as well as practical experience with commercial SP tools.
    
For example, the ASPIRE taxonomy includes an \emph{Obstacles} category that contains several concrete protections such as \emph{Control flow flattening} and \emph{virtualization} that the reverse engineers experience as obstacles. The taxonomy's \emph{Attack step} category includes multiple constructs that correspond to different ways to defeat those obstacles (i.e., to defeat the protections) by \emph{undoing} them, \emph{bypassing} them, \emph{working around} them, or \emph{overcoming} them in other ways. This clearly shows that reverse engineers experienced with protected software have advanced knowledge about the use of SPs. Likewise, the constructs \emph{Recognize similarity with already analysed protected application}, \emph{Reuse attack strategy that worked in the past} and \emph{Pattern matching} are included because over time reverse engineers learn to recognize fingerprints and weaknesses of protection tools and exploit that knowledge. The explanation of \cite{Ceccato2019} on their construct \emph{Choose the path of least resistance} even mentions that reverse engineers often choose their attack strategy based on speculation about likely used protections.

\paragraph{C3 - Application Domain Knowledge} Professional reverse engineers are often specialized in certain classes of applications. Cryptanalysts specialize in cryptography, others are specialized in game cheating, others in circumventing Digital Rights Management (DRM) protections in media players, etc. They know the typical architectures of such applications, and the features that can be exploited, e.g., to search the code for the relevant artifacts. Cryptanalysts know and exploit that decryption routines operate on high-entropy data, game cheaters know how to locate the memory locations that store resources such as a player's amount of gold or health and exploit that those are shown on screen and can be controlled by the player; they know how 3D image scenes are being rendered, etc.
    
In the ASPIRE taxonomy, the reliance on application domain knowledge shows in several constructs. The construct \emph{Recognizable library} refers to reverse engineers knowing which libraries are often used to implement certain functionality that is typically needed in the application at hand. The construct \emph{Identify sensitive asset} implies that reverse engineers know which valuable assets to locate in an application at hand. Clearly those assets are domain-dependent. More importantly, however, is how the professional pen testers were chosen in the ASPIRE experiments. \cite{Ceccato2019} note that their ``professional hackers have been explicitly selected by the industrial partners because of their specific expertise in attacking their applications.'' In their experiments, instead of using the industrial partner's commercial applications, mock-ups of a similar level of complexity were used (such that those partners did not need to share the source code of their commercial applications with other project partners). The above statement therefore means that the hackers were selected because of their domain knowledge, not because of their knowledge of a specific application. 
    
\cite{Votipka2020} observe how domain-specific beacons in the code (i.e., recognizable artifacts) are relied upon by reverse engineers, and that the type of beacons most easily recognized depends on their personal experience. Moreover, they observe how reverse engineers compare the code fragments they locate to reference implementations, which of course originate from specific domains.  
    
Beyond knowledge of benign application domains, malware analysts know which indicators of compromise to look for~\citep{Wong2021inside} and the types of analysis evasion techniques used by malware~\citep{Wong2021inside}.

\paragraph{C4 - Tool Usage} Professional reverse engineers know how to use RE tools to collect information about the software and to manipulate it, both statically and dynamically. They know how to search for and navigate between artifacts in the binary and the types of relations between artifacts that the tools can capture. They know how to configure the tools, how to exploit their flexibility by means of plugins, how to override standard heuristics, how to customize and extend them.

In the ASPIRE taxonomy, this is evidenced by the presence of 6 different constructs corresponding to 6 types of RE tools, 20 constructs denoting different types of software analysis that are supported by such tools, the already mentioned \emph{Customize/extend tool} construct, other related constructs such as \emph{Choose/evaluate alternative tool} and \emph{Customize execution environment}, and concrete attack step constructs such as \emph{Decompile the code}. 
    
Critically, professional reverse engineers know the limitations of their tools, and how to assist interactive tools to overcome those limitations. The most prominent example is that of interactive disassemblers and decompilers, which allow the users to override all kinds of information that the tools have collected about the software, such as function signatures, recovered data types, variable names, how low-level idioms are mapped onto high-level ones, etc. Professionals also understand where those tools' heuristics are unsound and where and how they are most likely to produce unreliable results, such as incorrect signatures, incorrect data types, etc. In the ASPIRE taxonomy, this is covered by the construct \emph{Tool limitations} in the \emph{Difficulty} category.
 
Other papers focusing on RE strategies also clearly identify the frequent use of static and dynamic analysis tools in the toolbox of reverse engineers~\cite{Votipka2020,Wong2021inside,Wong2024}
 
Several surveyed papers also focused on specific tools and their use by reverse engineers: \cite{yakdan2016} and \cite{matzen2021effects} focused on decompilers, \cite{henry2020sensorre} on tools for provenance support, \cite{2021inputoutput} on tools to help with de-obfuscation, \cite{Mantovani2022} on interactive disassemblers, and \cite{wagner2017knowledge} studied visualization tools for malware analysis.

\paragraph{C5 - Strategic thinking} Professionals know which RE strategies can work to reach certain end goals, such as circumventing DRM, stealing secret keys, etc. They know the likely attack surface, which sequences of attack steps can be most effective and most efficient to progress from one milepost to another; they know which hypothesis to formulate, check, and revise, etc.
    
In the ASPIRE taxonomy, this shows, amongst others, in the various constructs in the \emph{Prepare attack}, \emph{Build the attack strategy}, and \emph{Analyse attack result} categories, with the latter including \emph{Make hypothesis} and \emph{Confirm hypothesis}. Other researchers, such as \cite{Votipka2020} confirm the role of hypothesis formulation and validation/refutation, as well as the different high-level phases through which RE strategies iterate. They call those \emph{Overview}, \emph{Sub-component scanning}, and \emph{Focused experimentation}. \cite{Mantovani2022} study, among others, the effectiveness of different code browsing strategies (top-down or bottom-up) while RE binaries. 
    
Also for malware analysts, papers describe which strategies reverse engineers commonly deploy~\citep{Wong2021inside,Wong2024}.
    
\paragraph{C6 - Execution Environment} Reverse engineers know many relevant features of the environment in which the targeted software operates. They know the executable file formats, the system library APIs, relevant kernel APIs, etc. 

In the ASPIRE taxonomy, the construct \emph{Knowledge on execution environment framework} encapsulates this knowledge. In their research on malware evasion techniques and countermeasures, \cite{Wong2024} observe how malware uses system APIs as part of their evasion attempts.
    
\paragraph{Differences with Software Engineering}
Several papers also point out that the competencies needed for RE of binaries, and the activities performed thereto, differ to some extent from those that are useful in comparable SE tasks, such as source code comprehension. 

\cite{Votipka2020} observe that certain phases and activities in RE strategies are unique to RE of binaries, and do not take place when engineers aim for code comprehension on the basis of source code. That includes the \emph{Overview} phase, in which reverse engineers study the binary program as a whole, e.g., by executing it to observe its overall functionality, or by extracting meta-data from the binaries. Similarly, they state that scanning the code for beacons and comparing code fragments to reference implementations are activities only performed for binary RE, not for source code comprehension. 

\cite{hansch2018programming} observe that a considerable performance gap exists between experienced programmers and inexperienced ones for unobfuscated code comprehension, but also that that gap narrows considerably for obfuscated code. In other words, experience that helps with a common SE activity helps much less with a similar adversarial RE activity. 

\subsection{Design of Experiments}
\label{subsec:survey_experiment_design}
As students cannot be expected to have the same competencies, or the same level of competency, as experts, running experiments with students involves clear threats to external validity. Researchers can try to limit these threats by adapting their experimental design such that it limits the interaction between participant selection and the treatments. For that reason, it is useful to study how researchers may have adapted their experimental designs. A number of relevant aspects of the experiments covered in our literature study are listed in Table~\ref{tab:literature_review_part2}.

According to the Time column, most experiments have relatively short duration. The total task execution took half a working day or more ($>$4h) in only 9 experiments, and less than half a working day in 13 experiments. In only 2 experiments, participants had more than a day to complete their assignments. Real-world RE, by contrast, can often not be finished in such short times. For example, RE Skype took many months~\citep{skype,Biondi2006Silver}, and \cite{Ceccato2019} report RE experiments on applications designed to be representative of commercial product complexity running for a full month. 

In addition, we observe in the Size column that almost all experiments (that document this) feature very small programs, much smaller than real-world programs. This is of course in line with the short duration of the experiments. 

Clearly, both the small size and the short duration imply that the participants can only be given tasks with limited complexity. This complexity limitation can help to reduce the interaction between participant selection and treatment: If a participant's capabilities (e.g., strategic problem solving) to handle complexity is made less important or irrelevant because an experiment assignment does not involve complex problems, any such competency gap that may exist between experts and students will have less impact on how their execution of the assignment differs. In other words, limiting task complexity lowers the risk that the experiment results cannot be generalized due to competency gaps related to handling complexity. 

We also observed that in the experiments with students on protected software, only minimal obfuscation methods were used, often involving only identifier renaming or basic, single transformations rather than complex or layered techniques, to maintain task tractability~\citep{2014afamily,2014another,viticchie2016assessment,liu2017stochastic,hansch2018programming,kuang2018enhance,2020creative,2020splitting}. Together with the (sometimes minimal) training that the students got regarding the specific protections used in the experiments, this makes the gap in SP experience (C2) between students and experts less of a concern. However, deploying only a single obfuscation technique is not representative of real-world obfuscation deployment, where obfuscation recipes are used that prescribe which protections are synergistic and to be combined for protecting different types of assets~\citep{Liem08}. Generalization to real-world obfuscated software is hence anything but guaranteed.

Furthermore, in 7 of the 9 experiments involving obfuscated programs, the students worked on source code of the challenge applications (sometimes obtained from decompiled bytecode) instead of on binaries. This limits the impact of the gap that may exist with experts regarding their code reading competencies (C1). Moreover, in the experiments with decompiled Java code, the used obfuscations did not exploit the semantic gap between the Java source code and Java bytecode~\citep{java_obfuscation}, meaning that the decompiled Java source code is an accurate representation of the Java bytecode, not an unreliable one. This minimizes the impact of a lack of experience with low-level tools (C4). By contrast, commercial Java obfuscation tools used in production operate directly on Java bytecode
and intentionally exploit this semantic gap, which makes the decompiled Java
code incorrect or unreliable. This difference between the deployed obfuscations in experiments and those deployed commercially can obviously limit the external validity of the obtained results.

Beyond these foundational adaptations, some researchers provided concise textual documentation to explain the overall system architecture and purpose (e.g., what the Car Race application does and how to run it~\citep{2014afamily}, or how to run a client-server application~\citep{viticchie2016assessment}), allowing student participants to quickly understand the high-level design, thus alleviating to some extent the need for execution environment knowledge (C6). These summaries often helped students focus their RE efforts by clarifying the context of the challenge binaries. Additionally, providing clear, measurable goals or outcomes, for example, instructing participants to ``rename to meaningful names'' with a straightforward pass/fail metric~\citep{liu2017stochastic} or finding inputs that make the challenge binaries print ``success''~\citep{Mantovani2022}, prevents confusion and further supports effective evaluation. 

Several studies with students limited the freedom to operate (FTO) of the participants, and constrained them to familiar tools and environments on which they had been trained or had prior experience with, such as the Eclipse IDE~\citep{2014afamily,2020creative} or standard debugging tools like IDA Pro and OllyDbg~\citep{kuang2018enhance}. This approach ensured that students could adapt quickly to the experimental context without requiring additional training or encountering unnecessary complexity, thus reducing the relevance of the gap in tool usage skills (C4) with experts. \cite{Mantovani2022} developed their own very basic, web-based
interactive disassembler that the participants had to use. This tool allows subjects to view exactly one basic block of a CFG at a time, blurring out all others. This allows the tool to accurately measure how much time a participant spends reading the instructions in each basic block. The tool is severely limited compared to real disassemblers and lacks features such as artifact navigation, string searching, and string cross-referencing. A consequence of this strategy is that students and other participants are put on an equal footing by all having to use a previously unseen tool. Furthermore, the use of a restricted tool minimizes the potential issue that professionals with long-term experience with a particular tool are much more productive than students (C4). It also means that results are less generalizable to real-world disassembler usage scenarios.

None of the experiments with students required application domain knowledge, with the minor exception that one challenge of Mantovani et al.\ relied on understanding the standard C library APIs \texttt{listen()}, \texttt{bind()}, and \texttt{accept()}. Application domain knowledge (C3) beyond basic standard C library knowledge or knowing how to search through man pages was hence not required.

With these design choices, the interaction between the selected participants and the treatments is made as independent as possible from the skills that may distinguish professional reverse engineers from the students available in SE study programs. This clearly benefits the generalizability of the experiment's results from such students to professionals.

While it is clear that many researchers adapted their experiments to the participation of students, few explicitly discuss how their adaptations explicitly aim at reducing competency gaps between students and experts. \cite{2020splitting} state that they narrowed the gap in RE tool usage skills (C4) by providing source code to the participant instead of compiled code, noting that this presents a worst-case scenario, as if the participants had already been able to reconstruct and decompile the code (almost) completely, as experienced hackers could have done.

\subsection{Conclusions}

In the surveyed papers, we identified six motivations for conducting RE experiments. Specifically for assessing the strength of SPs applied to goodware by means of human experiments, students are a popular choice.

In that literature, the inability of many researchers to recruit students with extensive RE training resulted in those researchers minimizing the interaction between participant selection and treatment to rather extreme lengths, at the cost of increased interaction between setting and treatment. Commonly used ways in which this was done include short experiment duration; small challenge programs; single and simple obfuscations not exploiting the semantic gap between assembly code or bytecode on the one hand and source code on the other hand; starting from source code; limiting the FTO by requiring the use of specific, often basic tools; and requiring no domain knowledge. 

Almost no experiments have been reported where students use RE tools that professionals would rely on and where challenges are encountered of the complexity of real-world challenges. 

Most experiments with students relied on structured reporting and questionnaires to collect data, limiting the amount of collected data and potentially its accuracy. In the few experiments relying on event logging, most enforced the use of specific RE tools with logging capabilities onto the participants.

Few researchers actively managed the students' experience and skill diversity. 

All experiments with students fail to report fully to what extent the training of the student participants included the skills and knowledge that professionals would rely on for solving the same challenges. From the literature, we identified six such competencies, i.e., skills and knowledge domains: (C1) code reading, (C2) software protection knowledge, (C3) application domain knowledge, (C4) tool usage, (C5) strategic thinking, and (C6) execution environment knowledge.

Few papers involving students discuss privacy concerns, failing to argue why and how the participating students were consent freely, under no pressure.

 \section{Related Work}
\label{sec:related_work}

This section discusses relevant considerations expressed by other authors (not included in our RE literature review in Section~\ref{sec:literature}) regarding empirical SE and RE experiments, focusing mostly but not exclusively on the use of students.

\subsection{Use of Students in Empirical Software Engineering Research}
\label{sec:use_students_in_SE}
Within the domain of SE a debate exists on the use
of students vs. the use of professionals as subjects in controlled
experiments. There are several good reasons to involve students rather
than professionals, but also some reasons to be cautious 
\citep{falessi2018empirical}. In this section we will explore this debate. We
will focus on {\em software engineering} experiments, because ---outside the
papers surveyed in the previous section, which did not participate in this debate--- we found no related work
discussing the use of students in {\em reverse} engineering research.

Overall, we believe that most of the arguments and observations made in the debate on SE experiments also apply to RE experiments. All evidence we have in that regard will be presented in Section~\ref{sec:our-experience}. Where we believe counterarguments to exist without having strong evidence, we will add them to the debate in this section.

\subsubsection{Benefits of Using Students}

Compared to involving professionals, using students can offer a number of potential benefits \citep{falessi2018empirical,feitelson2022considerations} such as 
   easier accessibility,
   lower cost, 
   better adherence to treatment and instructions, 
   and higher motivation when experiments are a learning experience. 
Selecting subjects for inclusion in a study because they are easily 
available to the researcher and can be recruited at a lower cost
is known as {\em convenience sampling} \citep{Wohlin}.

However, the use of students has been criticized because it can lower the 
external validity of experiments, since students may 
   lack the experience, 
   knowledge, 
   and proficiency 
associated with professional software engineers. This can result in experiments with
students producing different results than the equivalent experiments done with professionals.
The situation is not clearcut, however, and \cite{berander2004} argues that students
can display higher levels of commitment and this can help make SE 
students comparable to SE professionals.

In the case of RE, in particular in experiments to assess the strength of SPs (M3) by measuring the effort required to defeat them, we fear it might be hard to make students more committed than professionals. By its very nature, such experiments have a capture-the-flag and cat-and-mouse-game nature, in which the participants try to ``defeat'' the protections deployed on the challenge binaries by the researchers, similar to how participants try to defeat each other in capture-the-flag and hacking competitions. This gives us a reason to believe, without hard evidence to back this up, that participating professionals aiming for personal satisfaction, prize money, or fame (depending on how an experiment with professionals is organized and results are publicized) might well be equally motivated and committed---if they actually participate in experiments at all, that is. 

\subsubsection{Benefits of Using Professionals}

Given the choice, no human subject research projects will prefer the use of
student subjects over professionals. Rather, students are seen as, at
best, convenient proxies of professionals to offset the difficulty of
recruiting professional subjects and to avoid the higher cost of
retaining their services.

Sj{\o}berg and Bergersen, however, argue that for well-designed and
well-executed experiments (which do come at at higher cost), it will
be possible to find professionals willing to participate and that they
will, in fact, be properly motivated \citep{Feldt2018}. This
nullifies some of the supposed internal validity benefits of relying
on students.

They also put forward the importance of assessing the
relevant experience of the participants, noting, however, that
experience is not equivalent to knowledge, skill, and motivation.

Even when professionals are available, based on their observations of
professionals, Sj{\o}berg and Bergersen state that it might not
always be better to choose professional subjects over students~\citep{Feldt2018}.
Caution is needed against relying on a few (outlier) studies to justify that
students are good proxies for professionals, and against the claim
that students would be good proxies for professionals in general.

As in any human subject study, sample size matters. Sj{\o}berg and
Bergersen warn that involving few participants risks the results
not being representative for a whole population~\citep{Feldt2018}. Thus, a larger sample
of student reverse engineers might be preferable to a smaller sample
of professionals.

\subsubsection{Assessing Population Diversity}
\label{subsec:assessing_diversity}

Using subjects from students vs. professional cohorts is not a
simple binary choice.  Rather, Sj{\o}berg and Bergersen argue that not
only variations in skill {\em between} different populations (students
versus professionals) matter, but also variations {\em within}
these populations \citep{Feldt2018}. 

Sj{\o}berg and Bergersen also put forward the importance of {\em
  assessing} the relevant experience of the participants, noting,
however, that experience is not equivalent to knowledge, skill, and
motivation. Hence they propose using pretests and possibly post-tests,
or dedicated instruments to predict performance, which in their case
is programming performance, not RE performance. In
the end, they urge the community to run more experiments with
professionals, and fall back on students only as a good alternative
that requires carefully and cautiously reporting and discussing the
limitations of student-based experiments, as well as addressing the
moderating effect of skill level on the benefit of treatment.

\cite{host2000} argue that senior (e.g., final year) software
engineering students can be relatively comparable to young software
development professionals because their overall experience overlap
quite much. This is because SE is the main subject taught in many study programs, and it is especially the case when the study program includes internships and large project work.

We do not believe this argument can be made for RE, as this is much more a specialty domain that involves activities~\citep{Votipka2020} for which the average computer science (CS), SE, or computer engineering (CE) graduate has not received much training.

\cite{feitelson2022considerations} and \cite{falessi2018empirical} similarly
advocate recruiting senior students whose skills and experience more
closely approximate that of early-career professionals, while
screening out those lacking the necessary competency or
motivation. \cite{feitelson2022considerations} furthermore argues that student
participants should never be used as experts.

Based on the existing literature, \cite{falessi2018empirical} conclude that
the picture is far from clear as to whether students are good proxies
for professionals, and that many more comparative studies are needed
to get a clearer picture. They report that experts in empirical
SE experiments agree (1) that one should think about
population and validity before conducting the experiment when
planning to use convenience sampling, and (2) that suitability and
representativeness of students as proxies for professional developers
change with different contexts and with different types of population.

Regardless of the subjects ultimately participating in an experiment, it is essential 
to properly document this population, such as whether students or professionals 
(or both) were used, and the educational and experiential background of each 
subgroup~\citep{Tonella2007}. 

\subsubsection{Assessing Participant Skillsets}

Selecting qualified participants is critical for ensuring valid and reliable results
in empirical research. \cite{RAINER2022107002} propose a framework for assessing participant credibility in SE studies. They define credible
participants as those they trust, believe or rely upon to provide valid external
information, i.e., about software phenomena that are external to the participant,
such as events occurring in a real-world setting, outside of the participant, which
the participant experiences, such as testing code. \cite{RAINER2022107002} high-
light factors such as domain knowledge and motivation, both of which are crucial
when working with students.

Additionally, pilot tests or pre-assessments can be used to balance
skill levels across groups~\citep{ko2015practical}. 

\subsubsection{Adjusting Task Complexity} \label{taskComplexity}

When it comes to adapting experiments for student participants in
SE, it is essential to align the complexity and
context of tasks with students' current expertise. At the same time,
experiments must reflect real-world scenarios so that they will
produce insightful data.

Adapting challenges is not a trivial task. Previous studies
on SE experiments with students~\citep{berander2004,falessi2018empirical,host2000,salman2015,feitelson2022considerations,ko2015practical} typically provide general guidelines rather
than detailed technical methods:
\begin{itemize}
   \item \cite{host2000} emphasize that tasks must be appropriate for the skills and
domain knowledge that students possess, so that the results remain
valid for their level of experience.  
   \item  \cite{berander2004}
states that tasks should also align with the students’ educational
needs or background so that learning objectives and research
objectives are combined.  
   \item \cite{salman2015} highlight that
participants’ previous familiarity (or lack thereof) with the approach
behind a task can shift outcomes, underscoring the need to match tasks
to the skills of participants. 
   \item \cite{ko2015practical} state that, concretely, researchers should use
problem contexts and programming languages students are familiar with
and avoid domain-specific knowledge beyond their training, and that pre-experiment training sessions
can help familiarize students with any new techniques.
   \item Finally, \cite{feitelson2022considerations} argues that, whenever possible, researchers should use tools and development environments that students already know or can learn
quickly. Feitelson also emphasizes that controlled experiments should, as far as possible, rely on participants who are sufficiently prepared for the task under study (through subject selection and/or targeted pre-experiment preparation), so that outcomes are not dominated by a lack of basic background knowledge.
\end{itemize}

Beyond matching tasks to student expertise, maintaining experimental
validity may also require careful calibration of task difficulty and
structured training.  \cite{feitelson2022considerations} underscores the importance of
ensuring tasks remain neither trivial nor overly complex, so that
measured outcomes genuinely reflect students’ ability to comprehend
the material rather than guess or apply shortcuts. Clear, well-scoped
objectives -- where students understand what they are trying to
accomplish in each task and how success will be measured -- help focus
their efforts and prevent confusion.

These measures not only enhance the educational value of the
experiments for the participating students, but also mitigate potential confounding factors, ensuring
that the outcomes more accurately reflect the effectiveness of the
task design rather than differences in prior experience. 

\subsubsection{Code Readability}
\label{sec:code-readability}

\cite{feitelson2022considerations} observes the importance of code readability for SE experiments, and argues that aspects of code readability should be controlled if they are not being studied. We agree and think this applies strongly to RE experiments as well. As discussed in Section~\ref{subsec:survey_experiment_design}, many researchers conducting RE experiments with students have tried to control this by letting them work on source code, which was in some cases obtained from correctly decompiled bytecode.

We actually believe the code reading competency gap between students and professionals to be more pronounced in RE of natively compiled software. Professional reverse engineers of native executables are familiar with optimized assembly sequences that implement common programming constructs~\citep{xchg_rax_rax,hackers_delight}. Students are typically not. In modern study programs with large SE components, low-level programming languages such as assembler, C, and C++ that feature constructs such as pointers and two's complement integer number representations often get little attention. This can impact the students' performance even when they do not need to read actual assembly code, but when they can instead rely on decompiled source code. For example, after (even light) optimization by a compiler, original source code such as \texttt{state = (value == 0 ? 9 : 4);} which everyone with some C programming experience can easily read, can be decompiled into \texttt{state = (ulong)(-(uint)(value == 0) \& 5) + 4;} which only experienced hackers read and interpret fluently. 

Controlling code readability by avoiding such constructs is not always feasible, however, as in some cases SPs cannot be compiled in other ways. The above constructs, e.g., are an integral part of the obfuscation technique known as control flow flattening~\citep{wangFlatteningTechReport}. Pre-experiment training to bridge the code reading competency gap, in line with \cite{ko2015practical}, can then provide a solution. 

In their study of decompiler fidelity issues, \cite{dramko2024decompiler} classify the above example as a \emph{readability issue} in which the decompiled code is semantically equivalent to the original code but communicated using different language features that are difficult to interpret. In addition, they identify \emph{correctness issues}. This involves cases in which the decompiler, by relying on unsound code analyses, produces decompiled code that is semantically different from the corresponding original code.

Such correctness issues obviously do not impact SE experiments in which participants only have to read original source code. In RE experiments in which binary executables are decompiled, by contrast, the possibility of correctness issues requires the participants to be skeptical about the decompiled code they are reading. Professional reverse engineers are aware of this and can handle this unreliability and uncertainty. This shows, e.g., in decompilers such as Ghidra offering plenty of interactive interfaces to reverse engineers to allow them to manually correct the tool's own analysis results. Unless they have been trained in this regard, we do not believe an average student will know to read the code with a skeptical eye, and neither will they know how to correct the issues. 

Controlling this form of code reading competency gap between students and professionals by avoiding such constructs in the assignments may not always be possible. For example, anti-disassembly and anti-function-construction obfuscations~\citep{linn2003obfuscation,jens21} specifically aim at preventing disassemblers and decompilers from obtaining correct code representations. If the strength of such protections is to be evaluated (M3), only pre-experiment training can bridge the code reading competency gap between students and professionals, as recommended by~\cite{ko2015practical}.

\subsubsection{Assessing Research Quality}

Runeson argues that in laboratory experiments, realism is of less
importance, because they focus on observing code behavioral processes, i.e., small components of SE processes~\citep{Feldt2018}. In such experiments, using students instead of
professionals might hence have an impact on the research quality, but
it should not be the sole criterion, only one factor. Shepherd
further argues, based on example experiments, that researchers should
consider the representativeness of participants, task, and setting,
including how they interact with each other \citep{Feldt2018}.

Although students may produce worse absolute results than
professionals, the different treatments of the used samples may have
the same effect on the performance of the students (e.g., slowing them
down with some factor) as on the performance of professionals~\citep{porter95,porter98}. 
Moreover, when they have to
apply new approaches, both students and professionals lack experience
with them, so students and professionals can perform similarly~\citep{salman2015}.

\subsection{Motivating Students}

Ensuring voluntary participation is not only an ethical obligation \citep{feitelson2022considerations} but also a requirement for regulatory compliance. \cite{caldwell2010strategies} highlight incentives like monetary rewards  (such as direct payments or prize draws) as effective motivators, along with educational strategies to enhance participants’ understanding about the value of research.

\subsection{Data Collection}
\label{subsec:datacollection_background}
Regarding the use of free-form reports and transcribed interviews to capture unstructured thought processes, it has been argued
that they depend on self-reporting and the subjects' willingness and ability to share details accurately, creating potential recall errors and social desirability biases, and that manual analysis of such data may be prone to subjective interpretations~\citep{lethbridge2005studying,Nunkoosing2005,podsakoff2012sources}.

Unlike the event logging methods observed in our survey (Section~\ref{subsec:survey_data_collection}) that all rely on specific tools being logged, Taylor's approach of tool-agnostic data collection~\citep{Taylor2022,taylor2019getting} is not limited to specific tools. Taylor's data collection software collects screenshots, keystrokes, mouse clicks, active window information, etc.\ during RE experiment sessions, regardless of the specific RE tools used. By analysing that data, the timeline of RE activities can be coded, i.e., annotated with higher-level descriptions of the actions the participants have executed. We used Taylor's system to collect data in our own experiments with students (as will be discussed in Section~\ref{sec:experience-datapipeline:revenge}).

Finally, some researchers have explored specialized hardware to capture fine-grained data. For instance, eye-tracking and electroencephalogram (EEG) devices have been used in broader software-engineering studies~\citep{Ishida2019}, offering insights into exactly which artifacts participants focus on and the types of mental tasks they perform (e.g., syntactic parsing, memorization). However, such experiments require expensive, specialized equipment which might not always be feasible in a classroom setting, and adds another level of privacy and ethics concerns.

\subsection{Privacy and Ethics Considerations}

When conducting experiments involving students, it is critical to protect their rights, privacy, and well-being. \cite{feitelson2022considerations} emphasizes that students should not feel coerced into participating in an experiment by their instructors, underscoring the importance of ensuring that participation is voluntary and free from undue influence. Additionally, researchers must take special care to prevent coercion and minimize stress, recognizing that a student’s primary role is educational and that research activities should not negatively impact their learning experience.
 
\cite{SingerVinson2002} note that enrolling one’s own students as participants can create a power imbalance. Students may be reluctant to refuse or withdraw for fear of academic repercussions. To address this, they recommend clearly separating research activities from grading to protect students’ anonymity and ensure genuine voluntariness.

A context in which similar questions about power imbalances have been discussed is the use of student participant pools in Psychology departments ~\citep{moorthy2020recruiting,walker2020opportunity}. Many departments rely on such pools to ensure adequate sample sizes: typically, students can earn a small amount of course credit (for example, up to 10\%) by taking part in one or more approved studies of their choice via a central portal. When carefully designed and overseen, these systems can satisfy ethical requirements for voluntary participation. At the same time, several authors have argued that the structural dependence of staff research output on these pools, combined with the link to course credit, can create perceptions of pressure and raise questions about how robust voluntariness really is ~\citep{moorthy2020recruiting,walker2020opportunity}.

\subsection{Reverse Engineering (of Executables)}

Last in this overview of related work, we note that some authors studied and discussed RE of executables and the involved processes and competencies without reporting actual experiments. 

\cite{bryant2011software} decomposed the functions of RE of binary executables to isolate the different information-processing and sensemaking subtasks. Their discussion of different data representations,  of situation awareness and understanding, of the RE sensemaking processes, and of knowledge modeling confirms the need for the six competencies discussed in Section~\ref{subsec:survey_human_competencies}. They also discuss several methods to elicit mental models, including various forms of cognitive task analysis (e.g., through document analysis and structured interviews) and of verbal protocol analysis (e.g., thinking aloud voice recording). 

\subsection{Software Complexity Metrics}
In non-adversarial contexts, i.e., on software not transformed specifically to prevent code comprehension, SE studies have been conducted on the validity of using software complexity metrics to estimate code comprehension difficulty~\citep{Feitelson23}.

Several software complexity metrics that originated from SE research, such as cyclomatic complexity~\citep{mccabe1976complexity} and various length metrics~\cite{halstead1977elements}, have also been proposed to evaluate the potency of SPs~\citep{collberg1997taxonomy}. \cite{desutter2024evaluation} observed software complexity metrics to be by far the most popular form of strength measurement in papers presenting new obfuscation techniques.

Yet not a single paper surveyed in Section~\ref{sec:literature} aimed to validate the use of software complexity metrics on protected software. The only two papers that studied how complexity metrics impact RE tasks~\citep{burk2022,sutherland2006reverse} did not involve protected software.

 \section{Our Experience with Student Reverse Engineering Experiments}
\label{sec:our-experience}

Over four consecutive course editions (2021--2024), we ran a series of empirical RE experiments embedded in our Software Hacking and Protection (SHP) course at Ghent University. Guided by the  existing literature on such experiments, and by broader methodological guidance on student-based experiments as discussed in Section~\ref{sec:related_work}), we describe the educational context, our experimental design and challenge binaries choices, and the main outcomes and lessons from these four years. We highlight recurring design concerns---keeping tasks well scoped, preparing participants, accounting for skill differences, and putting explicit ethical safeguards in place.

\subsection{Educational Context and Research Aims}
\label{sec:experience-context}

Our own experiments with students focus mostly on motivation M3, namely to study and assess the impact of novel protections that we designed in our own research groups on RE processes. In one experiment round, we also tried to replicate results on the best RE strategies (M1) obtained by \cite{Mantovani2022}. Our focus on M3 in all but one experiment round is important, as much of what we describe below on how we try to close the gap between the performance of students and professionals can only be justified for M3, not for M1 or M4.

For our experiments, we involve students enrolled in a Master-level, 6 credit, 12 week course on Software Hacking and Protection (SHP). To enroll for this course, students need to have already taken prior courses on computer architecture (the software/hardware interface -- assembler), on C programming, and on operating systems (process execution, file systems, memory protection). The SHP course features modules on binary file formats and assembly code, static binary RE (disassembly and decompilation), dynamic RE (debugging, tracing, fuzzing, pointer chain scanning), attack strategies (use cases: game cheats, crypto key extraction, passphrase and license key checks), tampering (interposition, code editing), SP (obfuscation, anti-tampering, anti-debugging), MATE risk management, empirical SP and RE research, side-channel and fault-injection attacks, and malware analysis and forensics. We therefore teach them the basics of C1, C2, C4, C5, and C6.

Every week, the students in our SHP course study written documentation as well as short demonstration videos, and in most weeks they perform self-study hands-on exercises in preparation of a 75 minute Q\&A or lecture, followed by a 3-hour lab in which they use actual techniques and tools. In the last two course editions, the used tools were Ghidra’s disassembler and decompiler (used in 6 labs), the GDB debugger (5 labs), ELF utilities (\texttt{objdump}, \texttt{readelf}, \texttt{strings}, 2 labs), \texttt{scanmem} (1 lab), \texttt{LD\_PRELOAD} (2 labs), \texttt{strace} and \texttt{ltrace} (3 labs), and the Tigress\footnote{\url{https://tigress.wtf/}} obfuscator (2 labs). In the first two editions, the students used IDA Pro (without decompiler) instead of Ghidra, and fewer labs were devoted to RE.  

The labs serve both as guided learning activities (TAs support students when they get stuck) and as an evaluation component; grading is largely participation-based, such that well-prepared students who engage seriously with the lab are expected to pass with a good grade even if they do not fully complete every task. To that extent, the assignments prescribe a clear path towards solving the ever more complex challenges, i.e., which attack strategies to deploy. The link between RE strategies and useful tools and techniques to reach the strategies' consecutive subgoals is emphasized, e.g., by giving the students diagrams that summarize the relations between the most relevant artifacts in the attacks to be executed and by explaining upfront how the features of different artifacts can be exploited when searching for the relevant artifacts and how to navigate between them. Towards the end of the semester, and hence leading up to our experiments, they hence have hands-on experience with many of the (basic) RE strategies, techniques, and tools that professional reverse engineers use, including all those strategies, techniques, and tools that they will have to deploy during the experiments, as well as with SPs.

Of course not all professionals use the same tools. For example, some will prefer IDA Pro or Binary Ninja over Ghidra. Moreover, our students have no hands-on experience with more complex tools such as Intel PIN for instrumentation, so their training does not at all imply full external validity. But because of the described training, we were able to provide the participants stripped binaries to which advanced obfuscations~\citep{jens21,jens2022flexible} had been applied, some of which exploit the semantic gap between source code and binary code~\citep{jens21}.

\subsection{Participants and Course Setting}
\label{sec:experience-participants}

The SHP course enrolled between 30 and 54 students annually in 2021--2024. Participation in our experiments is voluntary: students can complete the course without taking part in the experiments, and their grades are not influenced by experimental outcomes. These participation statistics are reported in Table~\ref{tab:course_details}. 

Across the four editions, participation was substantial, peaking in 2023 and remaining high in 2024 (Table~\ref{tab:course_details}). 

\subsection{Motivating Student Participation}
\label{sec:experience-motivating-participants}

Over the years, we introduced several measures to lower participation barriers and to increase the students' motivation to participate. We have no objective evidence to attribute changes in participation to the individual measures, but can report that we informally experienced positive feedback when the measures were executed and communicated (e.g., facial expressions, nodding, attentive attitude). We hence present these measures not as proven techniques, but rather as options that researchers can consider for future experiments with students. 

\begin{table}[t]
\centering
\renewcommand{\arraystretch}{1.2}
\setlength{\tabcolsep}{4pt}
\begin{tabular}{|c|c|c|c|c|}
\hline
\textbf{Edition} & \textbf{Students} & \textbf{Registrations} & \textbf{Drop-outs} & \textbf{Gift Card} \\
\hline
2021 & 54 & 17 (31\%) & 7 (41\%) & €50.00 \\
\hline
2022 & 52 & 36 (69\%) & 3 (8\%) & €50.00 \\
\hline
2023 & 30 & 30 (100\%) & 1 (3\%) & €100.00 \\
\hline
2024 & 51 & 40 (78\%) & 2 (5\%) & €50.00 \\
\hline

\end{tabular}
\caption{Participation details by course edition. \textbf{Students} is the number of students enrolled in the class. \textbf{Registrations} is the
number of students who registered for the experiment. \textbf{SDrop-outs} is the 
number of those registrants who withdrew from participation at some point
during the experiment.}
\label{tab:course_details}
\end{table}

\paragraph{Pseudonymous Participation}
One key strategy is to make participation pseudonymous and optional. Because the researchers---who also serve as teaching staff---cannot see which students opted in or out, students do not have to worry about potential consequences tied to their identities. By separating identity from experiment data, we remove a range of concerns that could otherwise deter participation. We discuss more about these ethical considerations in Section~\ref{sec:experience-privacy_and_ethics}. 

\paragraph{Integration into the Course}
We closely align the experimental tasks with the coursework they are already doing, to minimize the additional work required for participation and to reduce disruption to the course schedule. Concretely, the challenges in one round of experiments are exactly the ones of the weekly lab assignment for which the students are required to submit solutions as a part of the course, and our experimental session also coincides with our regular weekly lab session. If students agree to participate, the only additional step is downloading and activating the data collection software (discussed in Section~\ref{sec:experience-datapipeline}) which pseudonymously collects their data while working on the challenges. The participants then work the challenges and, when done, complete a brief
pseudonymous post-experiment questionnaire. In Section~\ref{subsec:pseudonymous}, we detail our specific protocol to ensure that student remain pseudonymous, even though students submit solutions to the same assignments for credit. 

\paragraph{Participation-based Grading}
During these lab sessions, when experiments also take place, we adopt a primarily participation-based grading approach, meaning students earn credits if they demonstrate genuine effort and preparation, even if they do not arrive at a fully correct solution. We hope this policy also helps to discourage cheating, thus making it feasible to include more challenging RE tasks that might exceed some students’ skill levels. Since we want to accurately analyze the steps students take to solve the challenges for those who agree to participate as part of the experiment, the teaching assistants (TAs) provide only essential technical support (e.g., troubleshooting software or environment issues) rather than direct guidance on how to solve the challenges. Importantly, the experimental challenges were designed to rely only on tools and tool features that students had already practiced in earlier labs, which reduced the need for substantive help during the experimental sessions.

For this reason alone, participation-based grading is a necessity. In any lab, there is always a tension between trying to have the lab be experienced as a learning opportunity versus seeing it as an evaluation~\citep{schinske2014teaching}. This tension grows larger if the students are not helped by TAs. By grading mostly based on participation~\citep{gillis2019reconceptualizing}, the lab/experiment can be experienced by the students as a self-study or training opportunity, rather than as a pure evaluation activity. 

At the same time, the fact that students receive a higher grade if they solve the challenges completely, helps us to motivate them to try to perform well, even with their reward for participating to the experiment being formally independent of their success. Because higher course grades are tied to fully solving the challenges, students have a clear incentive to try to perform well beyond mere participation, which helps reduce the risk that some participants would otherwise join mainly for the incentive and exert limited effort, an issue raised in prior student RE studies that used bonuses or prize draws as discussed in Section~\ref{subsec:literature_motivation}.

\paragraph{Teaching about Scientific Methodologies}
In the last two editions of the course, we included modules on scientific methodologies. We discussed how difficult it is to evaluate newly proposed protections~\citep{desutter2024evaluation}, how research into novel protections is a form of information technology design science research and hence  should apply proper evaluation methodologies~\citep{dsr}, how the domain suffers from a lack of standardized evaluation methodologies~\citep{Basile23}, how empirical research can produce new insights~\citep{Ceccato2019}, how research can suffer from various threats to validity~\citep{Wohlin}, and how awareness of such threats can benefit students in the course of their studies, e.g., for improving how their master's thesis dissertation will be received and graded.  

We also emphasized the benefits of contributing to the scientific community. We explain how real‐world research into RE can help identify common challenges, develop new protection strategies, and improve automated tools used in industry, thereby contributing to a safer software ecosystem.  By highlighting that their contributions help shape these important future developments, students can feel a sense of pride in supporting work that extends beyond the classroom. 

While we have no hard evidence that covering such topics in our course impacted the students' motivation to participate, we strongly feel it did as we got positive feedback on this part of the course.

\paragraph{Minimal Overhead}
For our experiments, we try to reduce the overhead for the students to the absolute minimum. Two forms of overhead are relevant. First, we minimize the burden on reporting: we ask the students to submit exactly the same report for their graded lab participation (under their own name) and for their experiment participation (using their pseudonym). Furthermore, we reduce the post-experiment questionnaires (discussed in Section~\ref{subsec:survey-new}) to the absolute minimum.

Second, we minimize the runtime overhead of the data collection software (which we will discuss in Section~\ref{sec:experience-datapipeline}). In the 2021 edition, some participants reported noticeable CPU usage and virtual machine slowdowns during the experiment. While students cannot anticipate such overhead beforehand, this kind of friction can increase frustration and can contribute to withdrawals once the session is underway. Between 2021 and 2022 we therefore optimized the software and logging configuration to reduce its performance impact. Consistent with these changes, the drop-out rate decreased substantially in the later editions (Table ~\ref{tab:course_details}), although multiple factors changed across years and we do not attribute this effect to any single measure.

\paragraph{Financial Motivation}
Finally, we offer a  gift card to each participant who submits a solution obtained with reasonable effort, which provides a tangible reward for students’ time and effort. Students claim their gift card pseudonymously, as will be explained in Section~\ref{subsec:giftcard}. 

As shown in Table~\ref{tab:course_details}, the combination of user-friendly data collection software, pseudonymity, financial incentives, and teaching about scientific methodologies coincided with increased participation. In 2021, only 17 of the 54 enrolled students participated (a 31\% registration rate) and 7 dropped out. By 2022, after improving the software and keeping a €50 gift card, we achieved a 69\% registration rate with an 8\% drop-out rate. In 2023, we further increased the gift card to €100, achieving a perfect registration rate of 100\% with minimal drop-outs. Of course this increased registration rate could also be due to other factors, but the fact that the registration rate dropped to 78\% in 2024 when we lowered the gift cards to €50 again (for tax compliance reasons as will be discussed in Section~\ref{subsec:giftcard}) suggest that the financial motivation is strong.

\paragraph{Ethics Approval}
Finally, we dedicate sufficient time and effort to explain to students how the ethics committee in our faculty operates and what the procedures are for obtaining ethical approval for experiments. Participation is based on informed consent: students are told what data are collected, how pseudonymity is maintained, and that they can withdraw at any time without academic penalty. We also give them access to the specific ethical protocol of the experiments and any other relevant documents. While we again have no proof that this helps to motivate the students, we are convinced that it helps them to understand that their rights are safeguarded and that they are absolutely free to participate or not. We can only hope that this has a positive effect on their willingness to participate. 

\subsection{Experimental Design Across Years}
\label{sec:experience-design}
In our experiments, we choose not to use participants’ prior course scores to assess skill levels because linking these scores to specific individuals would conflict with our pseudonymity protocols, as will be discussed in Section~\ref{subsec:pseudonymous}. Although we could measure skills pseudonymously before the experiment (e.g., through a skill test or self-reported data in a pre-experiment questionnaire), we avoid doing so in order to minimize overhead and avoid potential inaccuracy of self-reports.

Our experiment program evolved over four years. In 2021, we ran a single-round trial experiment that primarily served as a shakedown of our data collection and annotation pipeline~\citep{zhang2024reanalyst,taylor2019getting}. This experiment involved two challenge binaries that differed mainly in the use of static (one-way) versus dynamic (two-way) opaque predicates, and in how the opaque-predicate state  (i.e., the hidden run-time state that determines the predicate’s truth value) was encoded in program data structures~\citep{jens2022flexible}. The primary goal was to probe a within-session transfer  (learning) effect between two closely related challenges solved back-to-back, and to gauge how far we could increase protection strength before the tasks became infeasible for master students. This pilot also tested the practicality of deploying RevEngE in a classroom setting, including its impact on virtual-machine performance and data-upload reliability. 

From 2022 onward we adopted a two-round design per year. A first round with one primary treatment was used both to collect pilot data on challenge difficulty and to assess participant skill. In that first round, all participants received the same (or sufficiently similar) binaries. This design choice allows us to check whether the challenge difficulty is appropriate, and to estimate each participant’s relative skill level for use in the second round. A drawback is that we cannot control for random heterogeneity of participants \citep{wohlin2012experimentation} in the first round--—i.e., high- and low-skill participants may perform differently, potentially affecting the results. Nevertheless, with only one (or nearly uniform) treatment in the first round, we avoid conflating skill differences with multiple distinct treatments \citep{wohlin2012experimentation}, thereby mitigating threats to validity from unstratified skill levels. 

We then considered how many tasks each participant had completed successfully in the first round and how long it had taken them to do so to partition them into groups by skill level.

In the second round, we then addressed the heterogeneity by treating participant skill as a \emph{blocking factor}~\citep{wohlin2012experimentation}, meaning that from each skill group, we distributed the students randomly but evenly over the different treatments. As a result, each treatment, i.e., each variation of a challenge binary, was assigned the same (plus/minus 1) number of students from each skill group. This balanced design limits confounding effects of skill disparities between the different treatments. By comparing outcomes across skill‐blocked groups, we reduce confounding from skill disparities and make treatment comparisons more interpretable. This choice directly responds
to a limitation highlighted in our survey of prior student experiments in Section~\ref{subsec:student_participation}: many studies randomize without balancing for skill, while others give all participants the same binaries and therefore leave heterogeneity uncontrolled.

In 2022 and 2023, round~1 used a single baseline challenge configuration (i.e., one experimental condition) with relatively simpler binaries (often fewer protections or more transparent data representations), allowing us to assess individual skill and verify that the challenges and logging setup were working as intended. Round~2 then used more heavily protected binaries and multiple treatments (e.g., different protection configurations or tool setups). In 2024 we followed the same two‑round structure.

All experimental sessions were scheduled near the end of the semester, after students had completed all other SHP RE labs. Each round ran during a single 3–4 hour lab session, and Round~1 and Round~2 were always held on different days, one week apart. This timing ensured that students had already been exposed to the necessary tools and techniques, and avoided overloading them with back-to-back experimental tasks. 
 
\subsection{Design of Assignments}
\label{sec:experience-challenges}

RE of non-trivial applications, in particular of protected software, can require considerable effort. The longer-running experiments in Table~\ref{tab:literature_review_part1} are a testament to that, such as the month-long experiments conducted by \cite{Ceccato2019}. As they and \cite{Votipka2020} observed, professional reverse engineers execute complex attack strategies when they reverse engineer goodware. For example, they formulate hypotheses and then execute corresponding actions, including validation of those hypotheses. If a hypothesis is later refuted, through checks or because a dead end is reached, they backtrack to try again under alternative hypotheses. A typical example is that they choose their analysis tools based on assumptions about the protections that might be deployed in the software. Some engineers select more complex tools, such as dynamic analysis tools, assuming that static analysis tools will be thwarted by obfuscations, while others first try static analysis, only to move on to dynamic analysis if the static one fails to produce results. Another aspect of complexity is that full RE tasks consist of sequences (a.k.a.\ attack paths) of many different activities (a.k.a.\ attack steps) in which the engineers iteratively collect ever more information~\citep{Ceccato2019,Votipka2020}. In malware analysis complex strategies are used as well~\citep{Wong2021inside,Wong2024}. 
In short, real-world RE tasks come with a considerable FTO, and consist of complex sequences of activities. The (alternative) strategies with which such tasks are executed, as well as the executed attack steps and attack paths in them, are therefore modeled with attack graphs~\citep{attack_graphs} or attack trees~\citep{attack_trees} that often are quite complex. 

Conducting RE experiments with student participants while still reflecting real-world RE hence poses challenges. Due to the students limited availability, their limited skills and their limited knowledge, they cannot be asked to execute complex, complete real-world attack strategies (C5) in experiments. 
However, this should not prevent researchers from asking students to perform individual attack steps, i.e., concrete RE activities, or short chains of attack steps, that closely resemble the activities that expert reverse engineers would have to execute as part of their job. 

In our experiments, we do so (i) by explicitly limiting the FTO of the participants, by (ii) providing documentation and prepared analysis artifacts, and (iii) by describing the task such that its outcome corresponds to the goal of a specific attack step. Whereas complete attack graphs and trees (should) model all possible attack paths on an application or on a specific asset therein, limiting the participants' FTO comes down to forcing them to focus on specific attack paths. Providing documentation and prepared analysis artifacts regarding the challenges then allows the participants to skip certain information-gathering attack steps within those paths, thus allowing them to focus on specific attack subpaths. For example, in some experiments we prepared a pre-configured Ghidra project for the participants in which relevant artifacts had already been labeled. Or we gave them custom scripts to fuzz a specific computation in a program relevant for their assignment, as if an expert had already reached that point of custom tool development in their overall attack strategy. Others have also relied on this option to ``skip'' preparatory attack steps, such as the many experiments in which obfuscated source code was handed to the students, as if the decompilation step had already been executed (see Section~\ref{subsec:survey_experiment_design}). Providing documentation can also make up for holes in the student's application domain knowledge (C3) or execution environment knowledge (C6). Finally, choosing task outcomes that correspond to attack step results allows the participants to stop after the studied attack steps have been executed. For example, when attacking a license key checker, we did not ask the participants to tamper with the checker, we asked them only to identify true checks and decoy checks. 

Combined, these options relieve the participants of complex strategic thinking (C5) and of the need to have (extensive) a priori application domain knowledge (C3) and execution environment knowledge (C6), thus bridging to some extent one of the competency gaps that may exist between students and experts. 

These options to focus RE assignments on specific attack steps and attack subpaths are particularly useful in research on SP, because individual SPs typically aim to impact only specific attack paths. For example, data flow obfuscations aim to hinder data flow analysis. So to assess the strength of data flow obfuscations, one can use experiments that focus on attack steps that execute data flow analyses, or on specific later attack steps that depend on the outcome of a data flow analysis, such as some data localization or code localization step that depends on the precision of a data flow analysis result. 

This is in line with how \cite{schrittwieser2016protecting} classified the status of different obfuscations' resilience vis-\`a-vis the different possible attack step goals and vis-\`a-vis the different possible types of analysis (automated dynamic analysis, automated static analysis, pattern matching, and human-assisted analysis), and how numerous proposals for new obfuscation techniques have reported the impact of those techniques on the performance of concrete tools. For example, \cite{linn2003obfuscation} and \cite{jens21} tested how disassemblers were affected by their anti-disassembly obfuscations. 

\subsubsection{Reasons to Limit the Students' Freedom to Operate}

Apart from making the assignments doable for student participants, limiting their FTO can help ensure that valid (e.g., statistically significant~\citep{struts}) conclusions can be drawn from an experiment, because it can help to ensure that multiple participants perform the same (sub)task in the same way. Furthermore, by allowing them to skip attack steps unrelated to the SPs being studied, we can limit the amount of measurement noise by reducing the time spent on exploratory steps that are outside the scope of the intended measurement.

Another reason to limit their FTO is that reverse engineering lightly protected software, i.e., binaries protected with a limited number of obfuscations, is already hard enough for many students. Layering too many protections on top of each other risks making the challenges too difficult, for two reasons. First, there is the direct reason that the code simply becomes too confusing. Secondly, there is the indirect reason that students, while still taking a course on RE and SP, do not yet possess the mental capability to combine the information they have received about many different obfuscations. In our experience, their minds and working memories can be focused on a few specific obfuscations while prepping them for a specific experiment (as we will discuss in Section~\ref{sec:experience-training}), but they cannot be relied upon to recognize the fingerprints of a wide range of obfuscations. Hence, to make the challenges tractable within a limited time, both the number of deployed obfuscations and the extent of layering need to be limited.

Not layering as many protections as one would in practice~\cite{collbergbook,Liem08} may leave experimental challenges susceptible to alternative strategies that would otherwise be blocked, and these strategies can become unintended paths-of-least-resistance. To avoid participants pursuing out-of-scope strategies (which, in a deployed setting, would be mitigated by omitted layers), it is therefore also necessary to limit their FTO, which then in turn requires considering these constraints explicitly when interpreting and reporting results.

We conclude that in many cases, it is necessary to limit the students' FTO. We do not consider this a problem, as long as the researchers document this properly, as long as the imposed limitations make sense in light of the research questions and research hypotheses, and as long as the researchers then correctly limit the conclusions of the study to the allowed RE strategies.

Two options to limit their FTO are (i) to give the students concrete instructions on which attack steps to deploy and in what order; (ii) to forbid them to use certain attack methods, such as brute-forcing. Our experience is that most if not all students follow the given instructions.

\subsubsection{Our Challenge Designs}

In line with these observations and our motivation M3, we endeavored to study the strength of advanced SPs that---unlike the basic obfuscations studied in experiments reported in the survey in Section~\ref{sec:literature}---do aim at exploiting the semantic gap between assembly code and C source code~\cite{jens21} and that do aim at obfuscating program state by reusing complex data structures already used in a large, complex application~\cite{jens2022flexible}. Almost all of our experiment rounds featured one or more challenge binaries that have been protected with (different combinations of) such obfuscations. The tasks on those challenge binaries were always constructed with the goal of being solvable within a single 3--4-hour lab by well-prepared SHP students. The tasks were framed as well-scoped, objective goals with clear success criteria. As our experiments' outcomes will reveal in Section~\ref{sec:experience-lessons}, meeting the requirement of being solvable within a 3-4~hour time frame proved to be very hard to meet.

Tables~\ref{tab:our-experiments-part1} and~\ref{tab:our-experiments-part2} summarize the main characteristics of our 2021-2024 experiments. For each year and round we list how many tasks each student had to complete, i.e., how many binaries each student had to work on; the nature of those programs; the treatments, i.e., the applied obfuscations and other code transformations; the information and additional artifacts provided to the students to help them focus on specific RE subtasks; size indicators of the programs; the task they had to accomplish; and any limitations on their FTO, such as instructions to use a specific tool or technique, or instructions forbidding them to use some technique. Unless noted otherwise, all the binaries provided to students were stripped. 

\begin{table}[t]
  \centering
  \small
  \setlength{\tabcolsep}{3pt}

\begin{tabular}{l p{10cm}}
\specialrule{1pt}{1pt}{1pt}
\multicolumn{2}{c}{}\\[-0.5em]
\multicolumn{2}{l}{\emph{2021 - single round}}\\[-1.7em]
\multicolumn{2}{c}{}\\
\multicolumn{2}{c}{}\\
\textbf{\# Tasks}  & 2\\
\textbf{Programs}  & Crack-me C programs checking whether provided pair of license keys meet validity requirements\\
\textbf{Treatments} & differently implemented flexible opaque predicates~\citep{jens2022flexible}\\
\textbf{Size}     & SLoC post manual obfuscation: 134, 149, 146, 155 \\ 
\textbf{Information} & Format description and valid examples of first key for each binary; format of the description of the second key, one valid second key \\ 
\textbf{Task}      & Find another, significantly different valid second key\\
\textbf{FTO}       & Hint to use IDA Pro, gdb, Wolfram Alpha (for solving equations), and Linux tools such as \texttt{strings}, \texttt{objdump}, \texttt{ltrace}\\
\specialrule{1pt}{1pt}{1pt}
\multicolumn{2}{c}{}\\[-0.5em]
\multicolumn{2}{l}{\emph{2022 - round 1}}\\[-1.7em]
\multicolumn{2}{c}{}\\
\multicolumn{2}{c}{}\\
\textbf{\# Tasks} & 3\\
\textbf{Programs}  & 1) 2 warm-up C crack-me programs that each accept a unique key\\
                   & 2) \texttt{sqlite} database engine with embedded license key checker that checks whether provided key meets all validity requirements\\
\textbf{Treatment} & 1) two variants of crack-mes, obfuscated with different combinations of Tigress obfuscations (string encoding, function merging, control flow flattening)\\& 2) differently implemented license checkers, using different data structures for the flexible encoding of state~\citep{jens2022flexible}\\
\textbf{Information} & 1) Overall program design\\
                     & 2) Architecture of license checker, format of keys, hints on how to locate relevant artifacts with static and dynamic approaches \\
\textbf{Size}     & 1) 138 C SLoC, excluding generated data headers and before obfuscation\\
                   & 2) \texttt{sqlite}: +1M C SLoC\\
\textbf{Task}      & 1) Find the correct key\\
                   & 2) Identify data structure storing checker's internal state \& how it stores that a check failed; determine checks performed on the key and distinguish true checks from decoys\\
\textbf{FTO}       & 1) specific attack path instructed using specific IDA Pro functionality\\
                   & 2) no brute-forcing or (manual) fuzzing allowed for differentiating true checks from decoys\\
\midrule
\multicolumn{2}{c}{}\\[-0.5em]
\multicolumn{2}{l}{\emph{2022 - round 2}}\\[-1.7em]
\multicolumn{2}{c}{}\\
\multicolumn{2}{c}{}\\
\textbf{\# Tasks} & 2\\
\textbf{Programs}  & \texttt{verit} SMT solver with embedded license key checker that checks whether provided key meets all validity requirements \\
\textbf{Treatment} & same as 2) in round 1, but using different data structures for flexible state encoding\\
\textbf{Information} & same as 2) in round 1, plus information as to what aspects of the checker's implementation have not changed compared to round 1, source code of all performed checks (true ones + decoys), and an IDA Pro database in which relevant artifacts have already been labeled; \\ \textbf{Size}     & \texttt{verit}: approx.\ 52k C SLoC\\
\textbf{Task}      & same as 2) in round 1\\
\textbf{FTO}       & same as 2) in round 1\\
\specialrule{1pt}{1pt}{1pt}
\end{tabular}
\caption{Summary of our student RE experiments, part 1: 2021--2022.}
\label{tab:our-experiments-part1}
\end{table}

\begin{table}[t]
  \centering
  \small
  \setlength{\tabcolsep}{3pt}

\begin{tabular}{l p{10cm}}
\specialrule{1pt}{1pt}{1pt}
\multicolumn{2}{c}{}\\[-0.5em]
\multicolumn{2}{l}{\emph{2023 - round 1}}\\[-1.7em]
\multicolumn{2}{c}{}\\
\multicolumn{2}{c}{}\\
\textbf{\# Tasks} & 2\\
\textbf{Programs}  & Two RE-Mind crack-me C programs reused from \cite{Mantovani2022}\\
\textbf{Treatment} & binaries were not stripped\\
\textbf{Information} & none\\
\textbf{Size}     & 146 and 207 SLoC\\
\textbf{Task}      & identify conditions to trigger "success"\\
\textbf{FTO}       & none\\
\midrule
\multicolumn{2}{c}{}\\[-0.5em]
\multicolumn{2}{l}{\emph{2023 - round 2}}\\[-1.7em]
\multicolumn{2}{c}{}\\
\multicolumn{2}{c}{}\\
\multicolumn{2}{l}{Same as 2022 - round 2, but with less complex license check implementations}\\
\specialrule{1pt}{1pt}{1pt}
\multicolumn{2}{c}{}\\[-0.5em]
\multicolumn{2}{l}{\emph{2024 - round 1}}\\[-1.7em]
\multicolumn{2}{c}{}\\
\multicolumn{2}{c}{}\\
\textbf{\# Tasks} & 4\\
\textbf{Programs}  & 4 crack-me versions of the GNU Core Utilities program \texttt{head}, bit-manipulation on 16-byte command-line key values to produce compared value\\
\textbf{Treatment} & inlining, opaque predicates with different forms of bogus control flow, such as forward vs. backward branches into existing code; applied differently to create 4 versions of each crack-me\\
\textbf{Information} & Description of key format, recognizable label of top-level bit-manipulation function in Ghidra project\\
\textbf{Size}     & unobfuscated \texttt{head.c}: 786 C SLoC; linked binaries: approx.\ 24kB .text sections \\
\textbf{Task}      & identify keys that trigger "success"\\
\textbf{FTO}       & none \\
\midrule
\multicolumn{2}{c}{}\\[-0.5em]
\multicolumn{2}{l}{\emph{2024 - round 2}}\\[-1.7em]
\multicolumn{2}{c}{}\\
\multicolumn{2}{c}{}\\
\textbf{\# Tasks} & 3\\
\textbf{Programs}  & 3 small crack-me programs that manipulate lists of numbers and perform computations on elements in resulting list\\
\textbf{Treatment} & inlining, code layout randomization~\citep{jens21}, opaque predicates, control flow flattening; all applied and combined differently to create 3 versions of each program\\
\textbf{Information} & name of the one top-level function of interest (which was the only one not stripped from the binary)\\
\textbf{Size}     & 139, 109, 111 SLoC before obfuscation\\
\textbf{Task}      & identify inputs that make the computation evaluate to 42\\
\textbf{FTO}       & not spend more than one hour on the first two (simpler) tasks\\
\specialrule{1pt}{1pt}{1pt}
\end{tabular}
\caption{Summary of our student reverse engineering experiments, part 2: 2023--2024.}
\label{tab:our-experiments-part2}
\end{table}

All challenge artifacts, including source code, binaries, compilation scripts, documentation and assignment descriptions are available at \url{https://github.com/csl-ugent/RE-experiment-artifacts}.

Compared to existing experiments on obfuscated goodware as discussed in Section~\ref{subsec:survey_experiment_design}, the main difference is our experiments did not shy away from providing binaries to the students protected with the aforementioned types of advanced obfuscations, and that we were hence not restricted to small code samples that can be decompiled correctly but instead could study how the failure to correctly decompile code impacts the RE with state-of-the-art tools such as IDA Pro and Ghidra that are popular among expert reverse engineers.

\subsubsection{Narrow Focus}

To make the students focus on the specific RE tasks we are interested in, and to minimize the effort they need to invest in the
preceding but (to us) uninteresting tasks, we experimented with multiple options to provide them additional information and prepared artifacts throughout these experiments. 

The first option is to give them symbolic information, i.e., symbols that identify relevant artifacts in the binaries and that reveal relevant properties of them, such that the participants do not need to waste time finding those artifacts. We tried several methods to do so: 
\begin{enumerate}
    \item When the participants use specific analysis tools such as disassemblers or decompilers that store all the information they retrieved and reconstructed from a binary in a custom database, we can prepare such a database for the participants in advance and augment it with additional symbols, labels, notes, comments, bookmarks, etc. Interactive disassemblers and decompilers offer the necessary functionality to add such information, with the databases serving as a kind of notebook to their users, which can then query their notes easily to search and navigate through the artifacts of the binary. By giving the participants a database that has been augmented with useful information, we give them the same notebook that reverse engineers would have obtained by assembling and encoding that information themselves. 
    \item Rather than completely stripping the binaries from all symbols, it is possible to keep some symbols, such as the names of global variables or functions. Those symbols are then automatically added to the tools' databases, without the need for the researchers to add them manually.
    \item One can embed artifacts in the challenge binaries that are easy to find or trigger with the tools the participants will use. For example, one can let the program print recognizable strings when relevant code fragments get executed. If those strings are embedded statically, it is trivial to find those fragments through string searches and cross-references in static analysis tools, instead of having to use dynamic tools such as debuggers.  
    \item Rather than statically linking all code and data into a single binary, some artifacts can be separated into dynamically linked libraries, in which they can be referenced with information-carrying symbols.
\end{enumerate}
All of these worked fine in our experiments. The first two methods are of course the least intrusive, because they encode information in exactly the same way as reverse engineers would encode it. 

The second option we used is to provide additional information is to explain to the participants the high level designs of the license manager, key checker, ... that the students are targeting as part of the challenge. This mimics a scenario in which reverse engineers have already learned this information from previous tasks. This is a realistic scenario, which assumes that no security through obscurity is being relied upon but that, instead, the actual protection strength originates from the diversity in the deployment of the studied SPs.    
    
Another method we used to make the participants focus on specific, relevant tasks without wasting time on other tasks is to give them custom tools that reduce the effort needed for those other tasks as much as possible. As an example of a custom tool, we gave them a script in which they can copy-paste arithmetic expressions from decompiled source code to determine (with a brute-force approach) whether or not those expressions (likely) evaluate to constant values. This mimics the realistic scenario in which a professional reverse engineer has built and collected a range of scripts throughout their career to automate commonly recurring tasks. Of course it is the researchers' responsibility to only provide custom tools that can also be found in the toolboxes of professional reverse engineers, and for that reason the provisioning of such tools needs to be documented explicitly in the reports. 

The final method we used to avoid having inexperienced student participants waste time on minor mistakes on which professionals would (most likely) not waste time is to let the program explicitly print their inputs. Especially in capture-the-flag challenges where participants need to find the correct inputs that make a program print a ``success'' message or that make a license check succeed, this can help. It does so by reducing the chance that participants misunderstand or neglect how textual inputs provided on the command line or in files (e.g., ASCII strings representing hexadecimal numbers in some range) are interpreted and converted into internal data formats in the programs, such as signed integers. Similarly, we let the program produce error messages when the provided inputs do not meet format requirements, such as the length of valid license keys, even if those requirements were already highlighted in the assignment documentation.

Making the program print additional information to avoid minor mistakes needs to be done cautiously, however. For example, to print error messages when provided license keys do not meet the format requirements, the program needs to perform an actual check on the provided keys. If the error messages are not hidden in the static binary and the format check happens in the same code fragment that also prints the message, the participants can easily find the checking code, and thus obtain the address where the keys are stored in the program's address space. In that way, it might create a new path-of-least-resistance for the participants. Care has to be taken that such paths are not introduced unintentionally. 

\subsubsection{Avoiding Unnecessary Complexity}
In our own experiments with student participants, we opted to compile the challenge binaries with the open-source compilers LLVM and GCC at optimization levels -O1 and -O2 because those levels offer, in our view, the best balance between the generated code being representative of assembly code as found in real-world binaries on the one hand, and allowing us to control the complexity of the binary code given to student participants on the other hand. Less optimization (-O0) yields unrealistic code on which not even register allocation has been applied, and more optimization (-O3) often increases the gap between the original source code structure and the generated assembly code to such an extent that the optimization can be experienced as an obfuscation in itself. In several instances, we actually experimented with both levels -O1 and -O2, and then eventually selected the level that produced the most interesting code structures in relation to the objectives of our study. 

For example, for our 2024 experiment we generated samples by inserting bogus code into the original source code by means of opaque predicates. Some bogus code contained backward gotos, i.e., gotos that jumped to labels higher up in the source code, thus introducing unstructured (a.k.a.\ irreducible) control flow including loops. Compiling those samples at different compiler optimization levels produced different binaries, with different (but unstructured) control flow. The Ghidra decompiler, which, like other decompilers, has been tuned to work well on the patterns of control flow that compilers produce for structured code (with few exceptions such as gotos used for error handling), then also produced different decompiled source code for those different binaries. For some optimization level, the decompiled loop contained while loops and gotos; for another level it only contained gotos. As we were interested in measuring the impact on the required RE effort of having an obfuscator introduce control flow that appears legit at an adversary's first glance while actually being bogus, we opted for the level that gave us the code with loops. In general, we advise that researchers make the choice that is most aligned with their research questions and research hypotheses, avoiding unnecessary complexity when possible, and to report about their choices transparently.  

One interesting option to explore in future research would be the development of custom LLVM/GCC backends that avoid the use of exotic, rarely used instruction opcodes and instead only use the more basic instructions that most students are familiar with. The purpose is then of course to not confuse the student participants with unusual instructions that they haven't been taught and which would slow them down. To the extent professionals are familiar with such instructions, avoiding their use in the binaries would also help to minimize the gap in relevant code reading competencies (C1) between students and professionals with respect to the used challenges. Similarly, certain compiler optimizations that result in ``weird'' optimization idioms or name mangling could be disabled in a custom compiler configuration mode used for generating challenges for students.

\subsubsection{Quality Assurance}

In addition to selecting an appropriate level of complexity in the design of the challenges, quality assurance is needed to ensure that the actual implementation does not introduce unintended complexity or unintended simplicity. 

\paragraph{Avoiding Unintended Complexity}
Poorly-implemented challenges can increase complexity and effort for the student participants, potentially compromising study outcomes and participant motivation. One such issue emerged in one of our experiments due to unanticipated compiler behavior, specifically inlining, which increased the complexity of the tasks.

For our 2023 Round 2, the source code included 10 distinct checks participants needed to identify and analyze. However, after compilation, these checks expanded to 38 separate occurrences within the binary due to aggressive compiler inlining. This unexpected duplication elevated the complexity and workload for participants, requiring them to spend additional time deciphering repetitive logic unnecessarily. As we did not detect this issue in advance, by the time this oversight was discovered during the experiment, the students had already made substantial effort, creating frustration and potentially skewing experimental results.

This is just one example of what can go wrong when tools such as compilers and code rewriting tools are used to produce the actual binaries. To mitigate these risks, quality assurance of challenge design and implementation is critical. This includes eyeballing the generated binary code and pilot testing with tools the students will use to identify issues like unexpected transformations. This is work intensive, of course. Alternatively, one might be able to automate the checking by writing scripts that examine the generated assembly code. 

\paragraph{Avoiding Unintended Simplicity}
When software obfuscations are deployed by means of source-to-source rewriters such as Tigress, the most popular obfuscation tool for research on the obfuscation of C code according to a recent survey~\citep{desutter2024evaluation}, the risk exists that the compiler undoes certain obfuscations completely or partially, thus producing binaries in which the code is simpler than what the researchers aimed for. For example, optimizing compilers can simplify mixed-boolean-arithmetic expressions or subexpressions, they can compile out opaque predicates if they can determine their constant value, and they can compile out entire artifacts such as security checks if the compilers deem them not to contribute to the semantics of the program. 

In addition, the compilation tools can be misconfigured, for example by enabling the generation of debugging information or additional symbol information, or by forgetting to strip the binaries of all not strictly necessary such information. 

To mitigate this risk, the same quality assurance in the form of eyeballing the binaries and pilot testing them is necessary.

\subsection{Training and Preparation}
\label{sec:experience-training}

Right before the start of the RE experiments in which we give them such binaries, we further refresh their assembly and decompiled code reading skills (C1) with a focus
on the structures and idioms that they will encounter in the experiment challenges. We also present a refresher of some of the basic knowledge and tool capabilities (C4) on which they will have to rely during the experiments, discussing how certain tool functionality can be used for specific activities that they will need to perform for some of the challenge binaries, and we provide them with additional custom tools that professionals know how to develop quickly or would already have developed to increase their productivity (C4). 

For example, in the 2024 experiments students were asked to analyze binaries protected with techniques
similar to the stealthy protection integration of \cite{jens21}. To make their attacks feasible, the students received the following last-minute training (related to the different required competencies) before the experiment: 
\begin{itemize}
    \item C1 and C2: We explained the semantics of specific decompiled code fragments, structures, idioms, and patterns such as the ternary operator fragments mentioned in Section~\ref{sec:code-readability}, framing them in the context of flattened control flow to ensure that the students knew where to expect them and how to interpret them. 
    \item C4: We provided them with the source code for a simple fuzzing tool in which they can copy-paste predicate expressions from decompiled code to test whether or not those expressions likely are opaque predicates (that always produce the same value). 
    \item C4: We showed them how to edit bogus control flow paths (such as those introduced with opaque predicates) with Ghidra, reminding them that the decompiler is then invoked again, now producing deobfuscated code. 
    \item C4: We showed them how to correct Ghidra if it assigns incorrect meaning to control flow, such as when it misinterprets a jump instruction as a tail call. 
    \item C1 and C4: We discussed (unsound) heuristics that reverse engineers might use to identify unrealizable control flow paths, such as when control flow conditions involve uninitialized variables, or when Ghidra's decompiler has given variables names that reflect issues discovered during its data flow analysis. 
    \item C2: We revisited the obfuscation techniques that the students had previously studied in theory, and discussed some practical impacts of those protections on the disassembly and decompilation process. For example, we showed how the tools' heuristics can fail on code transformed with these protections and the fingerprints such failures can leave behind in the decompiled code.  We also presented the tool functionality with which the user can fix mistakes by the default heuristics. 
\end{itemize}

The 2023 edition included binaries that embedded a license key checker protected with the flexible SPs of \cite{jens2022flexible}. In this case, we gave the students an introduction that included the high-level design of that license key checker, i.e., how it can be seen as an automaton that operates on abstract states. We also discussed the meaning of the different states but did not reveal how they are implemented. With this introduction, we aimed at giving the students the same level of application domain knowledge that a professional might have (C3). 

In summary, we tried to ensure that much of the relevant knowledge and experience embedded in the long-term memory of professionals was refreshed and present in the short-term memory of the student participants at the start of their participation in the experiments. This way, we aimed to ensure that our student participants were sufficiently prepared for the experimental tasks, in line with Feitelson’s guidance \cite{feitelson2022considerations} as discussed in Section \ref{taskComplexity}.

Obviously, both the basic training as part of the course and the last-minute training and documentation in anticipation of the experiments, risks biasing the participants, in the sense that it steers their behavior during the experiments. We consider this acceptable, on the condition that we, the researchers, are transparent about the training of our participants and on the design of the challenges, such that others can assess the threats to external validity such training might induce. In short, if the participants are steered towards behavior that can be expected from professionals when those would replace the students, and this is reported transparently, this in our view poses no problem. 

We hence expect researchers to report on the types and amounts of relevant training that their participants received at the level with which we described our training above, instead of what we observed in existing literature, i.e., in the descriptions that are not more specific than those listed in Table~\ref{tab:literature_review_part1}. As supplemental material, a sufficiently detailed description of the training material should be made public, as well as the challenges themselves. As for the latter, the shared material should not only include the binaries, but also a description of what, e.g., the decompiled code looked like with the decompiler versions that the participants were using in the experiment. The reason is of course that it might be hard to reproduce this information years later when that tool version is no longer available, and that it is necessary to understand how that training corresponded to the exact structures, idioms, patterns, etc.\ that the participants faced in the experiment.

With an eye on reproducibility, the researchers should of course also share the entire pipeline that generates the challenge binaries, including the source code of the unobfuscated samples, the obfuscation script, the compilation/linking/stripping commands, the final binaries, the decompilation command + decompiled binaries (if relevant), etc.

\subsection{Data Collection and Analysis Pipeline}
\label{sec:experience-datapipeline}

All four years of student experiments used Taylor’s RevEngE framework for tool-agnostic data collection, combined with our reAnalyst tooling for semi-automatic annotation and analysis~\citep{zhang2024reanalyst}. RevEngE records time-stamped screenshots, keystrokes, mouse clicks, active window information, and process lists in participants’ virtual machines, while allowing them to pause logging or add free-text comments. reAnalyst then processes these logs to detect which tool is active, which functions and basic blocks are in view, and which annotations should be attached to each time interval, complementing any manual annotations or tool-specific logs. 

\subsubsection{Our Data Collection: RevEngE}
\label{sec:experience-datapipeline:revenge}
To balance the need for cheap, scalable, and detailed data collection with the goal of minimizing disruption to participants' normal workflows, we build on Taylor's data collection software RevEngE~\citep{Taylor2022,taylor2019getting}. Some technical details of our adaptions to RevEngE, as well as the rationale for using such software in RE experiments, have been discussed extensively in our prior work~\citep{zhang2024reanalyst}. Here we summarize its main characteristics. 

Unlike many prior approaches, RevEngE is tool-agnostic and unobtrusive: participants can use their preferred RE tools without noticeable interference, and system performance remains largely unaffected. This enables us to study real-world RE strategies at scale while minimizing intrusion into participants' workflows.  RevEngE captures time-stamped activities based on participants' interactions with their VMs. Periodically, it collects high-quality screenshots suitable for accurate optical character recognition (OCR). Additionally, it records process and thread details, including the CPU and memory usage of running processes, mouse inputs with x/y coordinates at each button press, and keyboard inputs. The collected data is then uploaded to our server and stored in a relational database. 

To deploy this data collection method in our student experiments, we followed a straightforward protocol designed for convenience and privacy. At the start of each experiment, participants download RevEngE onto their course VMs, which was provided by us at the beginning of the semester. Once installed, the software starts to record all interactions on their VMs, but participants can easily pause the recording at any time for their privacy. After the experiment is finished, they are requested to uninstall the software using a script we provide. 

\begin{figure}[t]
  \centering
  \small
    \begin{verbatim}
        14:37:48, Symbol: FUN_0010ed40, double click
        14:37:48, Entered Function: FUN_0010ed40
        14:38:09 - 14:37:57: Feature: Rename Function, Word: main
        14:39:52, Symbol: DAT_00288bb, single click
        14:39:54 - 14:40:02: Feature: Edit Label, Word: keyplus0x1000
        14:40:10, Symbol: keyplus0x1000, single click
        14:40:13, Symbol: Find References to keyplusOxl000, single click
        14:40:15 - 14:40:18: Feature: References to
        14:40:26 Entered Function: FUN_001A3A20
        14:40:26, Symbol: bVar8, single click
        14:40:28 - 14:40:29: Rename local variable, Word: license key
    \end{verbatim}
    \caption{\label{fig:sample_output} Annotations generated by reAnalyst from a ``locate the license key'' challenge performed by a Ghidra user.}
\end{figure}

\subsubsection{Our Data Analysis: reAnalyst}

In our first experiments with Taylor’s tools to annotate data logs collected with RevEngE, we found that annotating one minute of RE activity requires more than one minute of a researcher’s time. Such annotating needs to be performed by multiple RE researchers for reasons of accuracy and cannot be outsourced to, e.g., a Mechanical Turk because of the complex and domain-specific nature of the data to be annotated. We therefore concluded that in order to scale, Taylor’s method needs to be augmented with a semi-automated annotation process.

We therefore developed our own automated data annotation framework for RE experiments, reAnalyst \citep{zhang2024reanalyst,faingnaert2024tools}. reAnalyst processes the recorded stream of screenshots, keystrokes, active processes, and mouse click data collected by RevEngE, applying OCR and text processing techniques to analyze the data and extract meaningful annotations. These annotations include the order and timestamps of functions visited by a participant, buttons and URLs clicked on in the used GUI RE tools along with the type of click (single or double click), and words typed while using the specific disassembler features, as demonstrated in Figure~\ref{fig:sample_output}.  

The generated annotations can then help answer research questions by providing insights into how participants interact with various tools; how they traverse functions and their control flow graphs, navigating between artifacts and searching for them; and how and when they engage in debugging or deobfuscation tasks. In particular it supports research prompted by some of the motivations discussed in Section~\ref{sec:motivation}, including M1 by revealing evolving RE strategies, M2 by highlighting tool usage and limitations, and M4 by exposing the competencies linked to effective RE. 

In addition, reAnalyst's automatically generated annotations can help research\-ers assess whether participants followed experiment guidelines that might have restricted their FTO. For example, the tool can reveal whether debuggers, fuzzers or other dynamic analysis tools were used by a participant that was instructed to rely on static analysis techniques only. 

Another advantage is that the high resolution timing provided by RevEngE allows reAnalyst to reduce measurement noise. For example, consider an experiment in which the researchers wish to study the impact of certain protections on specific attack steps, but for practical reasons they can only record the total time required by the students to solve the challenge. All the time student participants then spend on other, preparatory attack steps or on peripheral issues, and the variations that students exhibit while handling those will introduce noise into the measurement. By excluding the time that participants spend on preparatory RE steps or on peripheral issues based on timestamps collected with RevEngE, reAnalyst can more accurately measure the amount of time the participants actually spend on a the specifically targeted attack steps.

By automating much of the annotation process, reAnalyst reduces the labor-intensive work required for manual data analysis, making it possible to study RE strategies and tool effectiveness on a much larger scale. reAnalyst has been tested and evaluated, demonstrating its reliability and efficiency, as evaluated in prior work~\citep{zhang2024reanalyst}. 

\begin{figure}[t]
\centering
\includegraphics[width=0.95\textwidth]{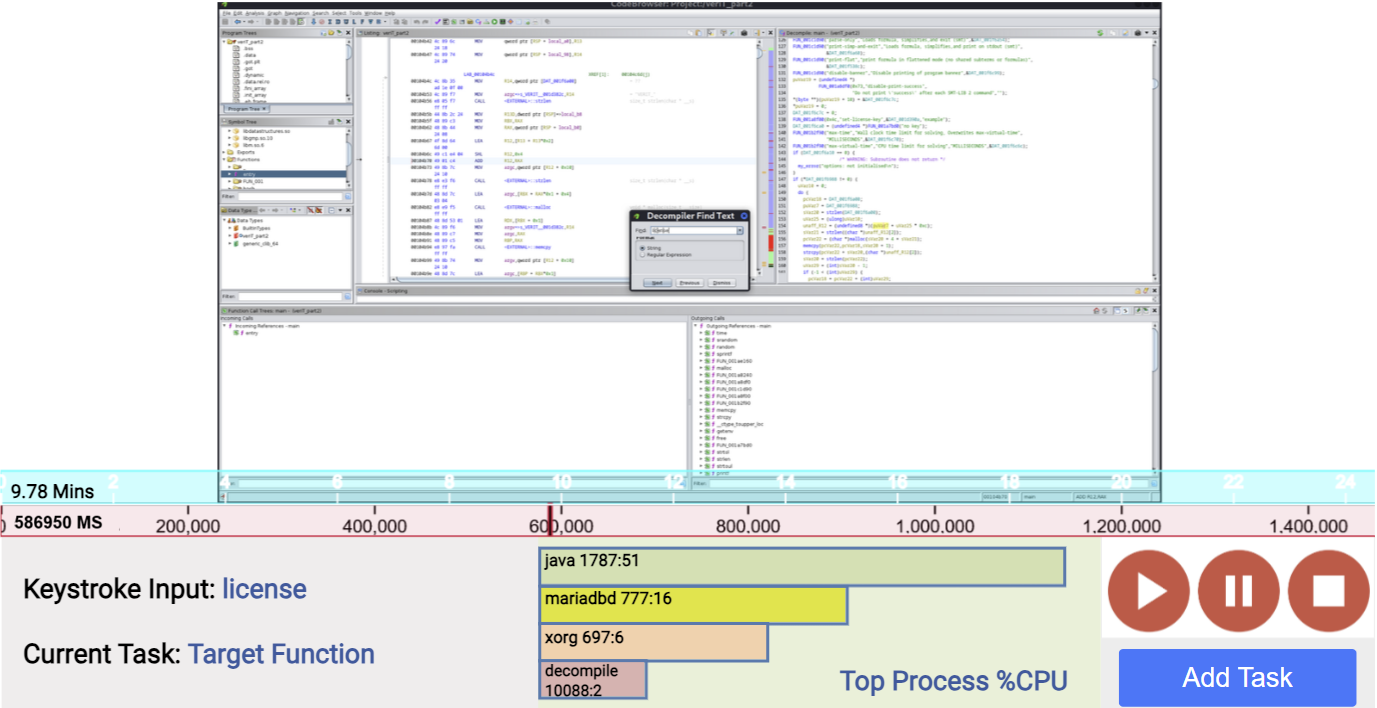}
\caption{\label{fig:timeline} Simulated animation view of RevEngE, which replays the collected screenshots along with recorded keystroke input (``license''), the current task annotation generated after function annotation (``Target Function''), and the top active processes and their CPU usages. One can use the buttons on the right to play, pause, or manually add task annotations to the timeline.}
\end{figure}

After annotations are generated automatically by reAnalyst and possibly refined manually, we upload the annotations to RevEngE. Its data visualization component then offers an animation view feature for researchers to view the collected data   along with annotations, as shown in Figure~\ref{fig:timeline}. Additional annotations can then still be added during session review. This supports a wide range of research activities: for example, it helps investigate M1 (strategy insights from session replays), M2 (identifying tool limitations), M3 (tracing the effort needed to bypass protections), and M6 (informing how RE might be more effectively taught).

The source code of RevEngE and ReAnalyst integrated into the complete system ReVeal, can be downloaded from \url{https://github.com/csl-ugent/reVeal}.

\subsubsection{Post-Experiment Questionnaire} \label{subsec:survey-new}

In addition to using data collection software, we ask participants to complete an online post-experiment questionnaire pseudonymously. This report includes their final solutions (e.g., the correct license key), as well as approximate completion times for each subtask and any RE tool features (e.g., bookmarks, comments, defined strings) they may have used. They can also share with us any comments or feedback. The list of survey questions from the experiment in 2024 is provided in Appendix~\ref{postsurvey2024}.

While questionnaires can provide valuable insights, they also introduce potential inaccuracies as discussed in Section~\ref{subsec:datacollection_background}. Self-reported completion times may---intentionally or unintentionally---be imprecise or students may forget the exact set of RE tools and tool features they used while solving the challenges. As stated above, reAnalyst is used to check and, if necessary, correct or augment the questionnaire responses. 

The final solution reports, for which the students simply copy and paste their answers from the report they already had to submit to the course lab, always provide a true reflection of participants' final outcomes. There is no reason for them to intentionally report an incorrect solution if they are able to solve it, and if they did not solve it, they simply will not be able to report a correct solution.  

We primarily use the information in the questionnaire for quick, early analyses or for categorizing students into different groups after the trial round. Additionally, we collect them (especially the final solutions) as a backup of the most important data in case the data collection software fails, though no such failures occurred after the 2021 experiment. For detailed analyses---such as studying how different protection techniques affect the time required to solve a challenge---we always cross-check self-reported data with the data collected by RevEngE and make necessary corrections. 

\subsection{Outcomes}
\label{sec:experience-lessons}

This subsection only highlights a few observations from each year that are most relevant for our methodological discussion. It does not attempt to summarise all measurements or statistical analyses from the experiments.

In 2021 we ran a single-round pilot with two opaque-predicate challenges. This mainly helped us test RevEngE in a course setting and see how small changes in data structures and opaque-predicate strategies affect how hard a challenge feels for students. Because only a limited number of students participated and our logging configuration slowed down the virtual machines, we were unable to draw strong quantitative conclusions about protection strength from this run. In 2022 we moved to a license key checker embedded in complex programs \texttt{sqlite} and \texttt{verit}, providing with pre-labelled variables for the latter, and prescribed use of IDA. This showed that we could integrate our experiments into a large existing code base, but the implementation of the license key checker itself turned out to be too complex for students to solve within a short lab session. We therefore treat the 2021 and 2022 editions mainly as sources of experience that informed the later experiment designs, rather than as datasets for detailed quantitative analysis.

In 2023, Round~1 replicated the two RE-Mind challenge binaries from~\cite{Mantovani2022}. Unlike the original RE-Mind experiment, we did not require participants to use the RE-Mind instrumented tool. Students performed the tasks in Ghidra, while we collected tool-agnostic interaction traces using RevEngE and analyzed them with reAnalyst. Based on these logs, we reconstructed the same kinds of visualizations as Figure~5 in their paper (time spent per basic block over the course of a session) and Table~2 (prevalence of function-level exploration strategies for novices and experts), and we observed qualitatively similar patterns, thus demonstrating that no reliance on existing tool functionality is required. However, compared to~\cite{Mantovani2022}, we observed differences in directional versus exploratory analysis strategies. We attribute these differences largely to tooling: our students used a full-featured RE tool (Ghidra) with search and cross-reference support, rather than the limited-purpose tool in the original RE-Mind study. In 2023, Round~2 reused the \texttt{verit} setup from 2022 with a less complex license checker implementation, but again too few students managed to complete the tasks within the lab session to support meaningful quantitative analysis, so we chose not to use this \texttt{verit} setup again in later editions. Nevertheless, the collected data such as screenshots and keystrokes from both \texttt{verit} rounds were valuable as development/training data for our reAnalyst pipeline~\citep{zhang2024reanalyst}.

In 2024, Round~1 focused on progressively heavier obfuscations of small bit-manipulation programs (from unobfuscated baselines to opaque predicates, forward and backward \texttt{goto}s, code splitting, and confused function boundaries). This confirmed how quickly the readability of decompiled code can become the dominant obstacle for students, even after targeted training on the relevant idioms. We also experienced that the PhD students and postdocs in our research group, all of whom have experience with low-level system software but several of whom do not conduct research on SP or RE, are much better than the average master's student in understanding bit-manipulation code, a.k.a.\ bit twiddling/fiddling. Concretely, the pilot tests we executed while designing the experiment with members of our research group on various candidates for the challenge binaries failed to reveal the difficulty that master students would have with them. This taught us the valuable lesson that we need to rely on pilot participants who are more comparable to the students in our course than our own research group members.

Round~2 then used three small list-processing programs whose observable behaviour could be described by a simple goal (e.g., “produce output~42”), but where different variants changed how the core computation was integrated (inlined vs.\ separate functions), how much layout randomization and bogus code was added, and which protections were present. We observed qualitative differences in which variants students tended to solve and how they approached them in our analysis, but a thorough quantitative comparison between these treatment variants is outside the scope of this article and is left for future work. 

Overall, our personal experience is that it is extremely hard to design RE challenges for students. For example, we have found it very hard to determine the appropriate level of SP to treat the samples that students need to reverse engineer. Students and their performance seems to be extremely sensitive to the level of deployed protection, and the delta between challenges that are too simple and challenges that are impossibly hard seems to be small.

Finally, many of the concrete recommendations that we will present in Section~\ref{sec:recommend} draw directly on these four years of experience. Our emphasis on pilot rounds and calibration, on constrained yet realistic tasks with controlled FTO, on targeted training to bridge experiment-specific competency gaps between students and experts, and on transparent documentation of protections and challenge design all grew out of the specific successes and failures observed in the student experiments described above.

\subsection{Privacy and Ethics Considerations}
\label{sec:experience-privacy_and_ethics}

As mentioned in Section~\ref{sec:experience-motivating-participants}, we organized our experiments to overlap with course labs to increase the number of participants and reduce time commitments: students already submit solutions to the assignment to gain course credits, so they can readily opt in to the experimental aspect. Therefore, our protocol must ensure that students can pseudonymously participate in the study and receive a gift card afterward, and still receive credit for their course assignment. To keep this arrangement practical and ethically sound, we introduced a dedicated ethical protocol and obtained formal IRB approval from our institution.

\subsubsection{Pseudonymous Participation}\label{subsec:pseudonymous}

All students complete the same graded lab assignment for the course. Opting in to the study is purely optional and has no effect on grading: participation only authorizes pseudonymous logging of RE activity during the lab and submission of a brief pseudonymous post‑experiment questionnaire. Grades are awarded independently of study participation, and graders have no access to research data. Conversely, researchers with access to research data are not involved in grading and cannot view course submissions.

To preserve pseudonymity while integrating the experiment into the lab, consenting students receive a randomly generated participant number before the study starts. They use this number only for experiment‑related actions (downloading/activating the data‑collection software and completing the post‑experiment questionnaire) and never link it to their course submission. Students are explicitly instructed not to reveal this number in any graded work. If a participant wishes to withdraw, request data deletion, or report technical issues, they do so pseudonymously via an online form that requires only the participant number.

Operationally, the two data flows are separated. Course submissions under students’ real names are handled on the course platform by a teaching staff member who is not part of the research team and who has no access to research data (including participant numbers and  collected data). Researchers who can access research data are prohibited from accessing the lab grading part of the online course platform, from accessing the students' grades, and from taking part in their grading; in practice, they are never involved in grading course assignments.

While, in principle, one could attempt to correlate experiment data with identities by comparing fine‑grained details of solutions, our protocol and strict role separation are designed to prevent such re‑identification. This setup lets students satisfy all course requirements and, if they wish, contribute to the study. If they do decide to contribute, they can be confident that participation neither affects their grade nor how they are perceived by their instructors.

Our protocol provides pseudonymity rather than true anonymity. As we already discussed in Section~\ref{sec:literature-privacy-ethics} under the European GDPR, \emph{anonymous} data is defined as data that cannot be linked to an individual by any technical or legal means~\citep{SRBvEDPS_T557_20}. In our case, although the risk of re-identification is low---given that only the separate teaching staff could theoretically match a name to a participant number---it is not zero. As the RE tasks also serve as lab assignments, we cannot make the data fully anonymous. Consequently, we do not qualify for the typical waiver of ethics committee approval that can be offered to studies collecting fully de-identified data. Instead, we have obtained explicit IRB approval for this protocol, treating all participant data with the same rigor as any personally identifiable information, even though no direct identifiers are collected within the research data itself.

\subsubsection{Privacy Controls During Data Collection}
Participants in our experiments work within a VM monitored by our custom data collection software, which captures all on-screen interactions as discussed in Section~\ref{sec:experience-datapipeline}. To protect their privacy and maintain pseudonymity, each participant can click a ``Pause'' button at any time to pause recording if they need to enter personal information, especially their passwords, or if their identity such as their name is revealed on the screen. The instructions and software user interface clearly inform participants of this option, and in our observations, no one has inadvertently disclosed sensitive personal credentials.

We have, however, noticed a small number of participants revealing their real names, typically by viewing the university course portal (which displays their name) within the VM rather than on their host machine, and without pausing. Although clicking ``Pause'' takes only a moment, some users may simply forget or remain unconcerned about pseudonymity. Others may not have read the instructions thoroughly and therefore misunderstand how pseudonymous participation works. Nevertheless, whenever we encounter a participant’s real name during analysis of collected data streams, we disregard it and do not link it to any data, including any summarized data or data produced by reAnalyst for further analysis. Moreover, the individuals analyzing the data have no role in grading the course assignments, as noted previously.

\subsubsection{Managing Physical Presence During Experiments}
Although our protocol protects pseudonymity in data handling and grading, a separate concern arises when teaching staff are physically present during lab sessions. Even if staff members do not recognize students by name, students may still worry that simply being seen working on the experiment compromises their pseudonymity. This issue is not unique to our setting. As we observed in Section~\ref{sec:literature-privacy-ethics} \cite{2014afamily} conducted student RE experiments in which participation was optional and only students who chose to take part attended. Their instructors, who provided lab assistance, could observe exactly who came to the experimental sessions, thus knowing which students did participate and which did not. 

By contrast, in our setup, \textit{all} students attend the regular lab hours--—whether or not they opt into the study. The professor does not attend these lab sessions, and the TAs provide only essential support. Although participants do use the data collection software (while nonparticipants do not), it typically runs in a minimized or background window, making it hard to distinguish them from nonparticipants unless one scrutinizes individual screens. As a result, unless there is a technical issue with the data collection software---at which point the TA might learn that the student is participating---TAs cannot readily identify participants just by seeing who is present. Moreover, TAs limit their involvement to clarifying assignment instructions or resolving major technical issues, providing no direct guidance on the challenge itself, and they do not take notes or record any observations during the session. This arrangement normalizes the lab environment for everyone and helps maintain the pseudonymity of those who choose to participate.

\subsubsection{Claiming Gift Cards}\label{subsec:giftcard}
Our student participants receive a gift card reward, but since participation is pseudonymous, we must ensure that only eligible participants receive their rewards, without knowing their identities. As reported in Section~\ref{sec:literature-privacy-ethics}, existing work on experiments with students has not addressed how to distribute these rewards pseudonymously while restricting them to eligible participants. 

In our protocol, as participating students submit their solutions for the research study, they are prompted to choose a strong passcode to decrypt their gift cards via the post-experiment questionnaire form, which they are required to complete using their pseudonymous participant number. In this form, they may also select their preferred gift card type from a list of options we offer (e.g., Amazon, IKEA). Participants are advised to keep their participant numbers and passcodes in a secure place so that others do not claim their card.\footnote{Notice that the researchers conducting the study also need to keep the participant numbers confidential. When stratification is done as discussed in Section~\ref{sec:experience-design}, the partitioning into groups should be communicated to students in a way that does not reveal any full participant numbers.}

Once we verify their successful participation, we publish all eligible gift cards to all students in the course in PDF format corresponding to their selected type. Each gift card is encrypted using AES-256~\citep{rijmen2001aes} with the strong passcode set by the corresponding participant, which is effectively infeasible to break. As a result, each eligible participant can anonymously locate their own gift card with their participant number and unlock their card with their passcode without revealing their identity to the research staff or other students. This procedure has been effective and secure, and to date, no participant has reported any issues with the gift card claiming process, such as difficulty finding or decrypting their gift card, or concerns about the security of their gift card files.

The only issue we encounter with giving out gift cards anonymously is compliance with tax regulations. Over time, the university's requirements have changed on the maximum value of gift cards we can distribute to study participants anonymously while remaining compliant with tax laws: compensation for participation to scientific experiments exceeding a limit now need to be handled as taxable income. As a result, we must either cap the value of each gift card per participant to that limit, or design some procedure to pass the participant's identities to the financial administration of our university. As we are the only research group that would need such a procedure, our university has no central administrative support for such a procedure. We would hence have to collect and pass those identities ourselves. Whatever mechanism we set up for doing this, we fear that this would lower the trust of the students in the pseudonymity of their participation. In the trade-off between financial motivation and trust in the ethical handling of the experiment, we have hence opted to lower the gift card values. 

\subsubsection{Course Organization and Content}
When faculty members integrate experiments in their courses like we did, 
a potential challenge is that they might be tempted to design their course not to optimize pedagogical outcomes, but to further their own research agenda. 

Psychology researchers also face this problem, which they often counter by having a collection of experiments running in the department. Faculty can in certain cases force students to pick one or two, or ask the students to complete alternative assignments towards obtaining the same competencies, but they cannot force the students to pick their own experiments. 

This option is not available in the typical CS or SE department, since they do not have enough experiments running at any one time, and it is certainly not an option for RE experiments, since we do not have enough faculty organizing RE courses and experiments. In theory, study program committees have oversight over the study programs and the courses being taught in them, and they approve the official course descriptions. In practice, the teaching professors have quite some leeway over how they organize their courses and which subjects and course material they include. This is especially the case for elective courses in master's study programs, which are to a large degree driven by the faculty's own research. University-level or faculty-level curriculum committees then often only rubberstamp the course descriptions and the detailed list of course topics. So in practice, the only solution to this problem is to ask the researchers to be aware of this challenge, to try to err on the safe side, and to be transparent.

In any case, integrating the experiments in a course, and then trying to train the students to be as representative of professionals as possible for the purpose of the experiments, implies that the teachers are selecting their course content based on the research hypotheses they want to test. 
One could argue that this is far from ideal, and that one may instead want to teach the students one set of topics, while testing a different set of hypotheses. 

We do not experience this as a problem, however. Elective courses at the master level such as our own course on Software Hacking and Protection almost always include two types of modules or topics. On the one hand, they include broad modules that provide the students an overview of the essential foundations and methodologies of the domain, and that train them to put those into practice. On the other hand, they include \emph{capita selecta} topics, i.e, advanced, specialized topics that are selected by the teacher among many possible topics to illustrate the complexity, depth, and strengths and weaknesses of parts of the state of the art rather than to offer a broad overview. It is in those specialized topic modules that the research-based nature of academic education shines, and it is there where faculty members have the most freedom to select the topics they consider exemplary for their domain.  
In such capita selecta modules in master programs, it can hence be perfectly fine to zoom in on precisely those topics, research questions, and hypothesis that are relevant for the planned experiments, thus achieving at once the pedagogical outcomes and interesting research outcomes. 

\subsection{Limitations and Possible Solutions}
We consider our approach to RE experiments with students as described in this section an interesting option to obtain valuable data and results that complements the types of results that can be obtained with the approaches surveyed in Section~\ref{sec:literature}. We also consider our approach to provide an interesting alternative to balance internal and external validity concerns when involving students in RE experiments. 

Nonetheless, it is important to note that our approach has significant limitations and drawbacks.
Most importantly, we consider this approach inappropriate for answering research questions regarding the best RE strategies (M1) and regarding the use of advanced tool functionality (M2) by experts. As discussed in Section~\ref{sec:experience-challenges}, the main area where our approach can contribute is the assessment of SP strength (M3), because individual SPs typically target specific RE attack steps, and experiments can be designed to let students perform only those steps. 

This hints at another limitation, namely that the approach might not be suited to study the real-world impact of \emph{layering} SPs. While layering a few obfuscating transformations was feasible for our experiments (if only because stripping binaries can be seen as an obfuscation compared to providing source code in which identifiers have not been obfuscated), studying advanced layerings of advanced protections is infeasible, as it would require too much time and too much strategic thinking for lab-time experiments with students. 

Overlapping the experiments with graded labs also introduces limitations to their design due to grading and time constraints, and it possibly creates a lack of intrinsic motivation by the students, thus introducing marked differences from a voluntary experiment conducted outside of a course setting. 

One possible solution to the student motivation problem is to make the experiment completely voluntary. From the academic year 2025-2026 onwards, the labs in our SHP course are no longer graded. They are now organized as pure learning activities, not as evaluation activities. For the evaluation, a practical exam is used in which the students have to demonstrate their practical SP and RE skills. This change was not driven by our desire to improve our experimental setting, but by feedback from students about the course organization, and their desire to have more unconstrained learning opportunities rather than being evaluated every week. Reorganizing the course as such required a considerable effort, due to which including additional experiments was not feasible this year. From next year onward, we plan to restart conducting experiments with our students, but then in labs that are not graded. As we will not help the students with the assignments during the experiments, those labs will then become training labs, rather than learning or evaluation activities, to which students can participate voluntarily in order to test and train themselves. 

 \section{Recommendations}
\label{sec:recommend}
Based on our own experience, we propose several recommendations aimed at enhancing the rigor, reproducibility, and external validity of future RE experiments with students.

\begin{itemize}
\item \textbf{Targeted Training for Bridging Competency Gaps.} Where a noticeable gap remains between the competencies of students and the skills a study assumes of professional reverse engineers, consider scheduling one or two  refresher sessions before the experiment.  These sessions can revisit compiler idioms, common obfuscation techniques, useful (often‑overlooked) tool features, or multi‑step attack planning---whichever topics the upcoming experiment actually rely on.  Depending on the RQs, such bridging sessions may be unnecessary, but when they are used their content should be reported in enough detail for others to judge external validity.
\item \textbf{Balanced Challenge Design.} Calibrate challenge difficulty through small pilot runs (e.g.\ with teaching assistants, PhD students, or volunteers from a previous cohort) and adjust the difficulty levels when necessary. Make sure that the pilot run participants match the eventual student participants' capabilities with respect to the type of code in the challenge binaries.
If two experimental rounds are planned, the first can be used as skill assessment that informs stratification in the second round.  The point is not to make tasks ever harder, but to ensure each challenge aligns with the study’s main RQs and with the realistic capabilities of the participant pool. Don't overestimate students' hacking and RE skills.
\item \textbf{Clearly Defined and Constrained Task Scope.}   Provide precise written objectives, success criteria, and constraints so that participants know exactly what a solution looks like.  When the RQs target a specific attack path, it can be useful to narrow the FTO (e.g., forbid brute‑force searching, restrict certain tools or features). To further limit the scope of tasks and to avoid that participants waste time on subtasks you do not aim to study, distribute helper artifacts such as pre‑annotated Ghidra/IDA databases and lightweight helper scripts.  

\item \textbf{Transparency and Detailed Documentation.} Make the complete set of research artifacts available, including: (i) the original source code, (ii) the exact build and protection toolchain with version numbers and used configuration options, and (iii) the generated binaries both before and after symbol stripping. Maintain transparency by documenting all aspects of the experiment setup, including participant selection criteria, training materials, task descriptions, tools provided, and any imposed operational constraints. Such documentation enhances reproducibility and allows for clearer assessment of external validity.

\item \textbf{Participant Motivation and Ethical Considerations.} Continue using ethically responsible motivation strategies such as offering appropriate incentives, ensuring pseudonymity, and integrating experimental tasks within course activities. These strategies not only improve student engagement and participation rates but also uphold ethical standards throughout the research process.
\end{itemize}

Implementing these recommendations, on top of recommendations from the field of SE applicable to RE experiments as discussed in Section~\ref{sec:use_students_in_SE}, can in our opinion and experience help to enhance the quality, reliability, and applicability of RE experiments involving students, contributing to more reliable findings and enhancing their relevance to real-world RE contexts.
 \section{Conclusions}
\label{sec:conclusions}

Empirical reverse engineering experiments with student participants can offer valuable insights into the effectiveness of software protections and the challenges faced by reverse engineers. We presented a systematic literature review on papers reporting RE experiments and user studies, as well as an overview of related work, mostly from the domain of software engineering research that also applies to adversarial reverse engineering research. In addition, we presented our experience with conducting such experiments in the context of a master-level software hacking and protection course, approaching the experiments in a way that allows us to collect data and insights that complement the experiments documented in the existing literature. On that basis, we formulated recommendations for future RE experiments with students.

While students offer a convenient and accessible participant pool, their varying skill levels and lack of professional experience require careful experimental design to maintain validity. Our work highlights the importance of aligning challenge complexity with student capabilities, providing targeted training, and minimizing unintended biases through quality assurance. By 
using tool-agnostic data collection methods like RevEngE and automated analysis tools like reAnalyst, researchers can obtain detailed, scalable insights without disrupting the natural workflow of participants.

The ethical considerations surrounding student participation, such as pseudo\-nymity, voluntariness, and motivation, are critical to maintaining the integrity of the research. Our protocols demonstrate that balancing educational objectives with research goals is achievable, provided transparency and ethical safeguards are prioritized. Future work should focus on developing standardized benchmarks and methodologies to further enhance the reproducibility and generalizability of empirical RE experiments. In summary, this paper has analyzed the complexities and considerations involved in conducting RE experiments with students, while recognizing the ongoing need to improve experimental design and participant training. 

 \section{Declaration}
\label{sec:declaration}

\subsection*{Funding}
We acknowledge support from The Research Foundation – Flanders (FWO) [Project nr.: 3G0E2318], from the Cybersecurity Research Program Flanders, and from the NSF under grants SATC/EDU-2029632 and SATC/TTP-1525820.

\subsection*{Ethical approval}
The Ethics Committee of the Faculty of Engineering and Architecture approved the experiments with master students reported in this paper.

\subsection*{Informed consent}
Prior to each experiment, participants signed an informed consent.

\subsection*{Author Contributions}
\begin{itemize}
    \item \textbf{Tab (Tianyi) Zhang}: Writing – original draft, Writing – review \& editing,  Validation, Software, Project administration, Methodology, Investigation,  Conceptualization, Visualization, 
    \item \textbf{Bjorn De Sutter}:  Writing – original draft, Writing – review \& editing, Supervision, Project administration, Methodology, Investigation, Conceptualization, Funding acquisition, Data Curation, Visualization
    \item \textbf{Christian Collberg}: Writing – review \& editing, Supervision, Methodology, Investigation, Conceptualization, Funding acquisition, Visualization
    \item \textbf{Bart Coppens}: Writing – review \& editing, Investigation, Supervision.
    \item \textbf{Waleed Mebane}: Writing – review \& editing, Validation, Methodology, Investigation
\end{itemize}

\subsection*{Data Availability Statement}

The artifacts of our experiment challenges are available at \url{https://github.com/csl-ugent/RE-experiment-artifacts}. The source code of the data collection and analysis software we used and developed is available at \url{https://github.com/csl-ugent/reVeal}. Data collected during the experiments can be made available upon reasonable request.

\subsection*{Conflict of Interest}
No conflict of interest.

\subsection*{Clinical Trial Number}
Clinical trial number: not applicable.

\bibliographystyle{spbasic}      \bibliography{references.bib}   \appendix

@article{Feitelson23,
author = {Feitelson, Dror G.},
title = {From Code Complexity Metrics to Program Comprehension},
year = {2023},
issue_date = {May 2023},
publisher = {Association for Computing Machinery},
address = {New York, NY, USA},
volume = {66},
number = {5},
issn = {0001-0782},
url = {https://doi.org/10.1145/3546576},
doi = {10.1145/3546576},
journal = {Commun. ACM},
month = apr,
pages = {52–61},
numpages = {10}
}

@article{caldwell2010strategies,
  title={Strategies for increasing recruitment to randomised controlled trials: systematic review},
  author={Caldwell, Patrina HY and Hamilton, Sana and Tan, Alvin and Craig, Jonathan C},
  journal={PLoS medicine},
  volume={7},
  number={11},
  pages={e1000368},
  year={2010},
  publisher={Public Library of Science San Francisco, USA}
}

@inproceedings{Biondi2006Silver,
 author = {Philippe Biondi and Fabrice Desclaux},
 title  = {{Silver needle in the skype}},
 booktitle = {{Black Hat Europe 2006 Proceedings}},
 pages = {25--47},
 year = {2006},
 url  = {https://blackhat.com/presentations/bh-europe-06/bh-eu-06-biondi/bh-eu-06-biondi-up.pdf}
}

@techreport{wangFlatteningTechReport,
	author = {Wang, Chenxi and Hill, Jonathan and Knight, John and Davidson, Jack},
	title = {Software Tamper Resistance: Obstructing Static Analysis of Programs},
	year = {2000},
	source = {http://www.ncstrl.org:8900/ncstrl/servlet/search?formname=detail\&id=oai%3Ancstrlh%3Auva_cs%3AUVA_CS%2F%2FCS-2000-12},
	institution = {University of Virginia},
	address = {Charlottesville, VA, USA},
}

@book{xchg_rax_rax,
    author = {xorpd},
    title = {{xchg rax,rax}},
    publisher = {CreateSpace Independent Publishing Platform},
    year = {2014}
}

@book{hackers_delight,
  title={Hacker's delight},
  author={Warren, Henry S},
  year={2013},
  publisher={Pearson Education}
}

@article{RAINER2022107002,
title = {Recruiting credible participants for field studies in software engineering research},
journal = {Information and Software Technology},
volume = {151},
pages = {107002},
year = {2022},
issn = {0950-5849},
OPTdoi = {https://doi.org/10.1016/j.infsof.2022.107002},
url = {https://www.sciencedirect.com/science/article/pii/S095058492200129X},
author = {Austen Rainer and Claes Wohlin}
}

@article{feitelson2022considerations,
  title={Considerations and pitfalls for reducing threats to the validity of controlled experiments on code comprehension},
  author={Feitelson, Dror G},
  journal={Empirical Software Engineering},
  volume={27},
  number={6},
  pages={123},
  year={2022},
  publisher={Springer}
}

@book{bryant2012understanding,
  title={Understanding how reverse engineers make sense of programs from assembly language representations},
  author={Bryant, Adam R},
  year={2012},
  publisher={Air Force Institute of Technology}
}

@inproceedings{wong2021inside,
  title={An inside look into the practice of malware analysis},
  author={Wong, Yong M. and Landen, M. and Antonakakis, M. and Blough, D. M. and Redmiles, E. M. and Ahamad, M.},
  booktitle={Proceedings of the 2021 ACM SIGSAC Conference on Computer and Communications Security},
  pages={3053--3069},
  year={2021},
  month={Nov}
}

@inproceedings{attack_graphs,
author = {Phillips, Cynthia and Swiler, Laura Painton},
title = {A graph-based system for network-vulnerability analysis},
year = {1998},
isbn = {1581131682},
publisher = {Association for Computing Machinery},
address = {New York, NY, USA},
url = {https://doi.org/10.1145/310889.310919},
OPTdoi = {10.1145/310889.310919},
booktitle = {Proceedings of the 1998 Workshop on New Security Paradigms},
pages = {71–79},
numpages = {9},
location = {Charlottesville, Virginia, USA},
series = {NSPW '98}
}

@article{attack_trees,
  title={Attack trees},
  author={Schneier, Bruce},
  journal={Dr. Dobb’s journal},
  volume={24},
  number={12},
  pages={21--29},
  year={1999}
}

@inproceedings {Votipka2020,
author = {Daniel Votipka and Seth Rabin and Kristopher Micinski and Jeffrey S. Foster and Michelle L. Mazurek},
title = {An Observational Investigation of Reverse {Engineers{\textquoteright}} Processes},
booktitle = {29th USENIX Security Symposium (USENIX Security 20)},
year = {2020},
isbn = {978-1-939133-17-5},
pages = {1875--1892},
url = {https://www.usenix.org/conference/usenixsecurity20/presentation/votipka-observational},
publisher = {USENIX Association},
month = aug
}

@book{wohlin2012experimentation,
  title     = {Experimentation in Software Engineering},
  author    = {Wohlin, Claes and Runeson, Per and H{\"o}st, Martin and
               Ohlsson, Magnus C. and Regnell, Bj{\"o}rn and Wessl{\'e}n, Anders},
  year      = {2012},
  publisher = {Springer},
  address   = {Berlin, Heidelberg},
  isbn      = {978-3-642-29044-2},
  doi       = {10.1007/978-3-642-29044-2}
}

@article{kuang2018enhance,
  title={Enhance virtual-machine-based code obfuscation security through dynamic bytecode scheduling},
  author={Kuang, Kaiyuan and Tang, Zhanyong and Gong, Xiaoqing and Fang, Dingyi and Chen, Xiaojiang and Wang, Zheng},
  journal={Computers \& Security},
  volume={74},
  pages={202--220},
  year={2018},
  publisher={Elsevier}
}

@book{collbergbook,
  title={Surreptitious Software: Obfuscation, Watermarking, and Tamperproofing for Software Protection: Obfuscation, Watermarking, and Tamperproofing for Software Protection},
  author={Nagra, Jasvir and Collberg, Christian},
  year={2009},
  publisher={Pearson Education}
}

@inproceedings{Mantovani2022,
  author    = {A. Mantovani and S. Aonzo and Y. Fratantonio and D. Balzarotti},
  title     = {{RE-Mind}: a First Look Inside the Mind of a Reverse Engineer},
  booktitle = {Proc. 31st USENIX Security Symposium (USENIX Security 22)},
  year      = {2022},
  pages     = {2727--2745}
}

@techreport{collberg1997taxonomy,
    author = {Collberg, Christian and Thomborson, C. and Low, Douglas},
    title = {A Taxonomy of Obfuscating Transformations},
    institution = {University of Auckland},
    year = {1997},
    month = {07},
    number = {148},
    url = {https://researchspace.auckland.ac.nz/handle/2292/3491},
}

@article{dongleprot_secmeasure,
    author = {Piazzalunga, Ugo and Salvaneschi, Paolo and Balducci, Francesco and Jacomuzzi, Pablo and Moroncelli, Cristiano},
    title = {Security Strength Measurement for Dongle-Protected Software},
    year = {2007},
    month = {12},
    doi = {10.1109/MSP.2007.176},
    publisher = {IEEE Educational Activities Department},
    volume = {5},
    number = {6},
    issn = {1558-4046},
    journal = {IEEE Security \& Privacy},
    pages = {32--40},
}

@inproceedings{2008twoardsexperimental,
    author = {Ceccato, Mariano and Di Penta, Massimiliano and Nagra, Jasvir and Falcarin, Paolo and Ricca, Filippo and Torchiano, Marco and Tonella, Paolo},
    title = {Towards Experimental Evaluation of Code Obfuscation Techniques},
    year = {2008},
    month = {10},
    doi = {10.1145/1456362.1456371},
    isbn = {9781605583211},
    booktitle = {Proceedings of the 4th ACM Workshop on Quality of Protection},
    publisher = {Association for Computing Machinery (ACM)},
    address = {New York, NY, USA},
    pages = {39--46},
    location = {Alexandria, Virginia, USA},
    series = {QoP '08},
}

@article{Tonella2007,
	author = {Tonella, Paolo and Torchiano, Marco and Du Bois, Bart and Syst{\"a}, Tarja},
	date = {2007/10/01},
	OPTdoi = {10.1007/s10664-007-9037-5},
	id = {Tonella2007},
	isbn = {1573-7616},
	journal = {Empirical Software Engineering},
	number = {5},
	pages = {551--571},
	title = {Empirical studies in reverse engineering: state of the art and future trends},
	url = {https://doi.org/10.1007/s10664-007-9037-5},
	volume = {12},
	year = {2007}}

@INPROCEEDINGS{2009assessment,
  author={Mariano Ceccato and Di Penta, Massimiliano  and Jasvir Nagra and Paolo Falcarin and Filippo Ricca and Marco Torchiano and Paolo Tonella},
  booktitle={IEEE ICPC}, 
  title={The effectiveness of source code obfuscation: An experimental assessment}, 
  year={2009},
  month={05},
  pages={178--187},
  doi={10.1109/ICPC.2009.5090041},
  issn={1092-8138},
}

@inproceedings{malware_visualcomp,
    title = {Visualizing compiled executables for malware analysis},
    author = {Quist, Daniel A and Liebrock, Lorie M},
    booktitle = {2009 6th International Workshop on Visualization for Cyber Security},
    pages = {27--32},
    year = {2009},
    doi = {10.1109/VIZSEC.2009.5375539},
}

@article{2014afamily,
	author = {Ceccato, Mariano and Di Penta, Massimiliano and Falcarin, Paolo and Ricca, Filippo and Torchiano, Marco and Tonella, Paolo},
	journal = {Empirical Software Engineering},
	number = {4},
	pages = {1040--1074},
	title = {A family of experiments to assess the effectiveness and efficiency of source code obfuscation techniques},
	volume = {19},
	year = {2014}
}

@INPROCEEDINGS{henry2020exploring,
  author={Henry, Wayne C. and Peterson, Gilbert L.},
  booktitle={2020 13th International Conference on Systematic Approaches to Digital Forensic Engineering (SADFE)}, 
  title={Exploring Provenance Needs in Software Reverse Engineering}, 
  year={2020},
  volume={},
  number={},
  pages={57-65},
  doi={10.1109/SADFE51007.2020.00008}}

@article{henry2020sensorre,
  title={{SensorRE}: Provenance support for software reverse engineers},
  author={Henry, Wayne C and Peterson, Gilbert L},
  journal={Computers \& Security},
  volume={95},
  pages={101865},
  year={2020},
  publisher={Elsevier}
}

@inproceedings {Wong2024,
author = {Miuyin Yong Wong and Matthew Landen and Frank Li and Fabian Monrose and Mustaque Ahamad},
title = {Comparing Malware Evasion Theory with Practice: Results from Interviews with Expert Analysts},
booktitle = {Twentieth Symposium on Usable Privacy and Security (SOUPS 2024)},
year = {2024},
isbn = {978-1-939133-42-7},
address = {Philadelphia, PA},
pages = {61--80},
url = {https://www.usenix.org/conference/soups2024/presentation/yong-wong},
publisher = {USENIX Association},
month = aug
}

@article{wagner2017knowledge,
  title={A knowledge-assisted visual malware analysis system: Design, validation, and reflection of KAMAS},
  author={Wagner, Markus and Rind, Alexander and Th{\"u}r, Niklas and Aigner, Wolfgang},
  journal={Computers \& Security},
  volume={67},
  pages={1--15},
  year={2017},
  publisher={Elsevier}
}

@article{Yamagishi2025,
	author = {Yamagishi, Rei and Fujii, Shota and Sato, Takayuki},
	date = {2025/05/22},
	OPTdoi = {10.1007/s10207-025-01059-3},
	id = {Yamagishi2025},
	isbn = {1615-5270},
	journal = {International Journal of Information Security},
	number = {3},
	pages = {137},
	title = {Cultivating skilled malware analysts: Clarification of practical malware dynamic analysis through interviews},
	url = {https://doi.org/10.1007/s10207-025-01059-3},
	volume = {24},
	year = {2025}}

@book{halstead1977elements,
  title={Elements of Software Science (Operating and programming systems series)},
  author={Halstead, Maurice H},
  year={1977},
  publisher={Elsevier Science Inc.}
}

@article{mccabe1976complexity,
  title={A complexity measure},
  author={McCabe, Thomas J},
  journal={IEEE Transactions on Software Engineering},
  number={4},
  volume={2},
  pages={308--320},
  year={1976},
  publisher={IEEE}
}

@article{baldwin2016,
	author = {Baldwin, Jennifer and Teh, Alvin and Baniassad, Elisa and van Rooy, Dirk and Coady, Yvonne},
	date = {2016/03/01},
	doi = {10.1007/s00766-014-0214-y},
	id = {Baldwin2016},
	isbn = {1432-010X},
	journal = {Requirements Engineering},
	number = {1},
	pages = {131--159},
	title = {Requirements for tools for comprehending highly specialized assembly language code and how to elicit these requirements},
	url = {https://doi.org/10.1007/s00766-014-0214-y},
	volume = {21},
	year = {2016}}

@INPROCEEDINGS{yakdan2016,
  author={Yakdan, Khaled and Dechand, Sergej and Gerhards-Padilla, Elmar and Smith, Matthew},
  booktitle={2016 IEEE Symposium on Security and Privacy (SP)}, 
  title={Helping Johnny to Analyze Malware: A Usability-Optimized Decompiler and Malware Analysis User Study}, 
  year={2016},
  volume={},
  number={},
  pages={158-177},
  doi={10.1109/SP.2016.18}}

@inproceedings {burk2022,
author = {Kevin Burk and Fabio Pagani and Christopher Kruegel and Giovanni Vigna},
title = {Decomperson: How Humans Decompile and What We Can Learn From It},
booktitle = {31st USENIX Security Symposium (USENIX Security 22)},
year = {2022},
isbn = {978-1-939133-31-1},
address = {Boston, MA},
pages = {2765--2782},
url = {https://www.usenix.org/conference/usenixsecurity22/presentation/burk},
publisher = {USENIX Association},
month = aug
}

@inproceedings{wettel2011,
author = {Wettel, Richard and Lanza, Michele and Robbes, Romain},
title = {Software systems as cities: a controlled experiment},
year = {2011},
isbn = {9781450304450},
publisher = {Association for Computing Machinery},
address = {New York, NY, USA},
url = {https://doi.org/10.1145/1985793.1985868},
doi = {10.1145/1985793.1985868},
booktitle = {Proceedings of the 33rd International Conference on Software Engineering},
pages = {551–560},
numpages = {10},
location = {Waikiki, Honolulu, HI, USA},
series = {ICSE '11}
}

@article{bryant2011software,
  title={Software Reverse Engineering as a Sensemaking Task.},
  author={Bryant, Adam R. and Mills, Robert F. and Peterson, Gilbert L. and Grimaila, Michael R.},
  journal={Journal of Information Assurance \& Security},
  volume={6},
  number={6},
  year={2011}
}

@article{bryant2013top,
  title={Top-level goals in reverse engineering executable software},
  author={Bryant, Adam R. and Mills, Robert F. and Grimaila, Michael R. and Peterson, Gilbert L.},
  journal={Journal of Information Warfare},
  volume={12},
  number={1},
  pages={32--43},
  year={2013},
  publisher={JSTOR}
}

@inproceedings{bryant2012eliciting,
  title={Eliciting a sensemaking process from verbal protocols of reverse engineers},
  author={Bryant, Adam R. and Mills, Robert F. and Peterson, Gilbert L. and Grimaila, Michael R.},
  booktitle={Proceedings of the Annual Meeting of the Cognitive Science Society},
  volume={34},
  year={2012}
}

@article{Feldt2018,
	author = {Feldt, Robert and Zimmermann, Thomas and Bergersen, Gunnar R. and Falessi, Davide and Jedlitschka, Andreas and Juristo, Natalia and M{\"u}nch, J{\"u}rgen and Oivo, Markku and Runeson, Per and Shepperd, Martin and Sj{\o}berg, Dag I. K. and Turhan, Burak},
	date = {2018/12/01},
	doi = {10.1007/s10664-018-9655-0},
	id = {Feldt2018},
	isbn = {1573-7616},
	journal = {Empirical Software Engineering},
	number = {6},
	pages = {3801--3820},
	title = {Four commentaries on the use of students and professionals in empirical software engineering experiments},
	url = {https://doi.org/10.1007/s10664-018-9655-0},
	volume = {23},
	year = {2018}}

@misc{skype,
  author       = {Ouanilo Medegan},
  title        = {Skype Reverse Engineering: The (long) journey},
  year         = {2012},
  howpublished = {\url{http://www.oklabs.net/skype-reverse-engineering-the-long-journey/}},
  note         = {Accessed: 2025-12-09}
}

@INPROCEEDINGS{2014another,
  author={Zhuang, Yan and Protsenko, Mykola and Muller, Tilo and Freiling, Felix C.},
  booktitle={2014 25th International Workshop on Database and Expert Systems Applications}, 
  title={An(other) Exercise in Measuring the Strength of Source Code Obfuscation}, 
  year={2014},
  volume={},
  number={},
  pages={313--317},
  doi={10.1109/DEXA.2014.69}
}

@inproceedings{viticchie2016assessment,
  title={Assessment of source code obfuscation techniques},
  author={Viticchi{\'e}, Alessio and Regano, Leonardo and Torchiano, Marco and Basile, Cataldo and Ceccato, Mariano and Tonella, Paolo and Tiella, Roberto},
  booktitle={2016 IEEE 16th international working conference on source code analysis and manipulation (SCAM)},
  pages={11--20},
  year={2016},
  organization={IEEE}
}

@inproceedings{2016comparing,
author = {Manikyam, Ramya and McDonald, J. Todd and Mahoney, William R. and Andel, Todd R. and Russ, Samuel H.},
title = {Comparing the Effectiveness of Commercial Obfuscators against MATE Attacks},
year = {2016},
isbn = {9781450348416},
publisher = {Association for Computing Machinery},
address = {New York, NY, USA},
url = {https://doi.org/10.1145/3015135.3015143},
OPTdoi = {10.1145/3015135.3015143},
booktitle = {Proceedings of the 6th Workshop on Software Security, Protection, and Reverse Engineering},
articleno = {8},
numpages = {11},
location = {Los Angeles, California, USA},
series = {SSPREW '16}
}

@inproceedings{obf_optvialangmods,
    author = {Liu, Han},
    title = {Towards Better Program Obfuscation: Optimization via Language Models},
    year = {2016},
    isbn = {9781450342056},
    publisher = {Association for Computing Machinery},
    address = {New York, NY, USA},
    url = {https://doi.org/10.1145/2889160.2891040},
    doi = {10.1145/2889160.2891040},
    booktitle = {Proceedings of the 38th International Conference on Software Engineering Companion},
    pages = {680--682},
    location = {Austin, Texas},
    series = {ICSE'16}
}

@inproceedings{Ceccato2017,
  author    = {Mariano Ceccato and Paolo Tonella and Cataldo Basile and Bart Coppens and De Sutter, Bjorn and Paolo Falcarin and Marco Torchiano},
  title     = {How professional hackers understand protected code while performing attack tasks},
  booktitle = {2017 IEEE/ACM 25th International Conference on Program Comprehension (ICPC)},
  year      = {2017},
  doi       = {10.1109/icpc.2017.2}
}

@article{2021inputoutput,
    author = {Zhao, Yujie and Tang, Zhanyong and Ye, Guixin and Gong, Xiaoqing and Fang, Dingyi and Tan, Zhiyuan},
    title = {Input-Output Example-Guided Data Deobfuscation on Binary},
    year = {2021},
    month = {01},
    doi = {10.1155/2021/4646048},
    journal = {Security and Communication Networks},
    issue_date = {2021},
    publisher = {John Wiley & Sons, Inc.},
    volume = {2021},
    issn = {1939-0114},
}

@article{moorthy2020recruiting,
  title={Recruiting psychology students to participate in faculty/department research: Ethical considerations and best practices},
  author={Moorthy, Gyan},
  journal={Voices in Bioethics},
  volume={6},
  year={2020}
}

@article{walker2020opportunity,
  title={The opportunity cost of compulsory research participation: Why psychology departments should abolish involuntary participant pools},
  author={Walker, Ruth},
  journal={Science and Engineering Ethics},
  volume={26},
  number={5},
  pages={2835--2847},
  year={2020},
  publisher={Springer}
}

@incollection{2020creative,
	OPTdoi = {10.1007/978-3-030-52581-1\_1},
	url = {https://doi.org/10.1007/F978-3-030-52581-1\_1},
	year = 2020,
	publisher = {Springer International Publishing},
	pages = {3--8},
	author = {Salsabil Hamadache and Malte Elson},
	title = {Creative Manual Code Obfuscation as a Countermeasure Against Software Reverse Engineering},
	booktitle = {AISC}
}

@article{2020splitting,
	author = {Viticchi{\'e}, Alessio and Regano, Leonardo and Basile, Cataldo and Torchiano, Marco and Ceccato, Mariano and Tonella, Paolo},
	journal = {Empirical Software Engineering},
	number = {1},
	pages = {1--48},
	title = {Empirical assessment of the effort needed to attack programs protected with client/server code splitting},
	volume = {25},
	year = {2020}
}

@InProceedings{exp_eval_obf_against_reveng,
    author = {BinShamlan, Mohammed H. and Alaidaroos, Alawi S. and Bin Merdhah, Mansoor H. and Bamatraf, Mohammed A. and Zain, Adnan A.},
    title = {Experimental Evaluation of the Obfuscation Techniques Against Reverse Engineering},
    year = {2021},
    doi = {10.1007/978-981-15-6048-4\_33},
    isbn = {978-981-15-6048-4},
    booktitle = {Advances on Smart and Soft Computing},
    editor = {Saeed, Faisal and Al-Hadhrami, Tawfik and Mohammed, Fathey and Mohammed, Errais},
    publisher = {Springer Singapore},
    address = {Singapore},
    pages = {383--390}
}

@article{podsakoff2012sources,
  title={Sources of method bias in social science research and recommendations on how to control it},
  author={Podsakoff, Philip M and MacKenzie, Scott B and Podsakoff, Nathan P},
  journal={Annual review of psychology},
  volume={63},
  number={1},
  pages={539--569},
  year={2012},
  publisher={Annual Reviews}
}

@article{Ceccato2019,
  author    = {Ceccato, Mariano and Tonella, Paolo and Basile, Cataldo and Falcarin, Paolo and Torchiano, Marco and Coppens, Bart and De Sutter, Bjorn},
  title     = {Understanding the behaviour of hackers while performing attack tasks in a professional setting and in a public challenge},
  journal   = {Empirical Software Engineering},
  volume    = {24},
  number    = {1},
  pages     = {240--286},
  year      = {2019},
  doi       = {10.1007/s10664-018-9625-6}
}

@INPROCEEDINGS{2019impact,
  author={BinShamlan, Mohammed H. and Bamatraf, Mohammed A. and Zain, Adnan A.},
  booktitle={2019 First International Conference of Intelligent Computing and Engineering (ICOICE)}, 
  title={The Impact of Control Flow Obfuscation Technique on Software Protection Against Human Attacks}, 
  year={2019},
  volume={},
  number={},
  pages={1-5},
  doi={10.1109/ICOICE48418.2019.9035187}
}

@ARTICLE{2019resilient,
  author={Zeng, Qiang and Luo, Lannan and Qian, Zhiyun and Du, Xiaojiang and Li, Zhoujun and Huang, Chin-Tser and Farkas, Csilla},
  journal={IEEE Transactions on Dependable and Secure Computing}, 
  title={Resilient User-Side Android Application Repackaging and Tampering Detection Using Cryptographically Obfuscated Logic Bombs}, 
  year={2019},
  volume={},
  number={},
  pages={1--1},
  doi={10.1109/TDSC.2019.2957787}
}

@inproceedings{berander2004,
  title={Using students as subjects in requirements prioritization},
  author={Berander, Patrik},
  booktitle={Proceedings. 2004 International Symposium on Empirical Software Engineering, 2004. ISESE'04.},
  pages={167--176},
  year={2004},
  organization={IEEE}
}

@inproceedings{salman2015,
  title={Are students representatives of professionals in software engineering experiments?},
  author={Salman, Iflaah and Misirli, Ayse Tosun and Juristo, Natalia},
  booktitle={2015 IEEE/ACM 37th IEEE international conference on software engineering},
  volume={1},
  pages={666--676},
  year={2015},
  organization={IEEE}
}

@article{host2000,
  title={Using students as subjects—a comparative study of students and professionals in lead-time impact assessment},
  author={H{\"o}st, Martin and Regnell, Bj{\"o}rn and Wohlin, Claes},
  journal={Empirical Software Engineering},
  volume={5},
  pages={201--214},
  year={2000},
  publisher={Springer}
}

@article{porter95,
  title={Comparing detection methods for software requirements inspections: A replicated experiment},
  author={Porter, Adam A and Votta, Lawrence G and Basili, Victor R},
  journal={IEEE Transactions on software Engineering},
  volume={21},
  number={6},
  pages={563--575},
  year={1995},
  publisher={IEEE}
}

@article{porter98,
  title={Comparing detection methods for software requirements inspections: A replication using professional subjects},
  author={Porter, Adam and Votta, Lawrence},
  journal={Empirical software engineering},
  volume={3},
  pages={355--379},
  year={1998},
  publisher={Springer}
}

@inproceedings{hansch2018programming,
  title={Programming experience might not help in comprehending obfuscated source code efficiently},
  author={H{\"a}nsch, Norman and Schankin, Andrea and Protsenko, Mykolai and Freiling, Felix and Benenson, Zinaida},
  booktitle={Fourteenth Symposium on Usable Privacy and Security (SOUPS 2018)},
  pages={341--356},
  year={2018}
}

@inproceedings{obf_googleplay,
    author = {Wermke, Dominik and Huaman, Nicolas and Acar, Yasemin and Reaves, Bradley and Traynor, Patrick and Fahl, Sascha},
    title = {A Large Scale Investigation of Obfuscation Use in Google Play},
    year = {2018},
    isbn = {9781450365697},
    publisher = {Association for Computing Machinery},
    address = {New York, NY, USA},
    url = {https://doi.org/10.1145/3274694.3274726},
    OPTdoi = {10.1145/3274694.3274726},
    booktitle = {Proceedings of the 34th Annual Computer Security Applications Conference},
    pages = {222--235},
    location = {San Juan, PR, USA},
    series = {ACSAC '18}
}

@INPROCEEDINGS{liu2017stochastic, 
    author={H. Liu and C. Sun and Z. Su and Y. Jiang and M. Gu and J. Sun}, 
    booktitle={2017 IEEE/ACM 39th International Conference on Software Engineering (ICSE)}, 
    title={Stochastic Optimization of Program Obfuscation}, 
    year={2017}, 
    volume={}, 
    number={}, 
    pages={221-231}, 
    doi={10.1109/ICSE.2017.28}, 
    ISSN={1558-1225}, 
    month={05},
}

@article{sutherland2006reverse,
  author    = {Iain Sutherland and George E. Kalb and Andrew Blyth and Gaius Mulley},
  title     = {An empirical examination of the reverse engineering process for binary files},
  journal   = {Computers \& Security},
  volume    = {25},
  number    = {3},
  pages     = {221--228},
  year      = {2006},
}

@INPROCEEDINGS{Ishida2019,
  author={Ishida, Toyomi and Uwano, Hidetake},
  booktitle={2019 IEEE/ACM 6th International Workshop on Eye Movements in Programming (EMIP)},
  title={Synchronized Analysis of Eye Movement and EEG during Program Comprehension},
  year={2019},
  volume={},
  number={},
  pages={26-32},
  doi={10.1109/EMIP.2019.00012}}

@article{Nunkoosing2005,
  title={The problems with interviews},
  author={Nunkoosing, Karl},
  journal={Qualitative Health Research},
  volume={15},
  number={5},
  pages={698--706},
  year={2005},
  publisher={SAGE Publications Sage CA: Los Angeles, CA}
}

@inproceedings{rijmen2001aes,
  author    = {Vincent Rijmen and Joan Daemen},
  title     = {Advanced Encryption Standard},
  booktitle = {Proceedings of Federal Information Processing Standards Publications, National Institute of Standards and Technology},
  year      = {2001},
  volume    = {19},
  pages     = {22}
}

@article{SingerVinson2002,
  title     = {Ethical Issues in Empirical Studies of Software Engineering},
  author    = {Singer, Janice and Vinson, Norman G.},
  journal   = {{IEEE} Transactions on Software Engineering},
  volume    = {28},
  number    = {12},
  pages     = {1171--1180},
  year      = {2002},
  publisher = {{IEEE}},
}

@article{falessi2018empirical,
  title={Empirical software engineering experts on the use of students and professionals in experiments},
  author={Falessi, Davide and Juristo, Natalia and Wohlin, Claes and Turhan, Burak and M{\"u}nch, J{\"u}rgen and Jedlitschka, Andreas and Oivo, Markku},
  journal={Empirical Software Engineering},
  volume={23},
  pages={452--489},
  year={2018},
  publisher={Springer}
}

@article{ko2015practical,
  title={A practical guide to controlled experiments of software engineering tools with human participants},
  author={Ko, Amy J and LaToza, Thomas D and Burnett, Margaret M},
  journal={Empirical Software Engineering},
  volume={20},
  pages={110--141},
  year={2015},
  publisher={Springer}
}

@article{lethbridge2005studying,
  title={Studying software engineers: Data collection techniques for software field studies},
  author={Lethbridge, Timothy C. and Sim, Susan E. and Singer, Janice},
  journal={Empirical Software Engineering},
  volume={10},
  number={3},
  pages={311--341},
  year={2005},
  publisher={Springer}
}

@inproceedings{alharbi2025designing,
  title={Designing a Reverse Engineering Tool for Functional Requirements Prioritization},
  author={Alharbi, Ohoud and Alrashed, Ashwaq and Albogobar, Sara},
  booktitle={2025 IEEE 5th International Conference on Software Engineering and Artificial Intelligence (SEAI)},
  pages={297--302},
  year={2025},
  organization={IEEE}
}

@ARTICLE{ebad2021measuring,
	author = {Ebad, Shouki A. and Darem, Abdulbasit A. and Abawajy, Jemal H.},
	title = {Measuring Software Obfuscation Quality–A Systematic Literature Review},
	year = {2021},
	doi = {10.1109/ACCESS.2021.3094517},
	journal = {IEEE Access},
	volume = {9},
	pages = {99024--99038},
}

@phdthesis{henry2020analytic,
  title={Analytic provenance for software reverse engineers},
  author={Henry, Wayne C},
  school = {Air Force Institute of Technology},
  year={2020}
}

@phdthesis{wong2022investigating,
  title={Investigating Collaboration in Software Reverse Engineering},
  author={Wong, Allison M},
  school = {Air Force Institute of Technology},
  year={2022}
}

@inproceedings{bergmann2025potential,
  title={Potential Analysis of Software Obfuscation to Protect Unmanned Systems against Forensic Analysis},
  author={Bergmann, N and Padilla, E and Bauer, J},
  booktitle={Proceedings of the 1st International Conference on Drones and Unmanned Systems},
  pages={266--272},
  year={2025}
}

@techreport{leger2021exploring,
  title={Exploring Explicit Uncertainty for Binary Analysis (EUBA)},
  author={Leger, Michelle A and Darling, Michael C and Jones, Stephen T and Matzen, Laura E and Stracuzzi, David J and Wilson, Andrew T and Bueno, Denis and Christentsen, Matthew and Ginaldi, Melissa and Hannasch, David and others},
  year={2021},
  institution={Sandia National Lab.(SNL-NM), Albuquerque, NM (United States)}
}

@techreport{matzen2021effects,
  title={Effects of Precise and Imprecise Value-Set Analysis (VSA) Information on Manual Code Analysis},
  author={Matzen, Laura E and Leger, Michelle A and Reedy, Geoffrey Edward},
  year={2021},
  institution={Sandia National Lab.(SNL-NM), Albuquerque, NM (United States)}
}

@phdthesis{Taylor2022,
  author    = {C. Taylor},
  title     = {Remotely Observing Reverse Engineers to Evaluate Software Protection},
  school    = {The University of Arizona},
  year      = {2022}
}

@inproceedings{taylor2019getting,
  author    = {Claire Taylor and Christian Collberg},
  title     = {Getting RevEngE: a system for analyzing reverse engineering behavior},
  booktitle = {14th International Conference on Malicious and Unwanted Software (MALCON)},
  year      = {2019}
}

@book{struts,
  title={Statistical rules of thumb},
  author={Van Belle, Gerald},
  year={2011},
  publisher={John Wiley \& Sons}
}

@inproceedings{Liem08,
author = {Liem, Clifford and Gu, Yuan Xiang and Johnson, Harold},
title = {A compiler-based infrastructure for software-protection},
year = {2008},
isbn = {9781595939364},
publisher = {Association for Computing Machinery},
address = {New York, NY, USA},
url = {https://doi.org/10.1145/1375696.1375702},
OPTdoi = {10.1145/1375696.1375702},
booktitle = {Proceedings of the Third ACM SIGPLAN Workshop on Programming Languages and Analysis for Security},
pages = {33–44},
numpages = {12},
location = {Tucson, AZ, USA},
series = {PLAS '08}
}

@InProceedings{java_obfuscation,
author="Batchelder, Michael
and Hendren, Laurie",
editor="Krishnamurthi, Shriram
and Odersky, Martin",
title="Obfuscating Java: The Most Pain for the Least Gain",
booktitle="Compiler Construction",
year="2007",
publisher="Springer Berlin Heidelberg",
address="Berlin, Heidelberg",
pages="96--110",
isbn="978-3-540-71229-9"
}

@article{zhang2024reanalyst,
title = {reAnalyst: Scalable annotation of reverse engineering activities},
journal = {Journal of Systems and Software},
pages = {112492},
year = {2025},
issn = {0164-1212},
doi = {https://doi.org/10.1016/j.jss.2025.112492},
author = {Tab Tianyi Zhang and Claire Taylor and Bart Coppens and Waleed Mebane and Christian Collberg and Bjorn {De Sutter}}
}

@inproceedings{faingnaert2024tools,
  author    = {Faingnaert, T. and Zhang, T. and Van Iseghem, W. and Everaert, G. and Coppens, B. and Collberg, C. and De Sutter, B.},
  title     = {Tools and Models for Software Reverse Engineering Research},
  booktitle = {Proceedings of the 2024 Workshop on Research on Offensive and Defensive Techniques in the Context of Man At The End (MATE) Attacks},
  year      = {2024},
  month     = {November},
  pages     = {44--58},
  url = {https://doi.org/10.1145/3689934.3690817},
  doi = {10.1145/3689934.3690817}
}

@inproceedings{nyre2022task,
  title={A task analysis of static binary reverse engineering for security},
  author={Nyre-Yu, Megan and Butler, Karin and Bolstad, Cheryl},
  year={2022},
  booktitle = {Proceedings of the 55th Hawaii International Conference on System Sciences},
  pages = {2187--2196}
}

@article{2018_diversification_and_obfuscation,
  title={Diversification and obfuscation techniques for software security: A systematic literature review},
  author={Hosseinzadeh, Shohreh and Rauti, Sampsa and Laur{\'e}n, Samuel and M{\"a}kel{\"a}, Jari-Matti and Holvitie, Johannes and Hyrynsalmi, Sami and Lepp{\"a}nen, Ville},
  journal={Information and Software Technology},
  volume = 104,
  year={2018},
  month={12},
  publisher={Elsevier},
  doi={10.1016/j.infsof.2018.07.007},
  url = "http://www.sciencedirect.com/science/article/pii/S0950584918301484",
}

@article{schrittwieser2016protecting,
	author = {Schrittwieser, Sebastian and Katzenbeisser, Stefan and Kinder, Johannes and Merzdovnik, Georg and Weippl, Edgar},
	title = {Protecting Software Through Obfuscation: Can It Keep Pace with Progress in Code Analysis?},
	journal = {ACM Computing Surveys (CSUR)},
	issue_date = {July 2016},
	volume = {49},
	number = {1},
	month = {04},
	year = {2016},
	issn = {0360-0300},
	pages = {4:1--4:37},
	articleno = {4},
	numpages = {37},
	url = {http://doi.acm.org/10.1145/2886012},
	OPTdoi = {10.1145/2886012},
	acmid = {2886012},
	publisher = {ACM},
	address = {New York, NY, USA}
}

@article{SRBvEDPS_T557_20,
  author = {{General Court (Eighth Chamber, Extended Composition)}},
  title = {{Single} {Resolution} {Board} v {European} {Data} {Protection} {Supervisor}},
  year = {2023},
  month = {4},
  note = {{Case T-557/20; [2024] 1 C.M.L.R. 46}},
  journal = {Common Market Law Review},
  volume = {1},
  pages = {46},
  publisher = {General Court of the European Union},
  url = {https://eur-lex.europa.eu/legal-content/EN/TXT/HTML/?uri=CELEX:62020TJ0557_RES}
}

@techreport{menlo-report,
  author = {Dittrich, D and Kenneally, E},
  title = {{The Menlo Report: Ethical Principles Guiding Information and Communication Technology Research}},
  institution = {U.S. Department of Homeland Security},
  year = {2012},
  month = {August},
  doi = {https://catalog.caida.org/paper/2012_menlo_report_actual_formatted}
}

@article{desutter2024evaluation,
author = {De Sutter, Bjorn and Schrittwieser, Sebastian and Coppens, Bart and Kochberger, Patrick},
title = {Evaluation Methodologies in Software Protection Research},
year = {2025},
issue_date = {April 2025},
publisher = {ACM},
address = {New York, NY, USA},
volume = {57},
number = {4},
issn = {0360-0300},
url = {https://doi.org/10.1145/3702314},
doi = {10.1145/3702314},
journal = {ACM Comput. Surv.},
month = dec,
articleno = {86},
numpages = {41}
}

@article{Basile23,
title = {Design, implementation, and automation of a risk management approach for man-at-the-End software protection},
journal = {Computers \& Security},
volume = {132},
pages = {103321},
year = {2023},
issn = {0167-4048},
doi = {https://doi.org/10.1016/j.cose.2023.103321},
url = {https://www.sciencedirect.com/science/article/pii/S0167404823002316},
author = {Cataldo Basile and Bjorn {De Sutter} and Daniele Canavese and Leonardo Regano and Bart Coppens}
}

@article{dsr,
  title={Design science in information systems research},
  author={Hevner, Alan R and March, Salvatore T and Park, Jinsoo and Ram, Sudha},
  journal={MIS quarterly},
  pages={75--105},
  year={2004},
  publisher={JSTOR}
}

@book{Wohlin,
	author = {Wohlin, C. and Runeson, P. and H\"ost, M. and Ohlsson, M.C. and Regnell, B. and Wessl\'en, A.},
	publisher = {Kluwer Academic Publishers},
	title = {Experimentation in Software Engineering},
        edition = {2nd},
	year = {2024}
}

@article{jens21,
    author = {Van den Broeck, Jens and Coppens, Bart and De Sutter, Bjorn},
    title = {Obfuscated integration of software protections},
    year = {2021},
    month = {02},
    doi = {10.1007/s10207-020-00494-8},
    journal = {Int'l Journal of Information Security},
    pages = {73--101},
    volume = {20},
}

@inproceedings{linn2003obfuscation,
  title={Obfuscation of executable code to improve resistance to static disassembly},
  author={Linn, Cullen and Debray, Saumya},
  booktitle={Proceedings of the 10th ACM conference on Computer and communications security},
  pages={290--299},
  year={2003}
}

@article{jens2022flexible,
  title={Flexible software protection},
  author={Van den Broeck, Jens and Coppens, Bart and De Sutter, Bjorn},
  journal={Computers \& Security},
  volume={116},
  pages={102636},
  year={2022},
  publisher={Elsevier}
}

@InProceedings{CybersecurityCurricularGuidelines,
author="Bishop, Matt
and Burley, Diana
and Buck, Scott
and Ekstrom, Joseph J.
and Futcher, Lynn
and Gibson, David
and Hawthorne, Elizabeth K.
and Kaza, Siddharth
and Levy, Yair
and Mattord, Herbert
and Parrish, Allen",
editor="Bishop, Matt
and Futcher, Lynn
and Miloslavskaya, Natalia
and Theocharidou, Marianthi",
title="Cybersecurity Curricular Guidelines",
booktitle="Information Security Education for a Global Digital Society",
year="2017",
publisher="Springer International Publishing",
address="Cham",
pages="3--13",
isbn="978-3-319-58553-6"
}

@inproceedings {dramko2024decompiler,
author = {Luke Dramko and Jeremy Lacomis and Edward J. Schwartz and Bogdan Vasilescu and Claire Le Goues},
title = {A Taxonomy of C Decompiler Fidelity Issues},
booktitle = {33rd USENIX Security Symposium (USENIX Security 24)},
year = {2024},
isbn = {978-1-939133-44-1},
address = {Philadelphia, PA},
pages = {379--396},
url = {https://www.usenix.org/conference/usenixsecurity24/presentation/dramko},
publisher = {USENIX Association},
month = aug
}

@book{ACMguidelines,
author = {Joint Task Force on Cybersecurity Education},
title = {Cybersecurity Curricula 2017: Curriculum Guidelines for Post-Secondary Degree Programs in Cybersecurity},
year = {2018},
isbn = {9781450389198},
publisher = {Association for Computing Machinery},
address = {New York, NY, USA}
}

@article{schinske2014teaching,
  title={Teaching more by grading less (or differently)},
  author={Schinske, Jeffrey and Tanner, Kimberly},
  journal={CBE—Life Sciences Education},
  volume={13},
  number={2},
  pages={159--166},
  year={2014},
  publisher={American Society for Cell Biology}
}

@article{gillis2019reconceptualizing,
  title={Reconceptualizing participation grading as skill building},
  author={Gillis, Alanna},
  journal={Teaching Sociology},
  volume={47},
  number={1},
  pages={10--21},
  year={2019},
  publisher={SAGE Publications Sage CA: Los Angeles, CA}
}

@article{chikofsky2002reverse,
  title={Reverse engineering and design recovery: A taxonomy},
  author={Chikofsky, Elliot J. and Cross, James H},
  journal={IEEE software},
  volume={7},
  number={1},
  pages={13--17},
  year={2002},
  publisher={IEEE}
}

@article{kitchenham2009systematic,
  title={Systematic literature reviews in software engineering--a systematic literature review},
  author={Kitchenham, Barbara and Brereton, O Pearl and Budgen, David and Turner, Mark and Bailey, John and Linkman, Stephen},
  journal={Information and software technology},
  volume={51},
  number={1},
  pages={7--15},
  year={2009},
  publisher={Elsevier}
}
\appendix
\section{Post-Experiment Questionnaire (2024)}
\label{postsurvey2024}
This is the post-questionnaire used for the student experiment conducted in 2024. Questionnaires from other years are similar, with mostly identical questions, although some variations exist depending on the experiment task and tools used.

\begin{enumerate}
    \item \textbf{Pseudonymous Participant Number:}\\
    Please provide your assigned participant number here.
    
    \item \textbf{Text Submitted on Ufora:}\\
    Copy and paste the text you submitted on Ufora for all tasks below. For each of the four binaries you tackled, include the following lines in your report (replacing the placeholder fields as indicated):

    \begin{verbatim}
    <task_number> start: <start_time>
    <task_number> end:   <end_time>
    <task_number> command-line: <command>
    \end{verbatim}

    Where:
    \begin{itemize}
        \item \texttt{<task\_number>} is replaced by your task number (e.g., 1a, 1b, 2a, etc.).
        \item \texttt{<start\_time>} is replaced by the time (in hours and minutes) when you started working on that binary.
        \item \texttt{<end\_time>} is replaced by the time (in hours and minutes) when you finished working on that binary.
        \item \texttt{<command>} is replaced by the exact command line you used to execute the binary on your best solution.
    \end{itemize}

    If your best solution prints \texttt{success}, simply provide that command line. If you did not find a complete solution (and gave up), provide the command line with the input you think is closest to an actual solution (e.g., the attempt that yields the most correct bits in the resulting value).

    \item \textbf{Difficulty Rating (1--10):}\\
    On a scale of 1 (very easy) to 10 (extremely difficult), how hard was this lab?

\item \textbf{Use of Ghidra Features:}\\
Which features of Ghidra did you use during the experiment? Select all that apply.
\begin{itemize}
    \item Bookmarks
    \item Comments
    \item Defined strings
    \item Variable renaming
    \item \dots
\end{itemize}
(Note: This is a sample list. The actual survey includes a more comprehensive set of features.)

    \item \textbf{Optional Feedback:}\\
    Please use this space to share any additional feedback about the experiment. Feel free to include comments on the tasks, tools, instructions, or anything else you would like us to know.
\end{enumerate}

\end{document}